\documentclass[11pt,a4paper]{article}
\usepackage[utf8]{inputenc}
\usepackage[T1]{fontenc}
\usepackage[shortlabels]{enumitem}
\usepackage{amsmath}
\usepackage{fullpage}
\usepackage{hyperref}
\usepackage{authblk}
\usepackage{graphicx}

\renewcommand{\max}{\mathrm{max}}
\newcommand{\edges}{m}
\newcommand{\colors}{k}

\title{Conflict Optimization for Binary CSP Applied to\texorpdfstring{\\}{} Minimum Partition into Plane Subgraphs and Graph Coloring}

\author[1]{Loïc Crombez}
\author[2]{Guilherme D. da Fonseca}
\author[3]{Florian Fontan}
\author[1]{Yan Gerard}
\author[2]{Aldo Gonzalez-Lorenzo}
\author[1]{Pascal Lafourcade}
\author[1]{Luc Libralesso}
\author[3]{Benjamin Momège}
\author[4]{Jack Spalding-Jamieson}
\author[3]{Brandon Zhang}
\author[5]{Da Wei Zheng}

\affil[1]{LIMOS, Université Clermont Auvergne, France}
\affil[2]{LIS, Aix-Marseille Université, France}
\affil[3]{Independent Researcher}
\affil[4]{David R. Cheriton School of Computer Science, University of Waterloo, Canada}
\affil[5]{Department of Computer Science, University of Illinois at Urbana-Champaign, USA}

\date{}

\begin{document}

\maketitle

\begin{abstract}
CG:SHOP is an annual geometric optimization challenge and the 2022 edition proposed the problem of coloring a certain geometric graph defined by line segments. Surprisingly, the top three teams used the same technique, called conflict optimization. This technique has been introduced in the 2021 edition of the challenge, to solve a coordinated motion planning problem. In this paper, we present the technique in the more general framework of binary constraint satisfaction problems (binary CSP). Then, the top three teams describe their different implementations of the same underlying strategy. We evaluate the performance of those implementations to vertex color not only geometric graphs, but also other types of graphs.
\end{abstract}

\section{Introduction} \label{s:intro}

\begin{figure}[p]
  \centering
  \includegraphics[scale=.25]{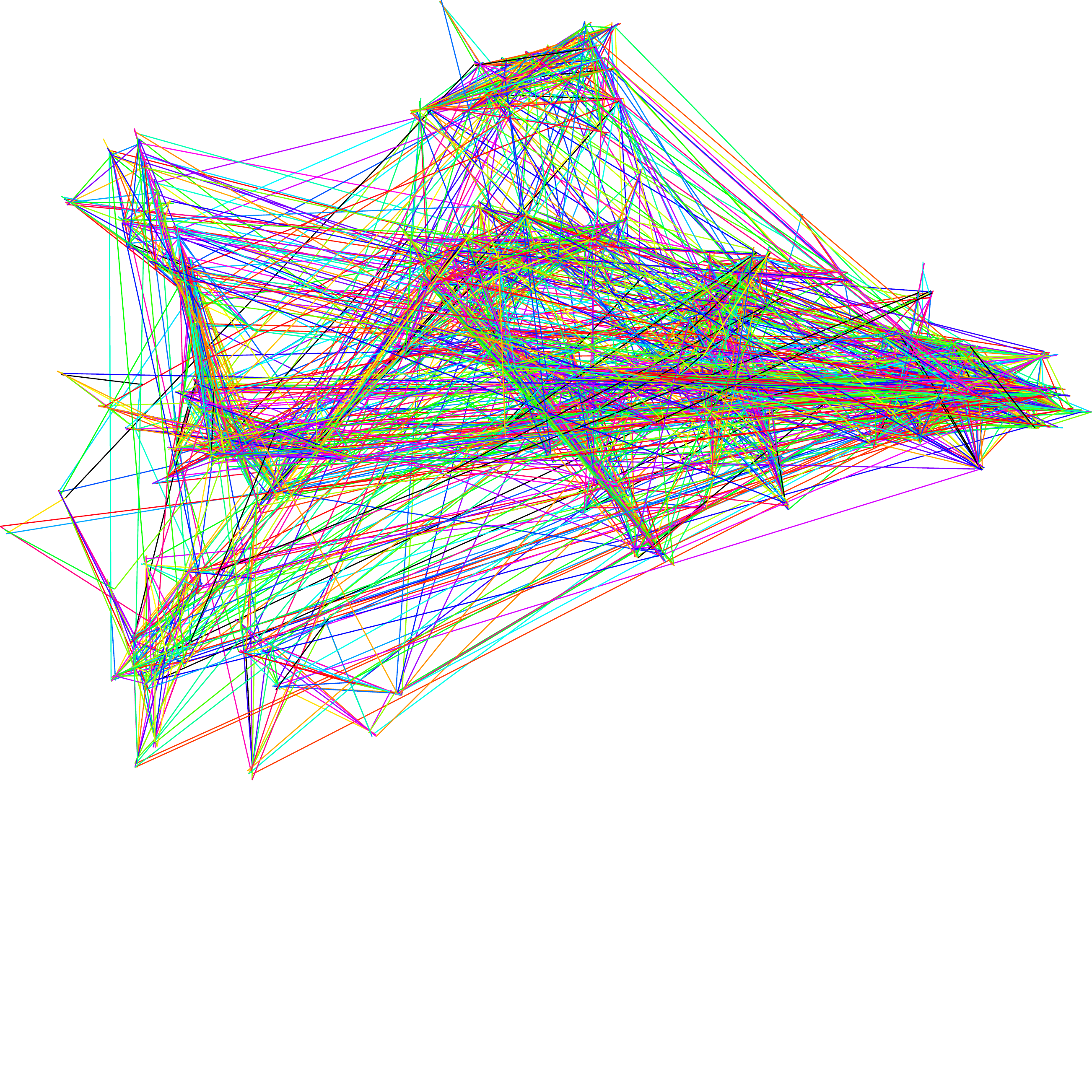}
  \hspace{5mm}
  \includegraphics[scale=.25]{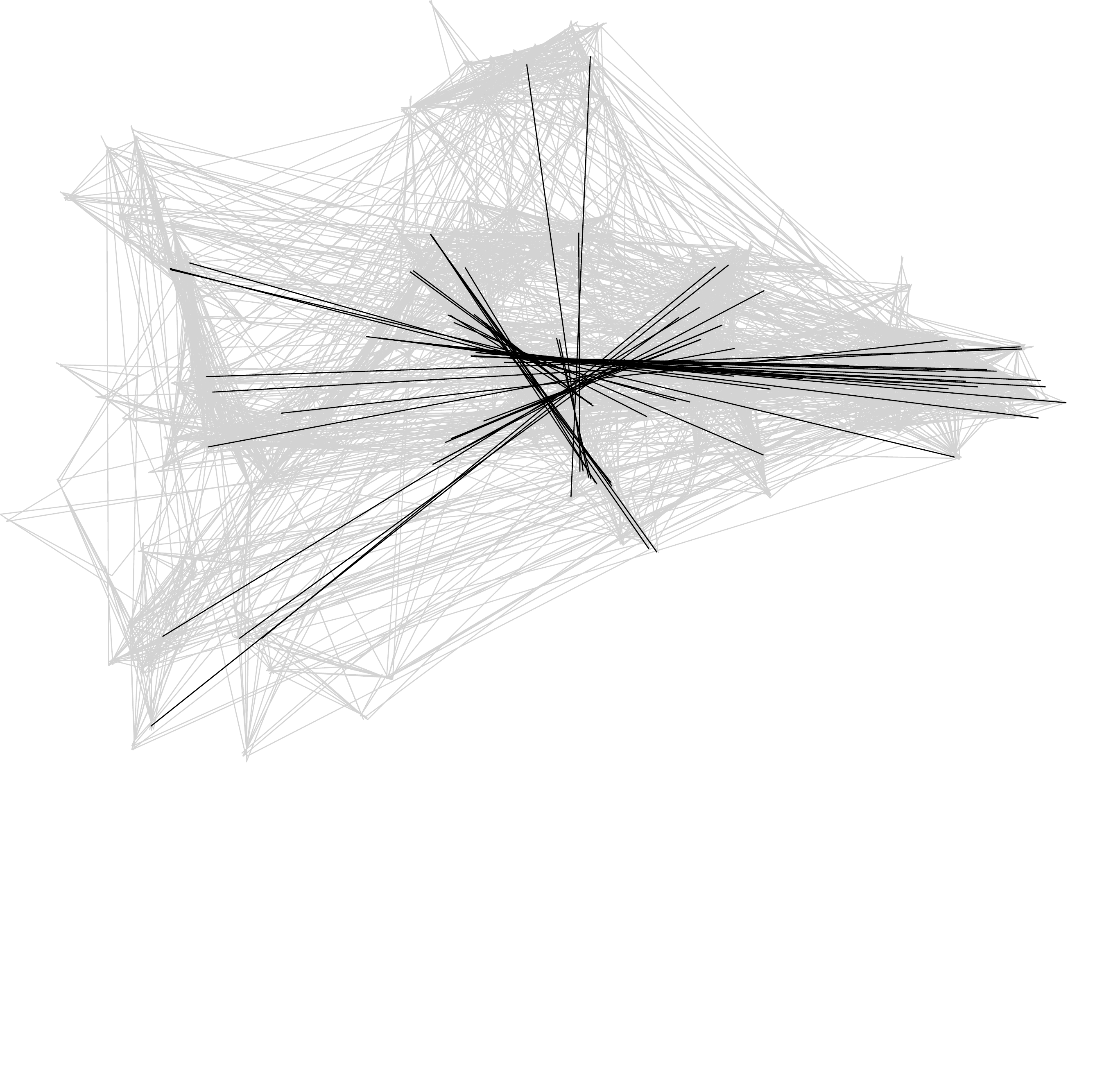}
  
  \includegraphics[scale=.063]{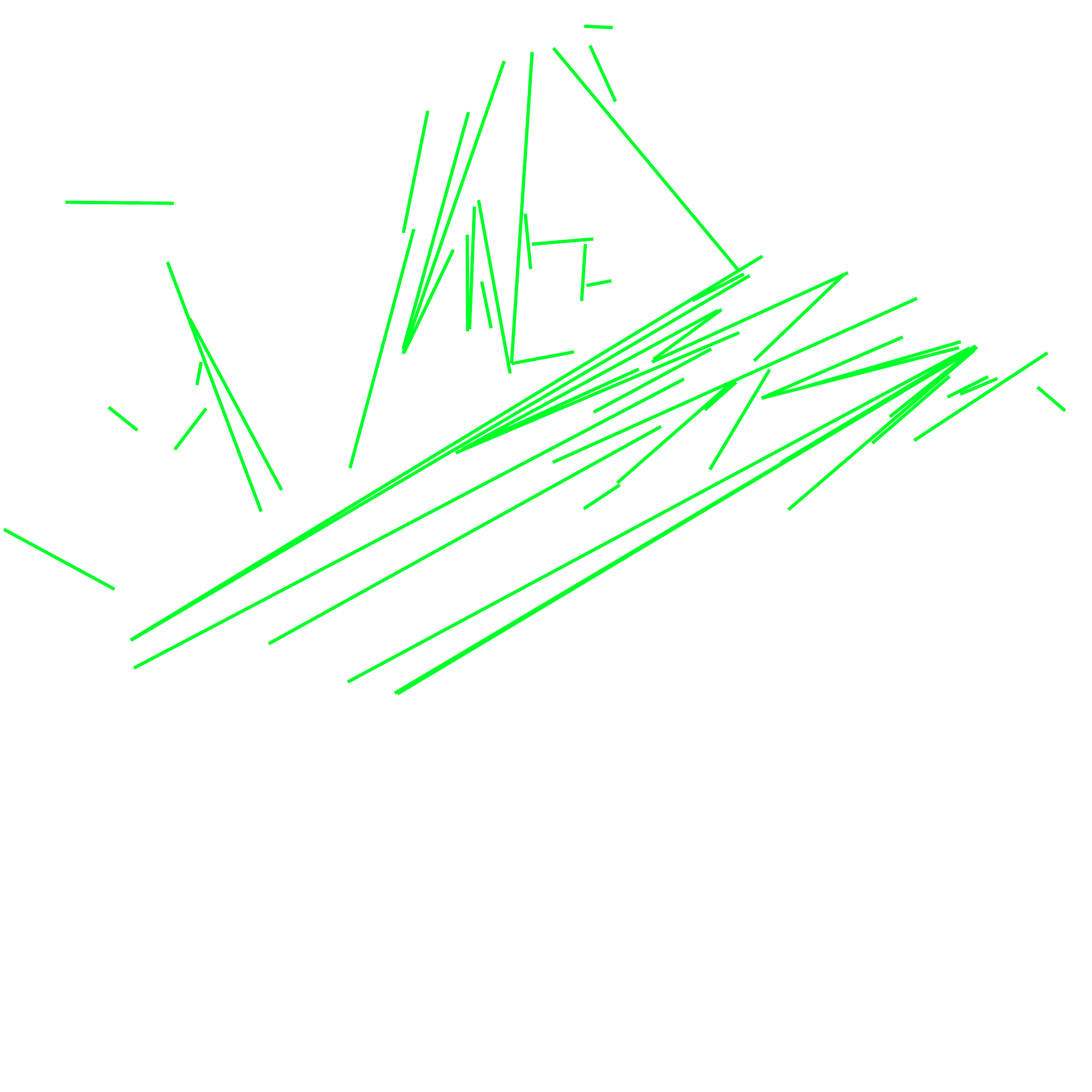}
  \includegraphics[scale=.063]{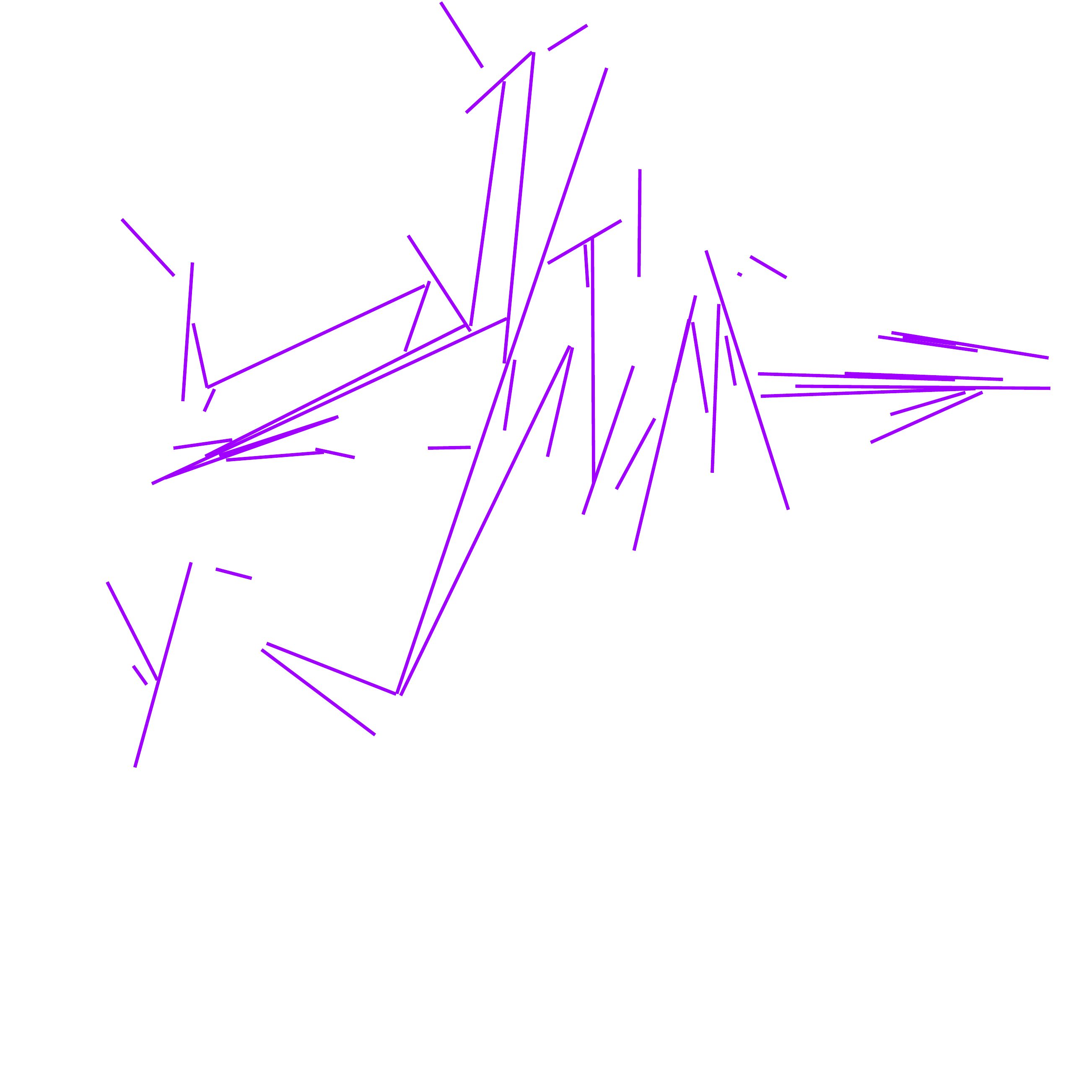}
  \includegraphics[scale=.063]{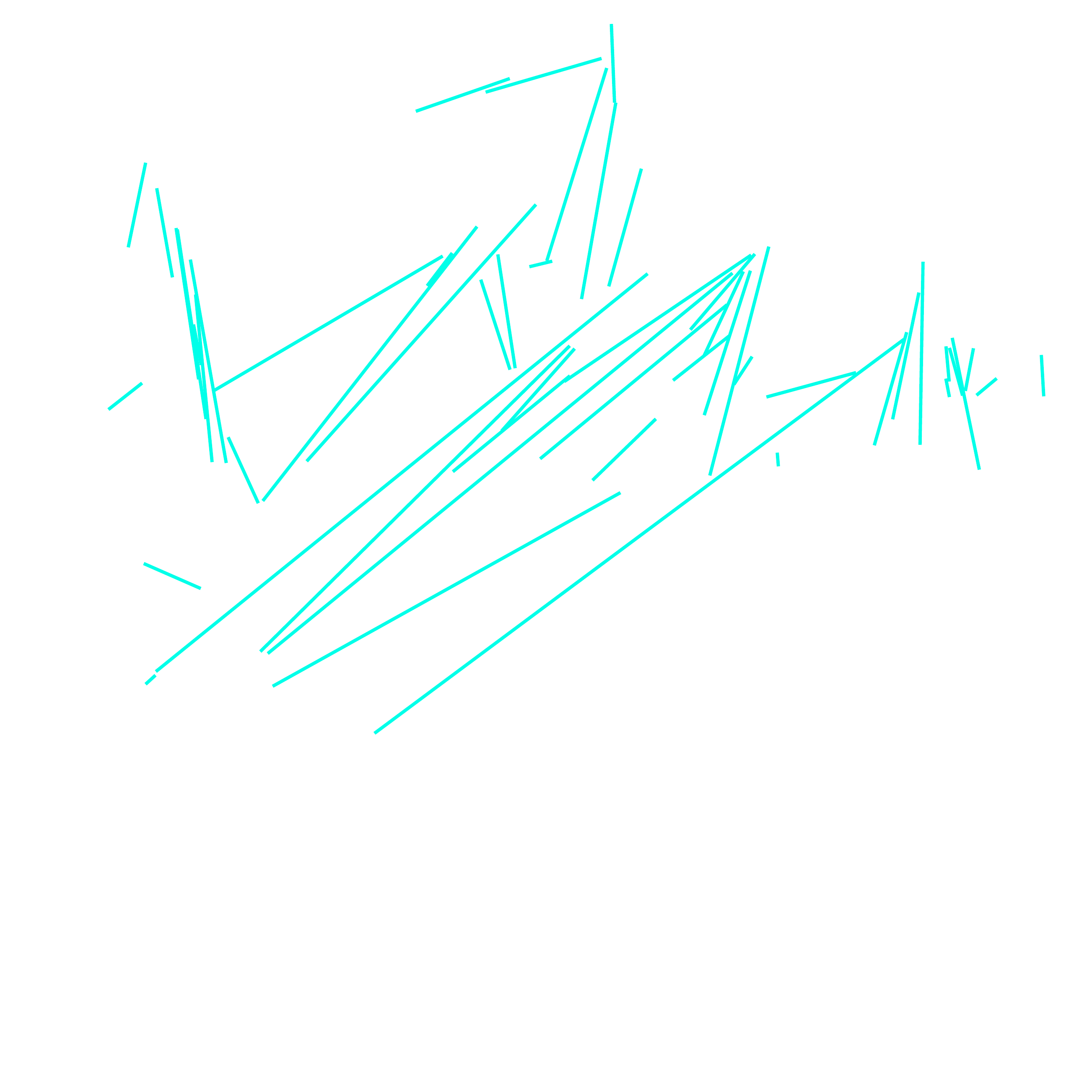}
  \includegraphics[scale=.063]{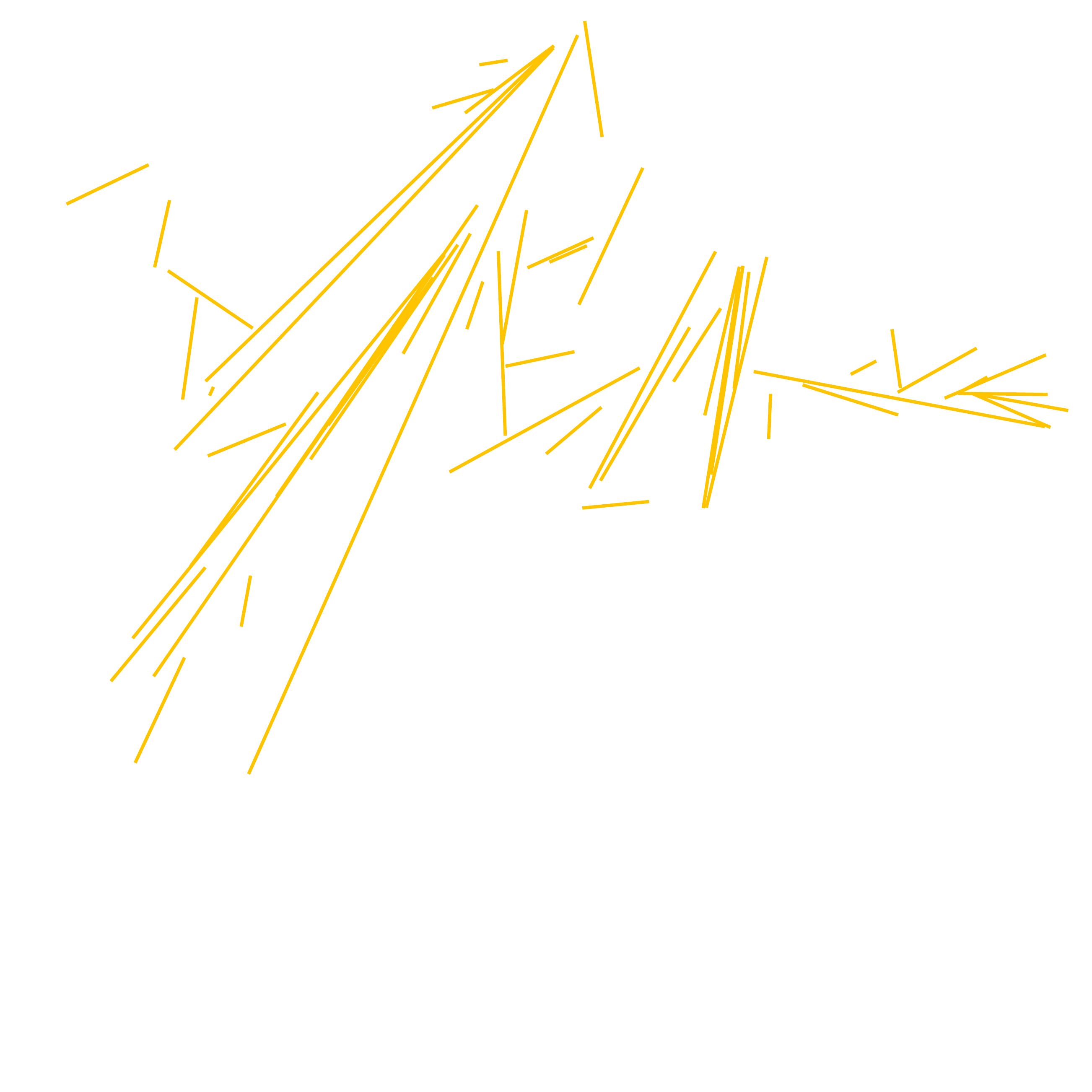}
  \includegraphics[scale=.063]{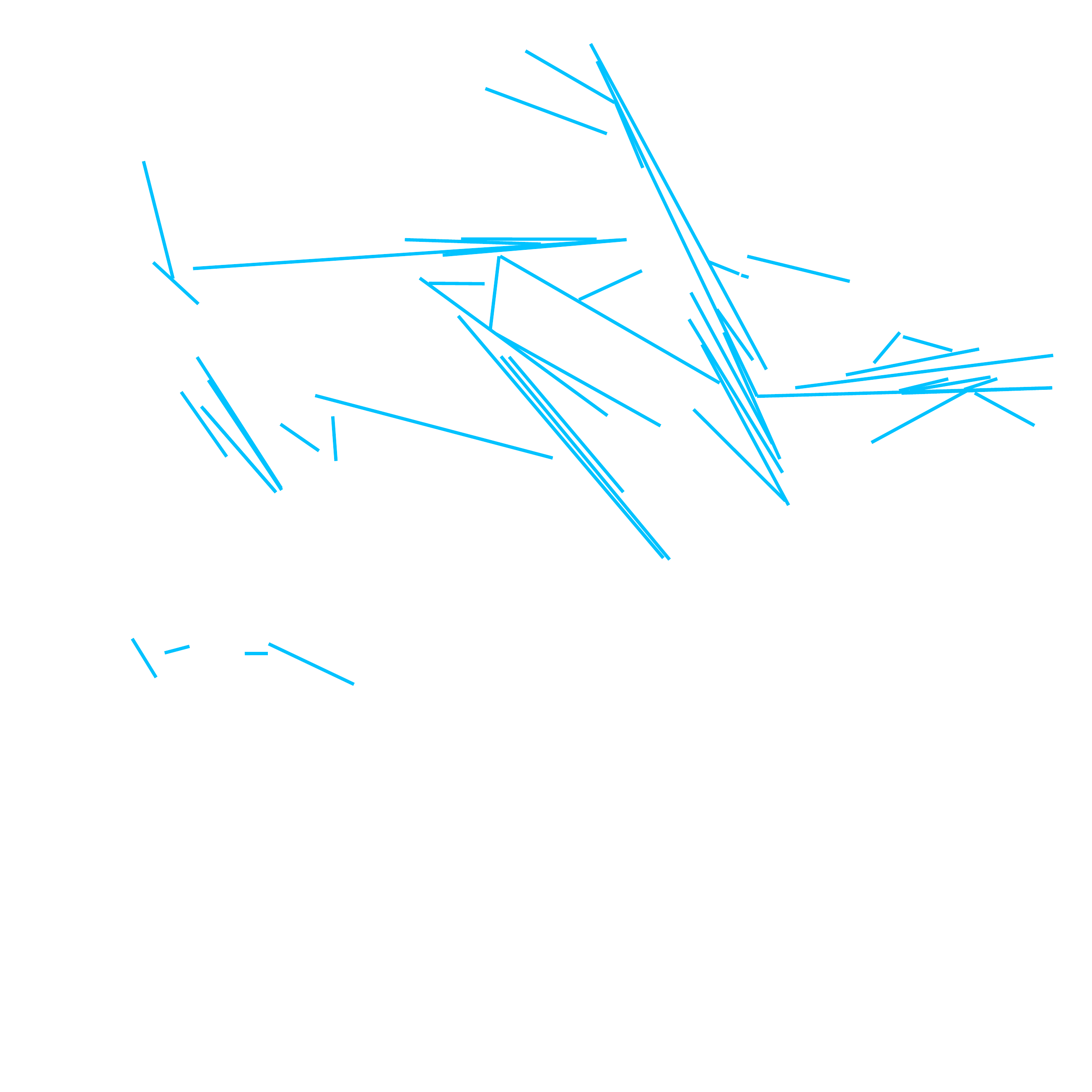}
  \includegraphics[scale=.063]{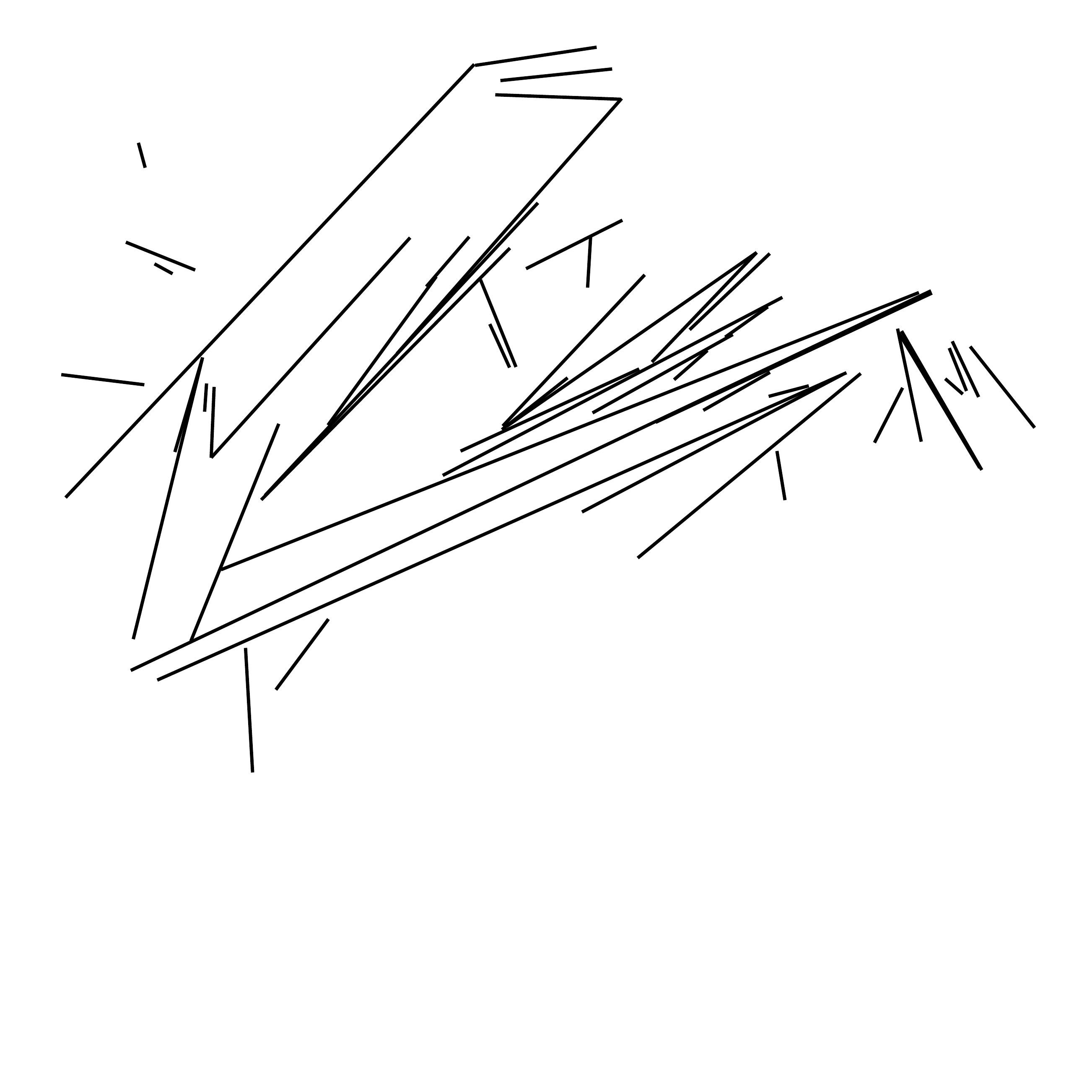}
  \includegraphics[scale=.063]{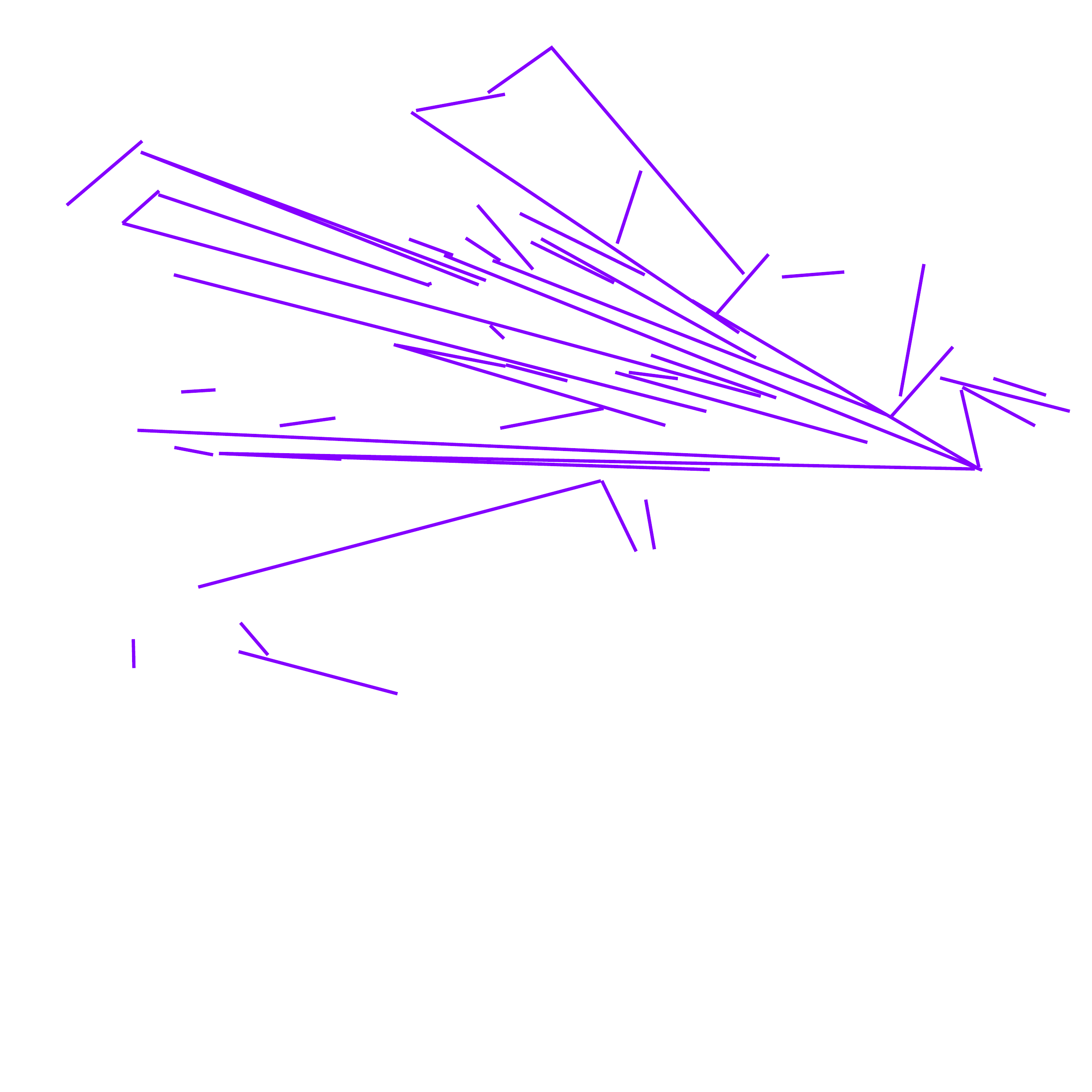}
  \includegraphics[scale=.063]{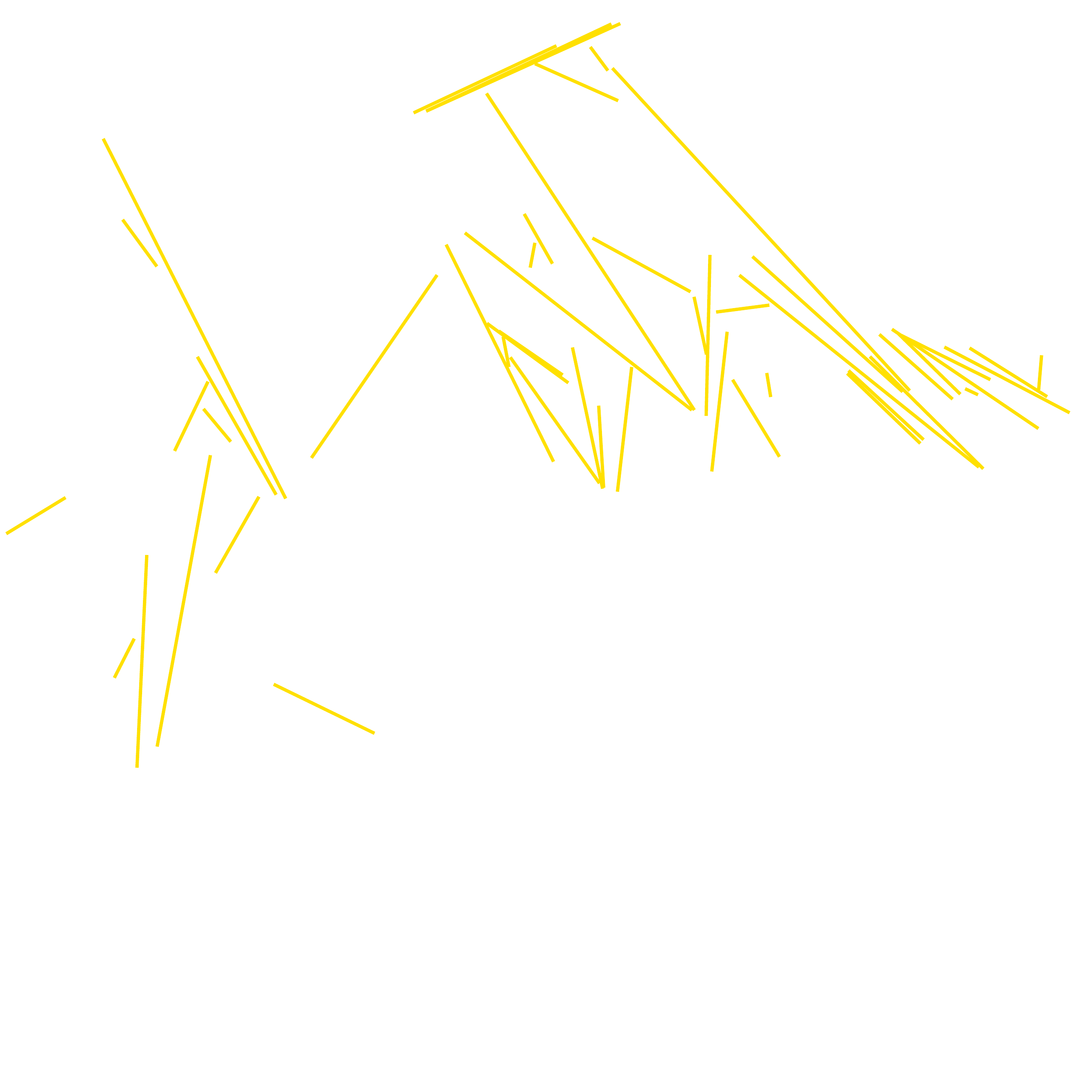}
  \includegraphics[scale=.063]{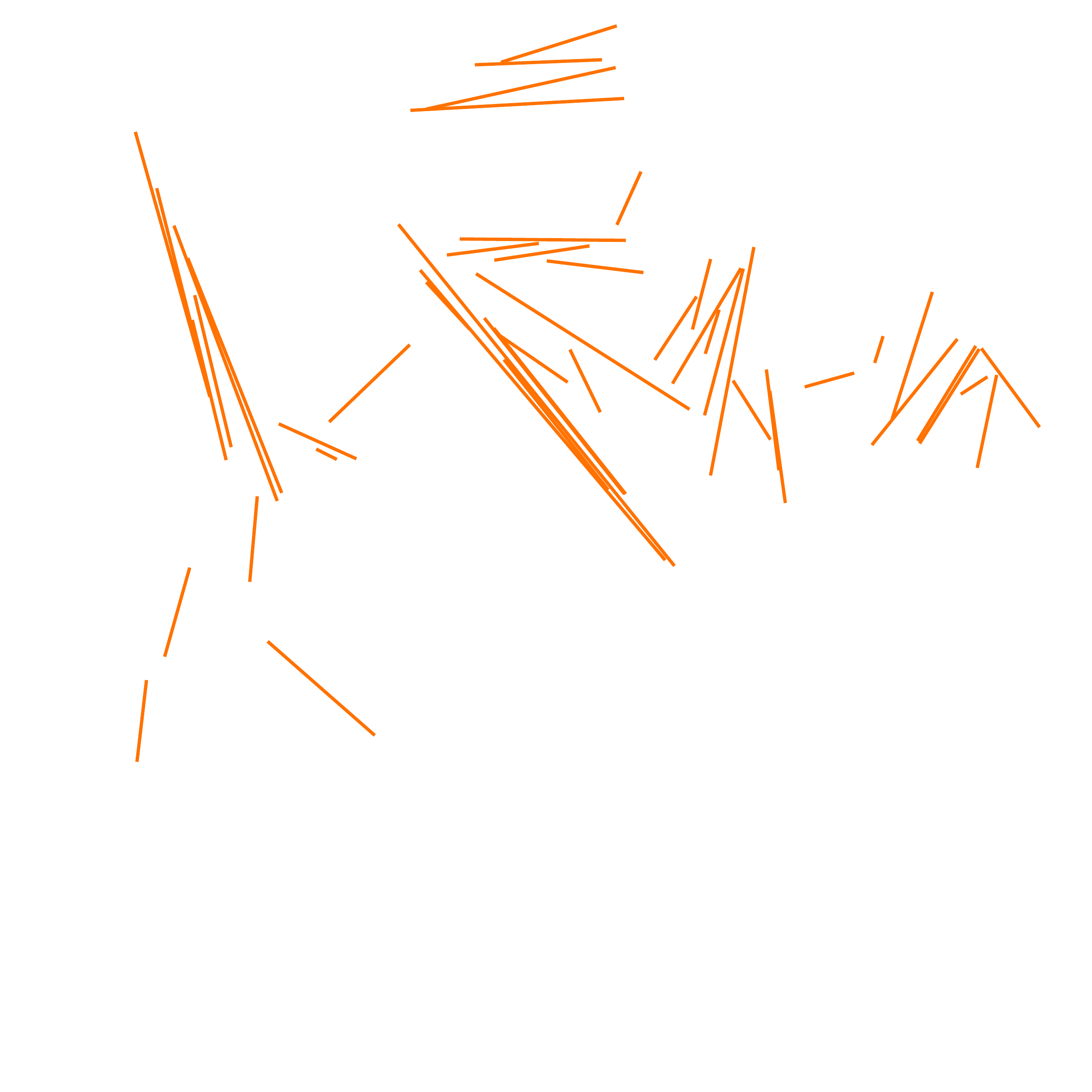}
  \includegraphics[scale=.063]{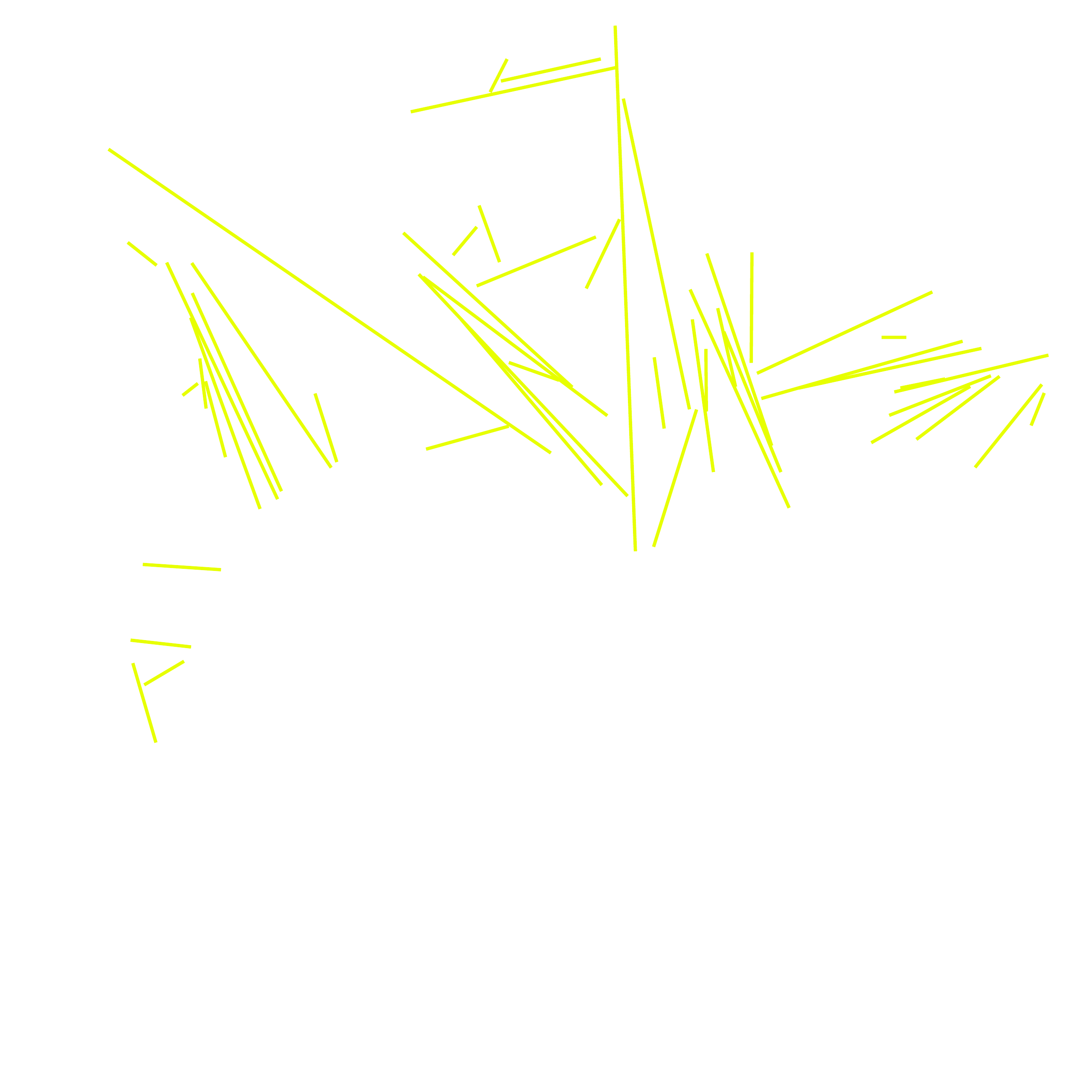}
  \includegraphics[scale=.063]{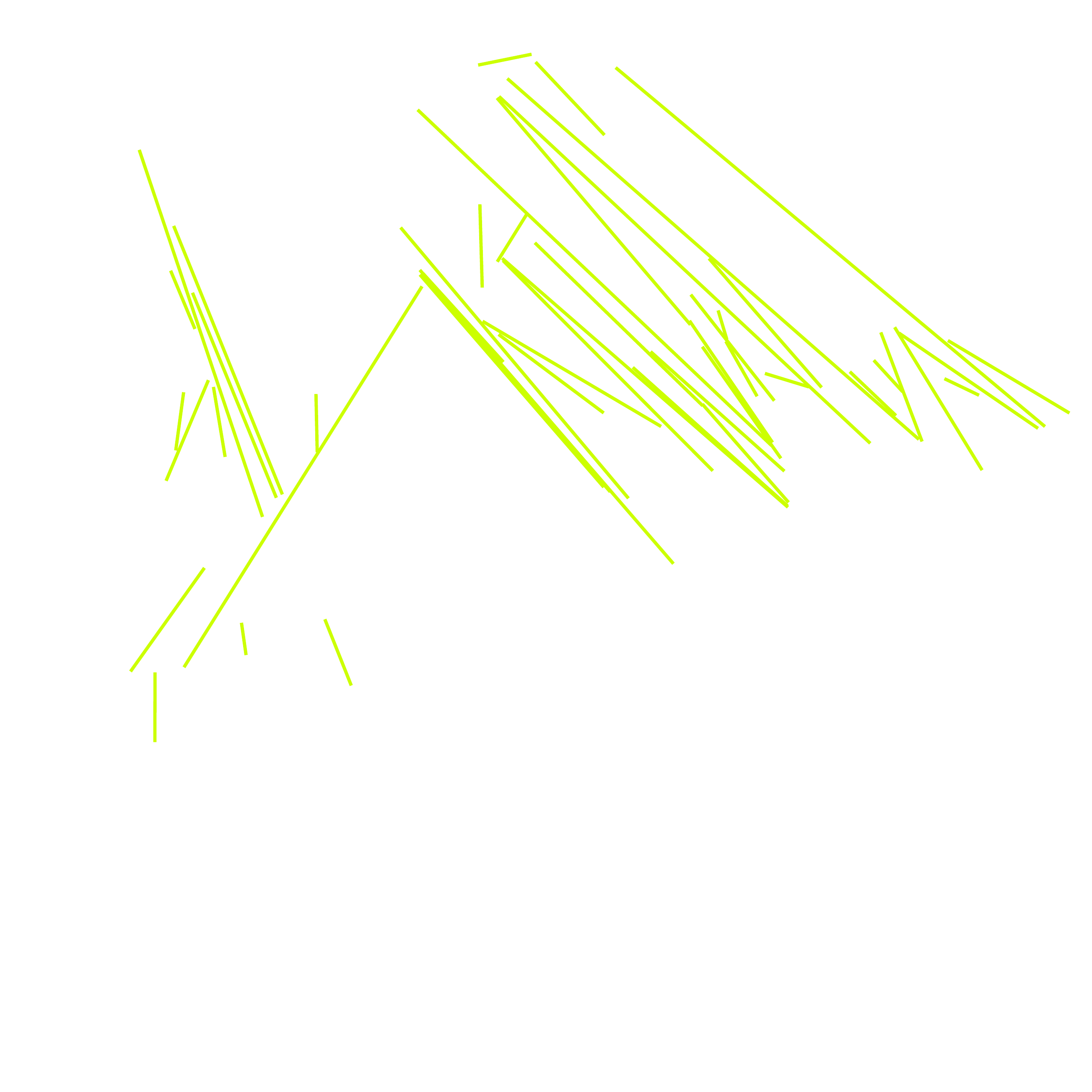}
  \includegraphics[scale=.063]{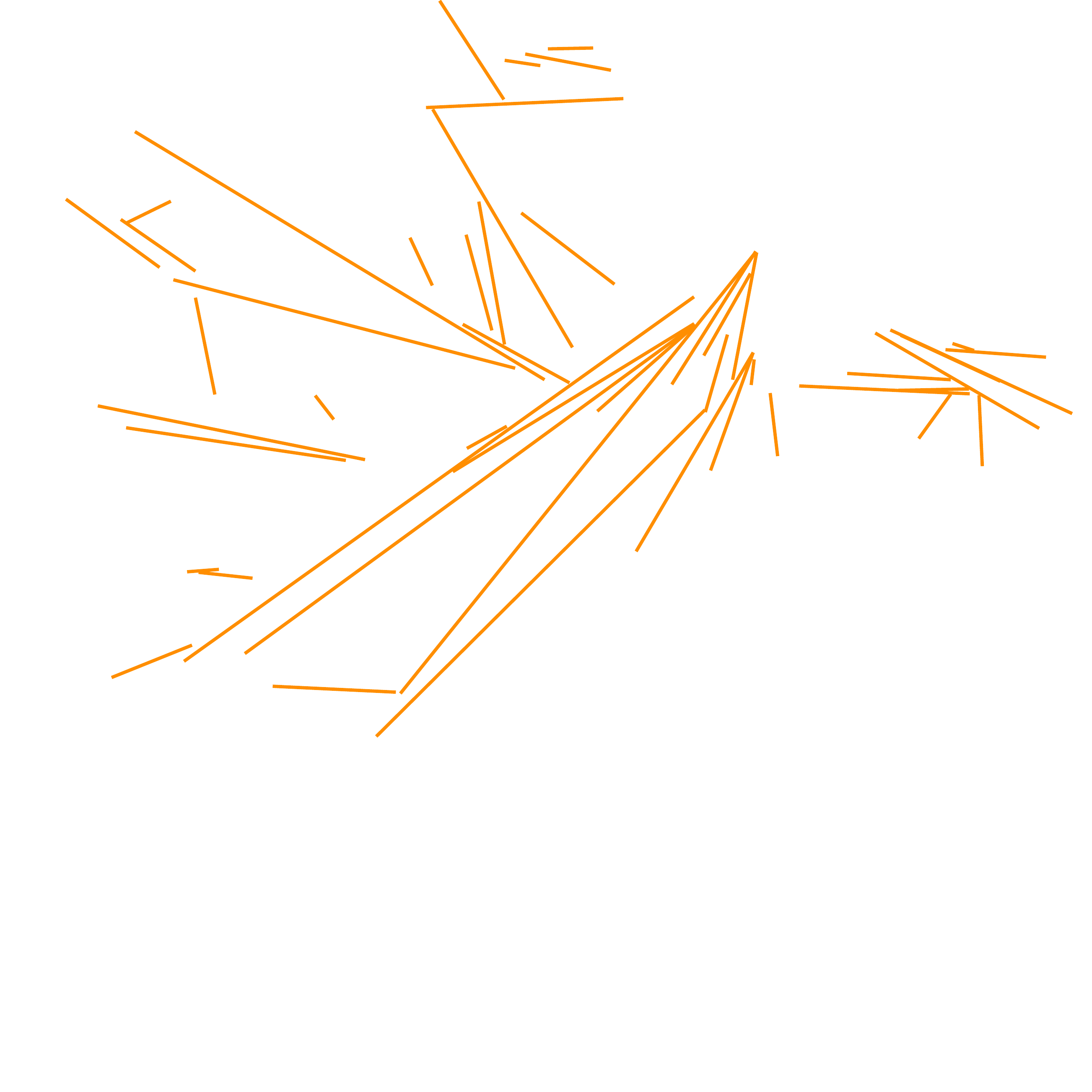}
  \includegraphics[scale=.063]{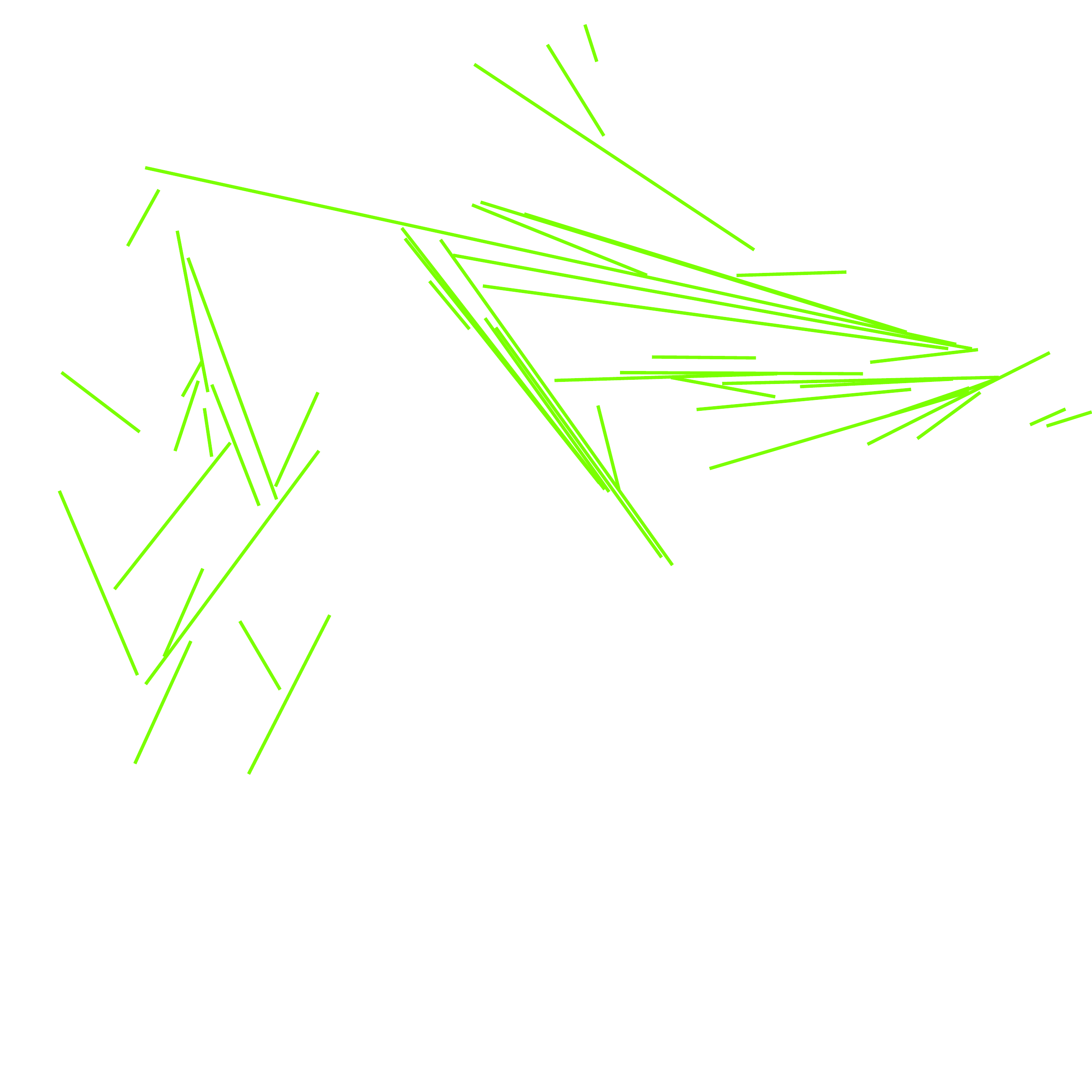}
  \includegraphics[scale=.063]{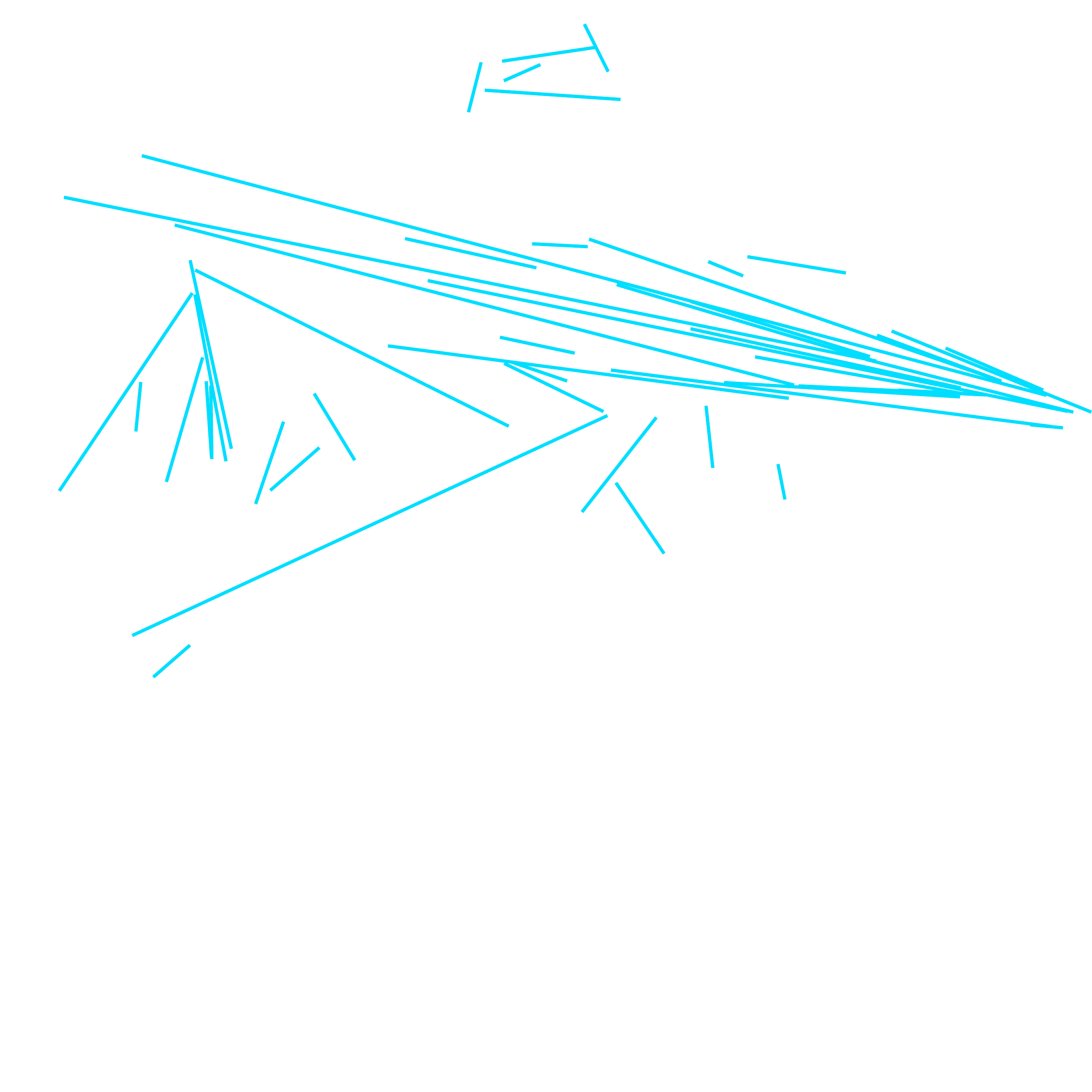}
  \includegraphics[scale=.063]{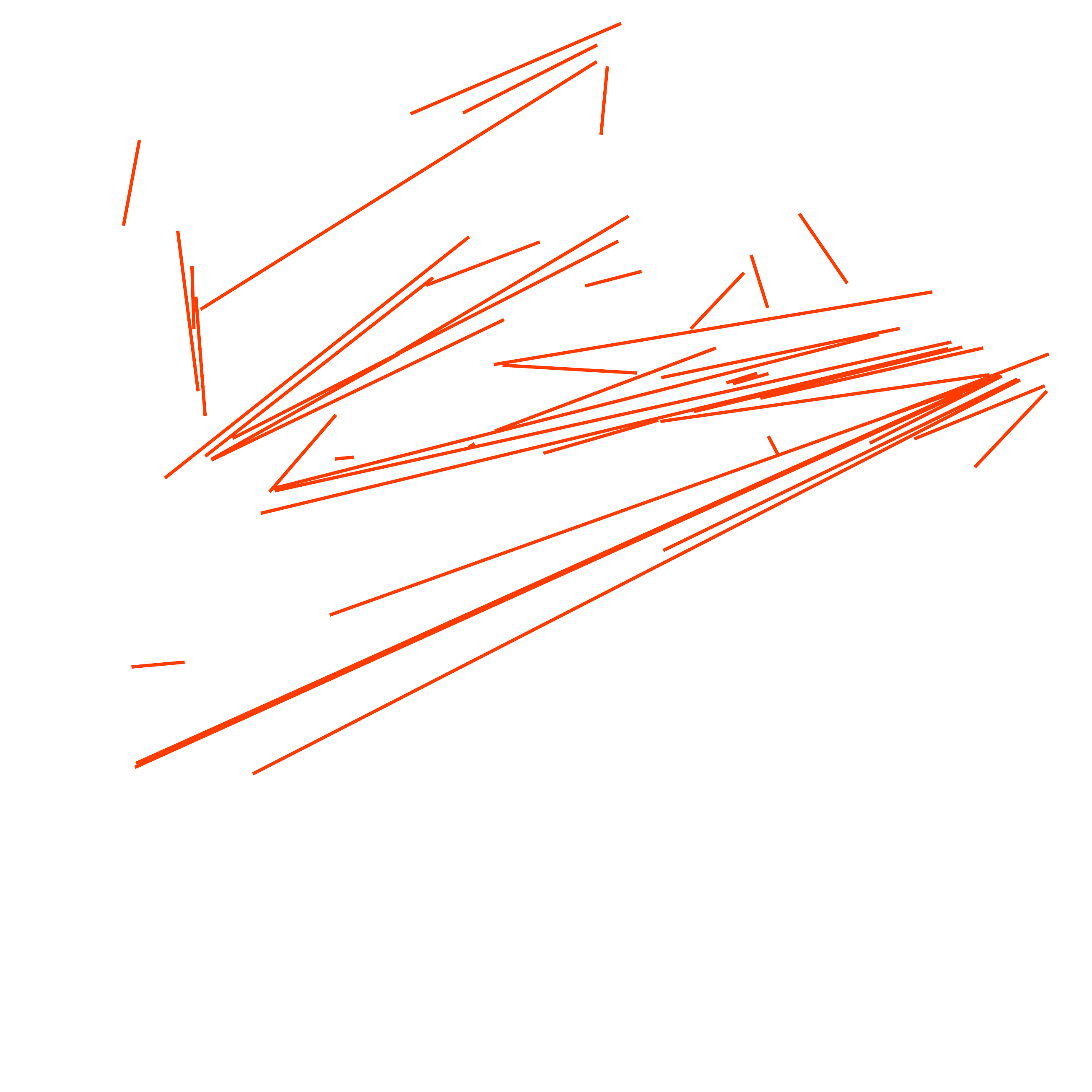}
  \includegraphics[scale=.063]{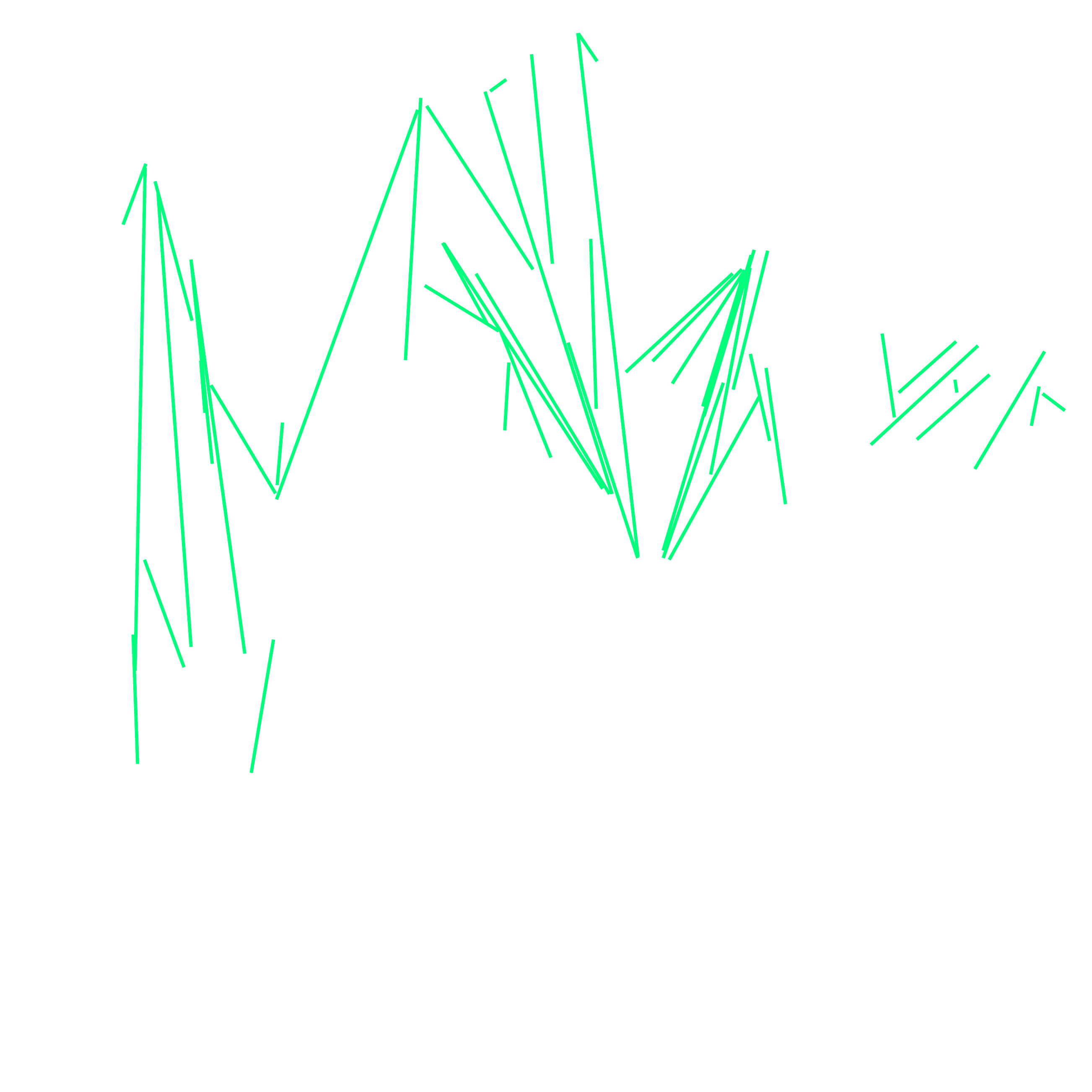}
  \includegraphics[scale=.063]{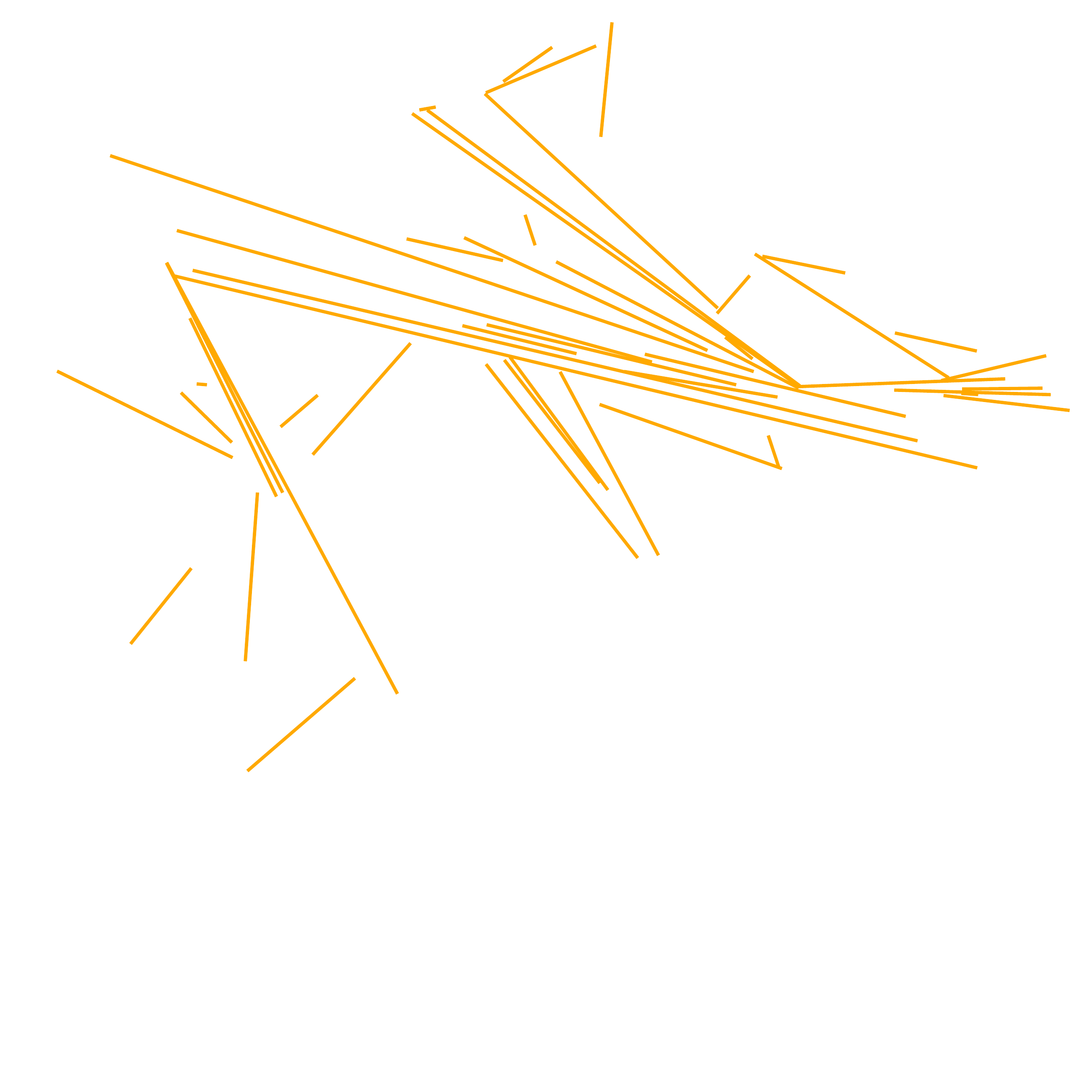}
  \includegraphics[scale=.063]{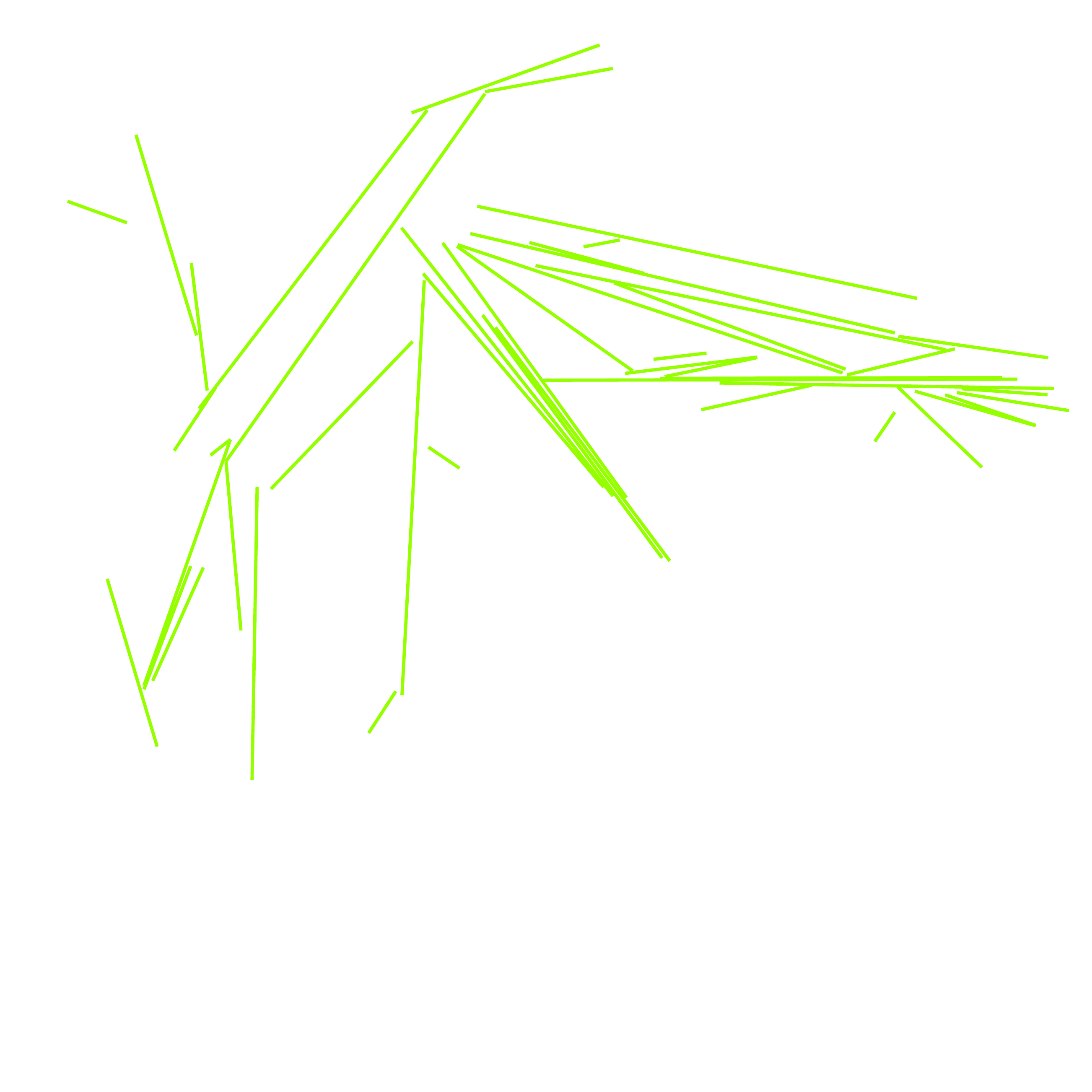}
  \includegraphics[scale=.063]{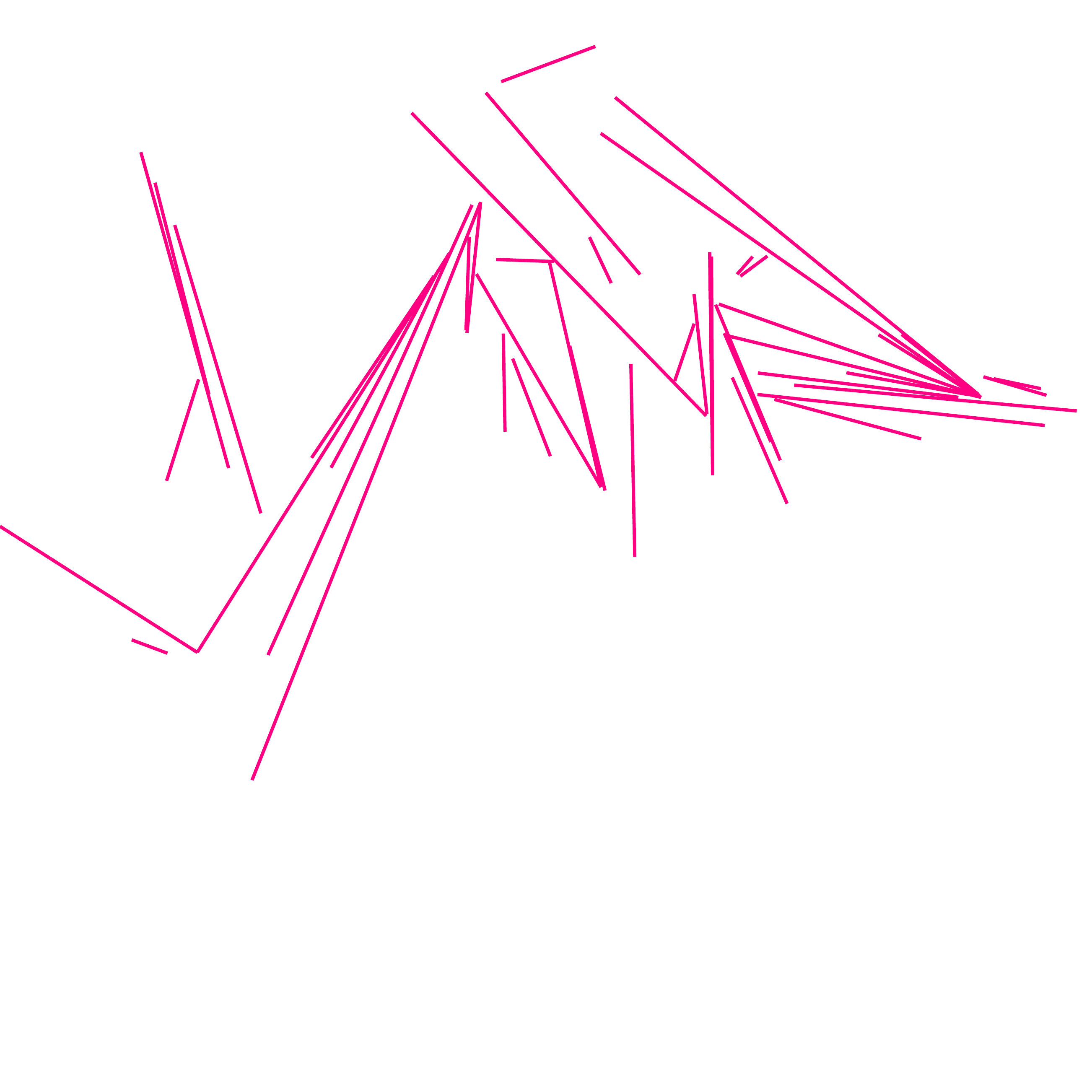}
  \includegraphics[scale=.063]{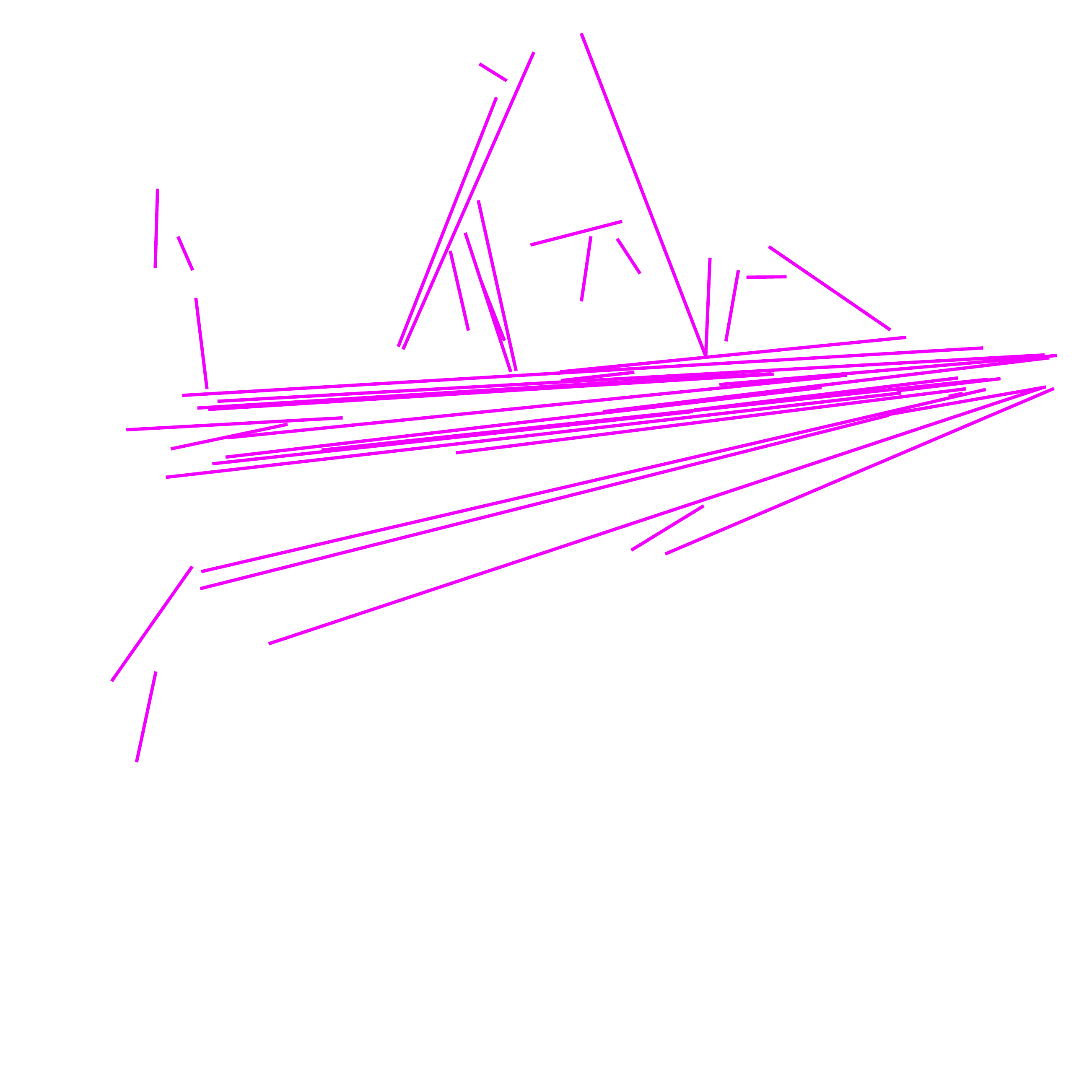}
  \includegraphics[scale=.063]{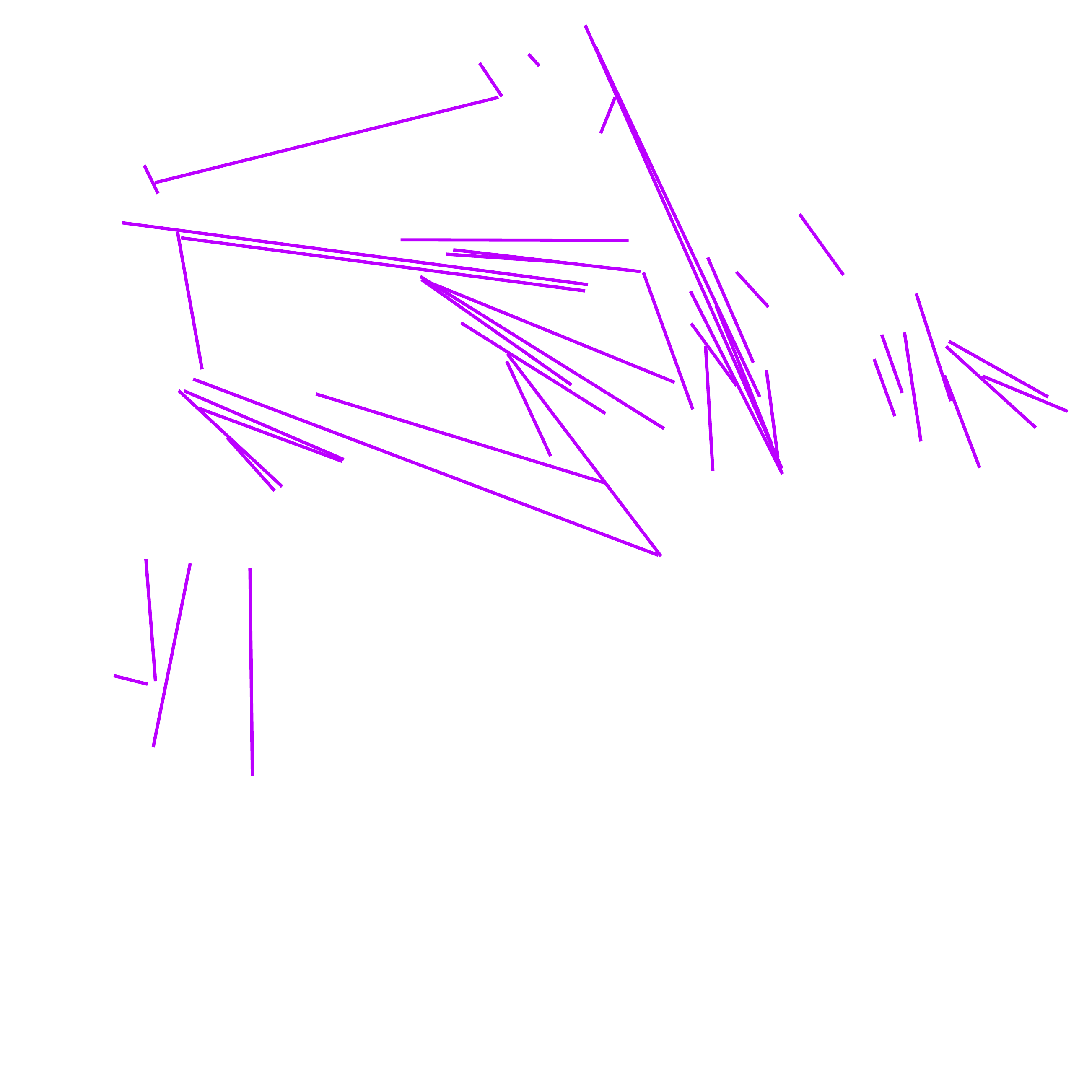}
  \includegraphics[scale=.063]{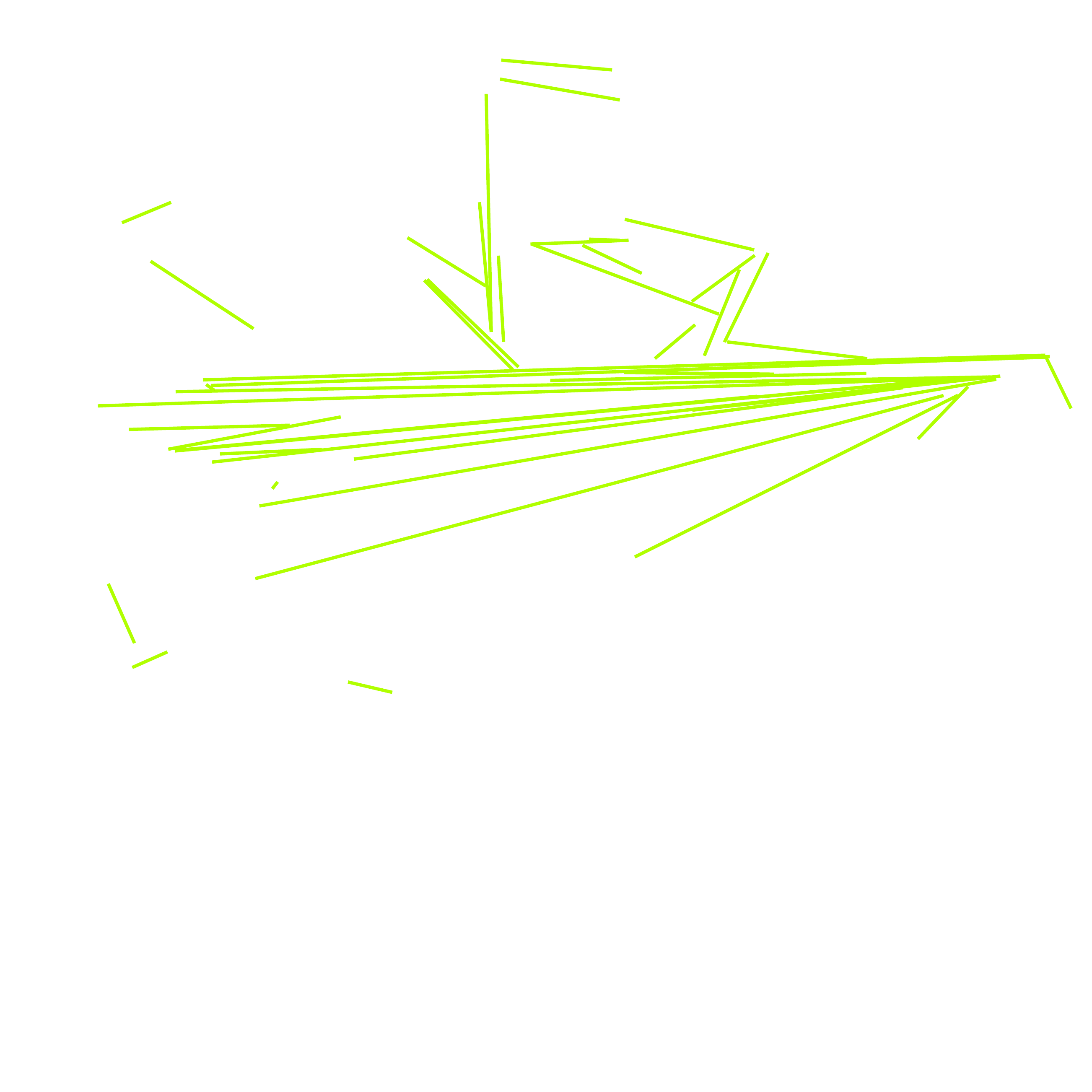}
  \includegraphics[scale=.063]{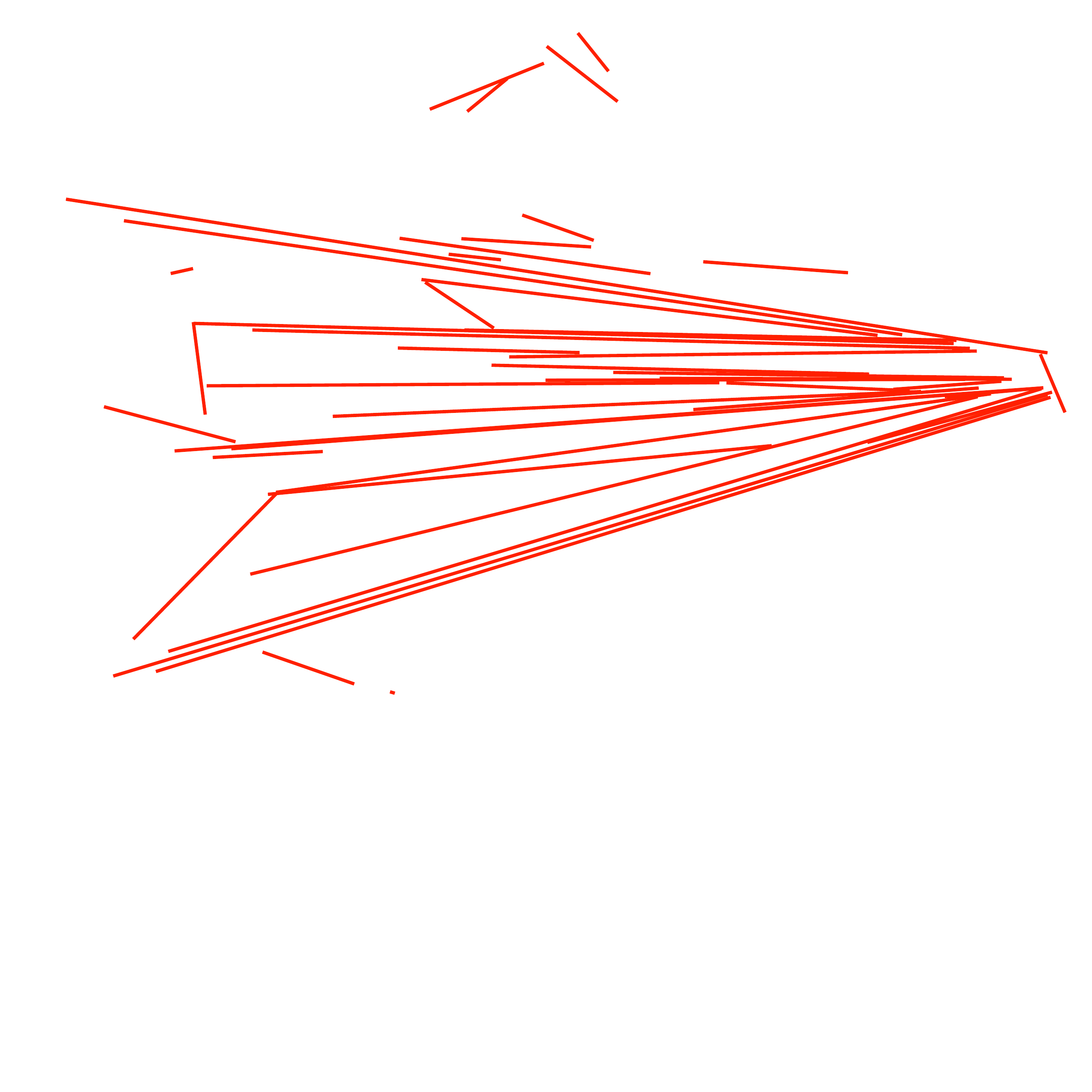}
  \includegraphics[scale=.063]{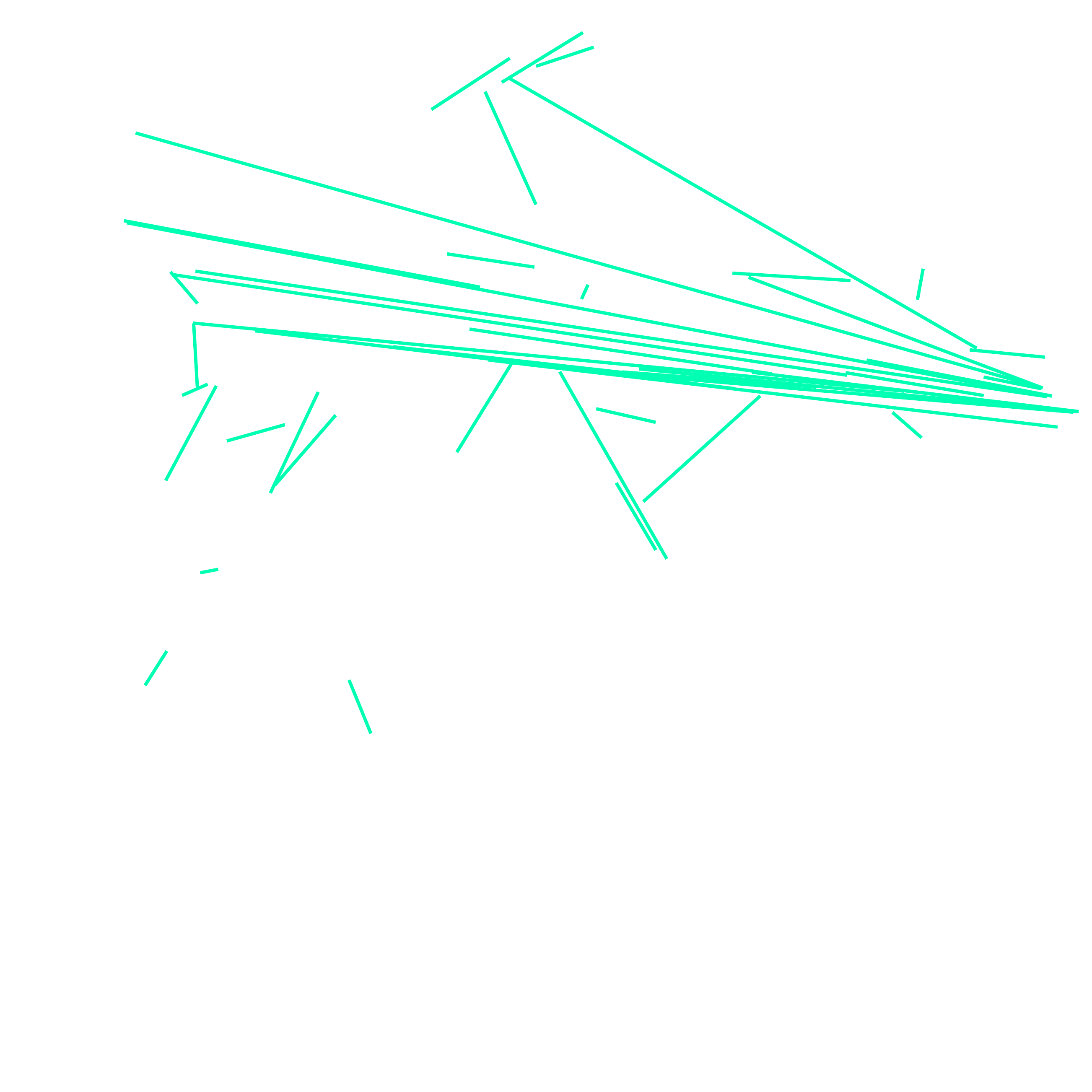}
  \includegraphics[scale=.063]{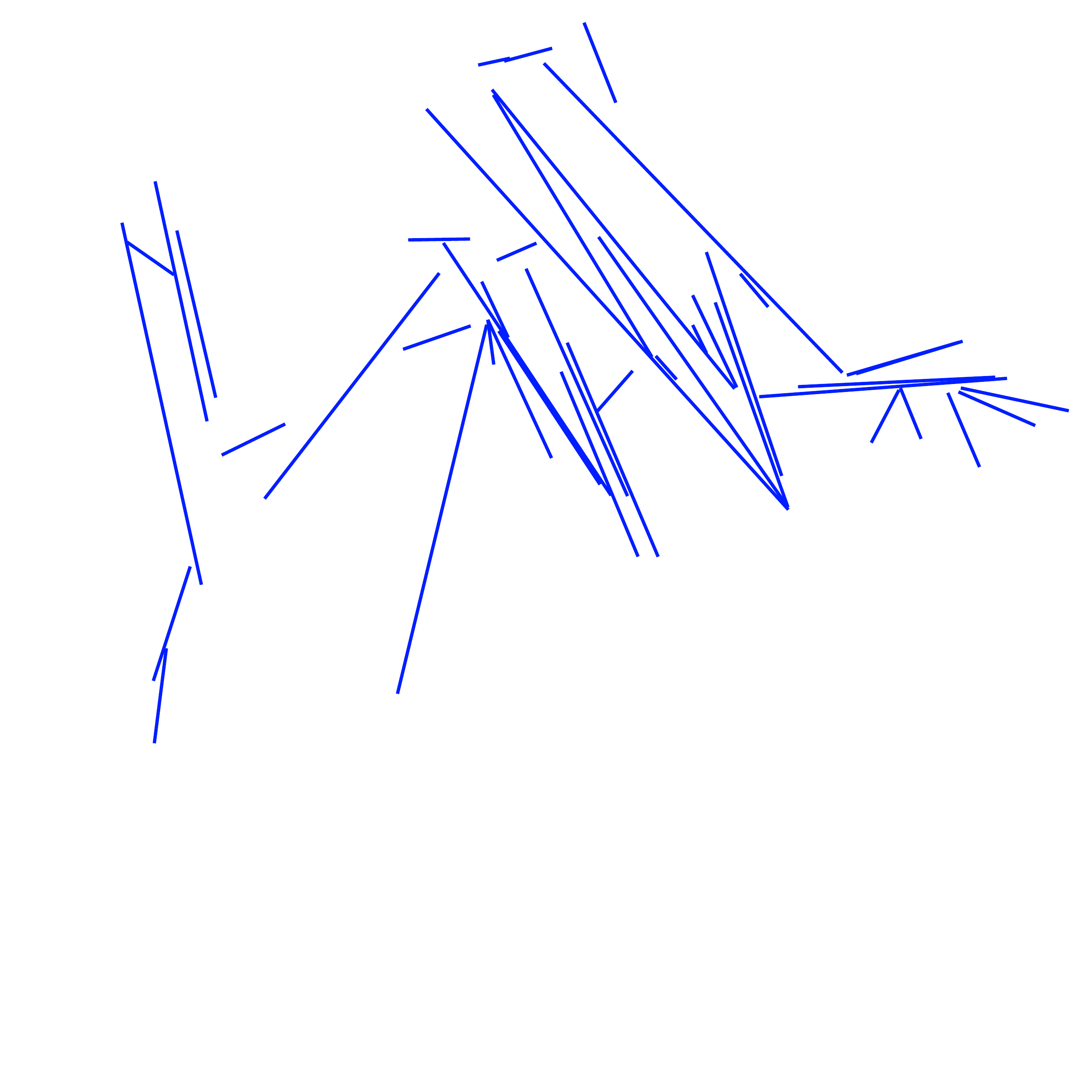}
  \includegraphics[scale=.063]{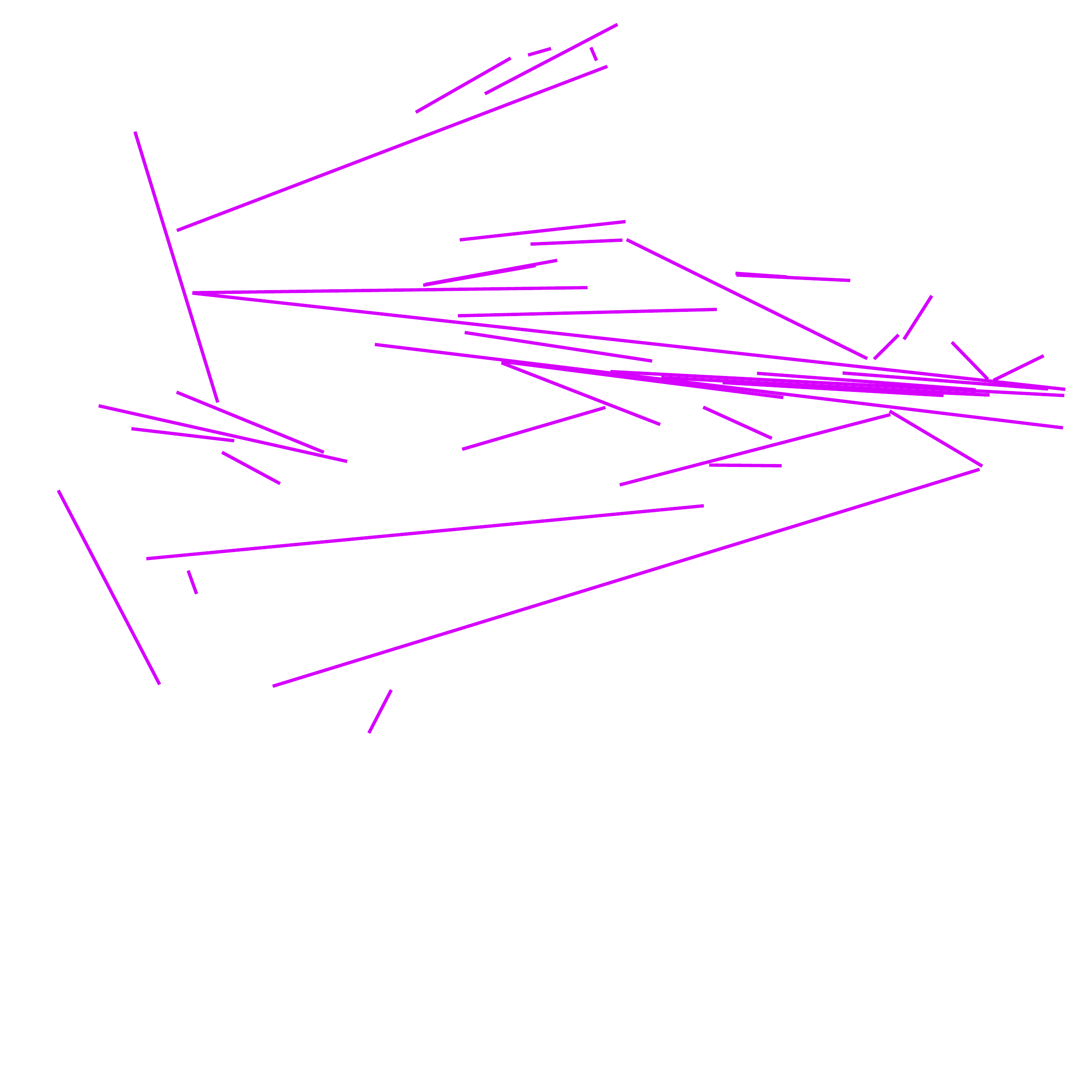}
  \includegraphics[scale=.063]{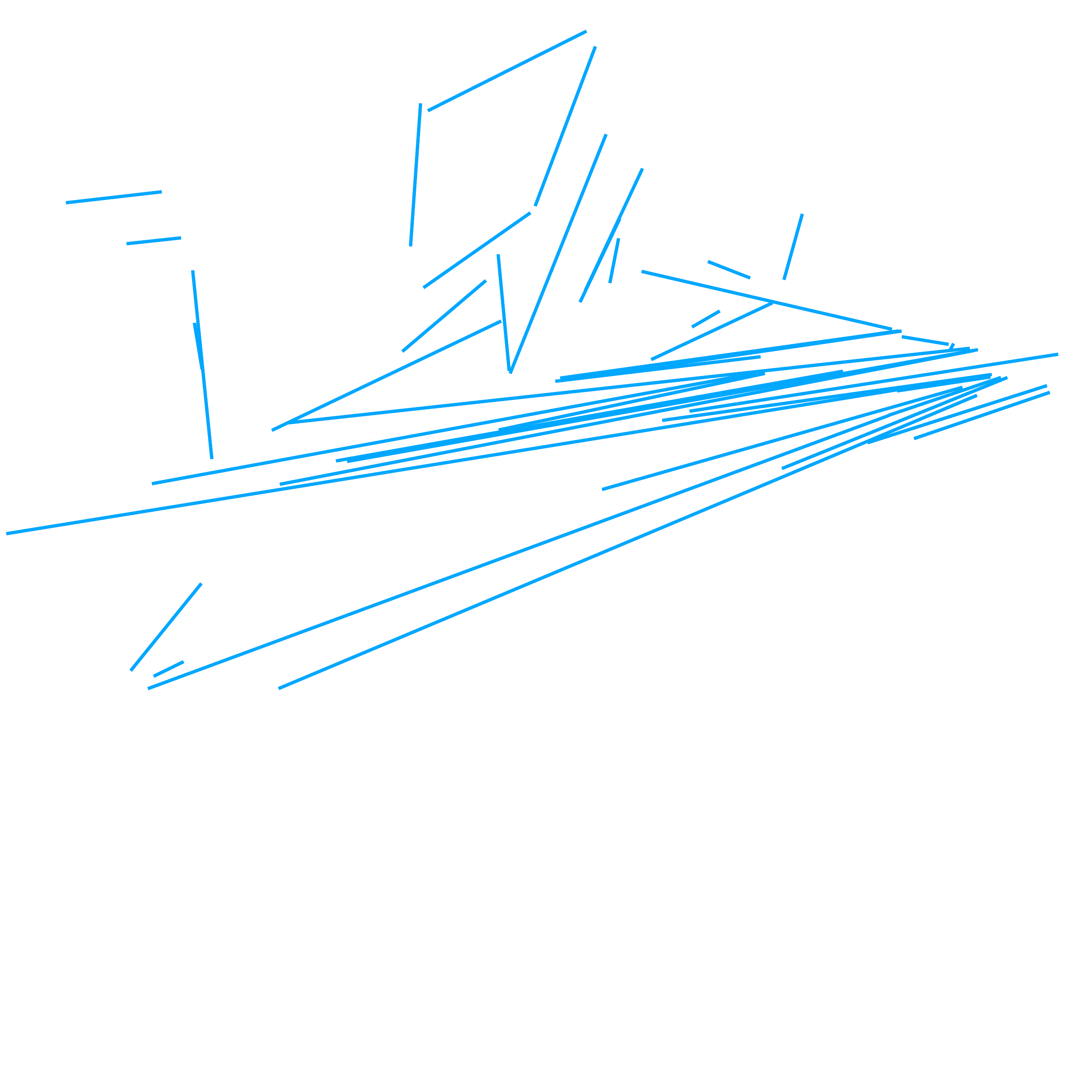}
  \includegraphics[scale=.063]{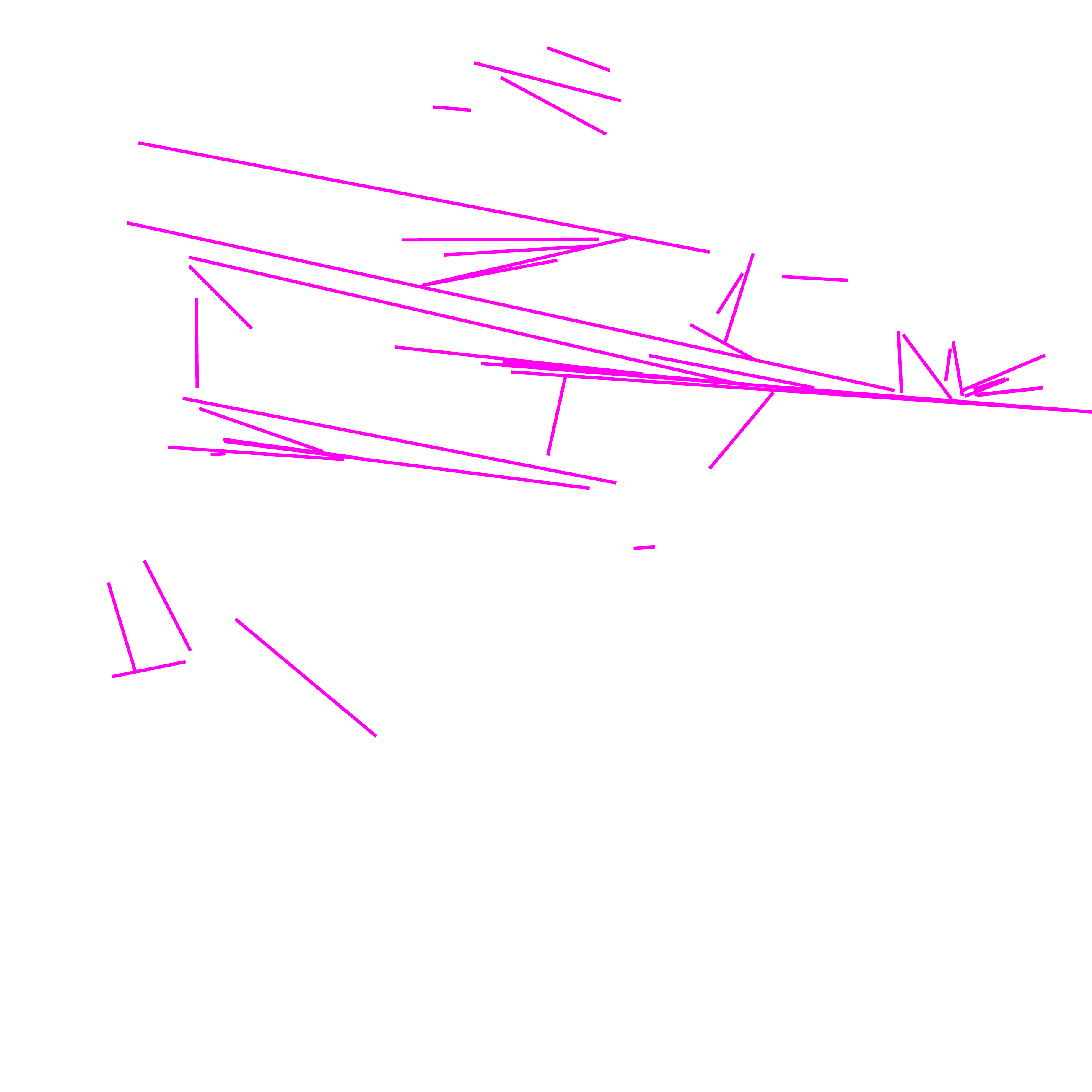}
  \includegraphics[scale=.063]{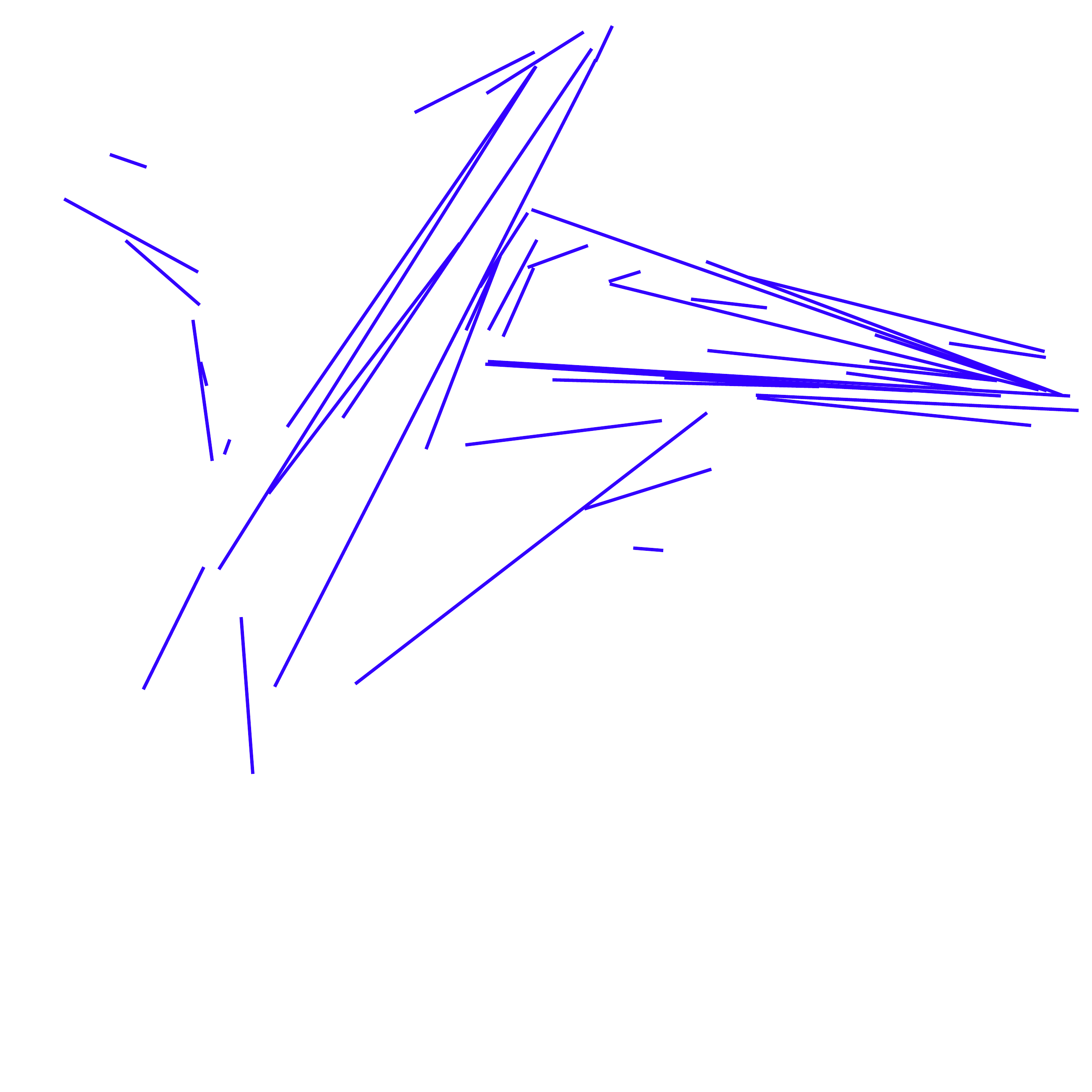}
  \includegraphics[scale=.063]{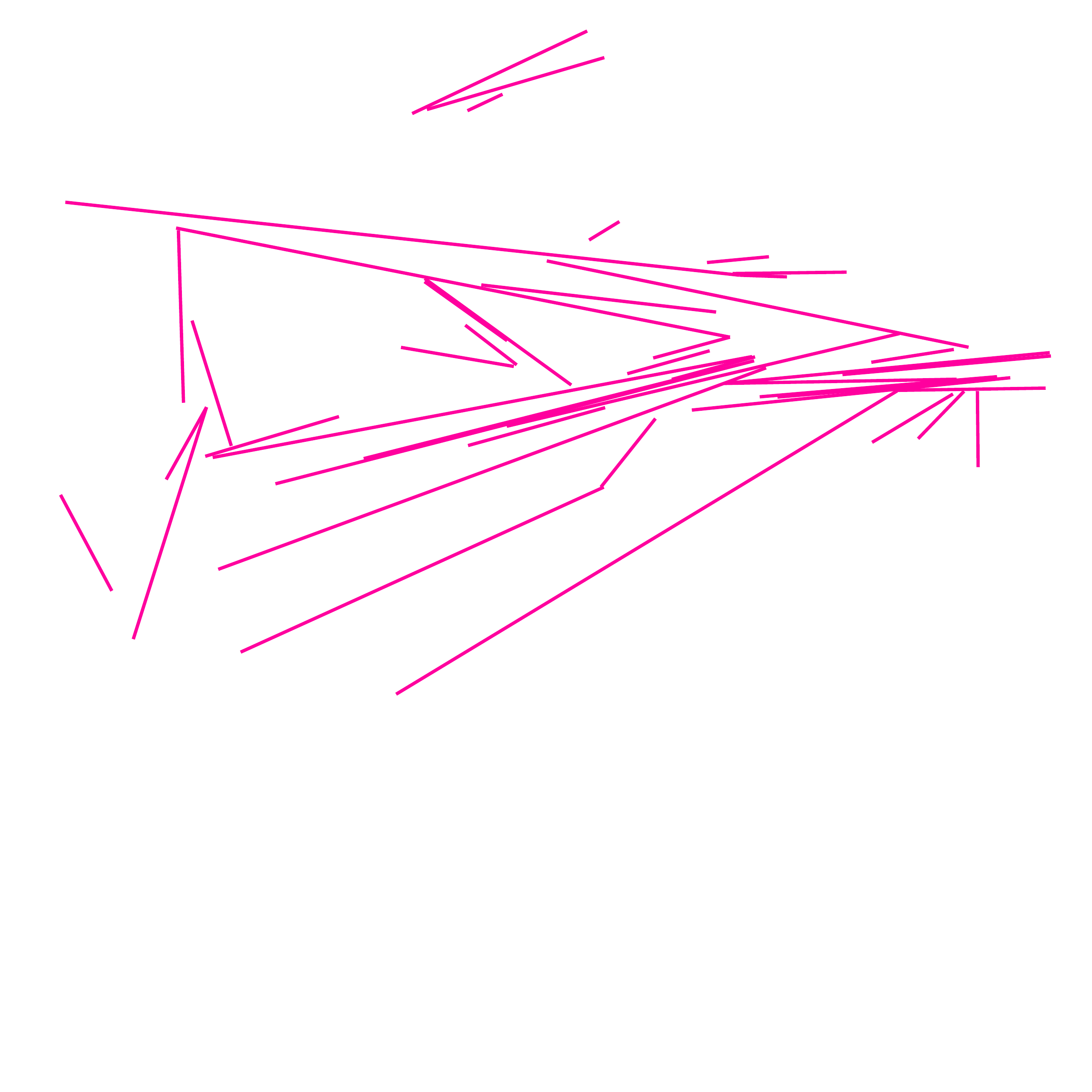}
  \includegraphics[scale=.063]{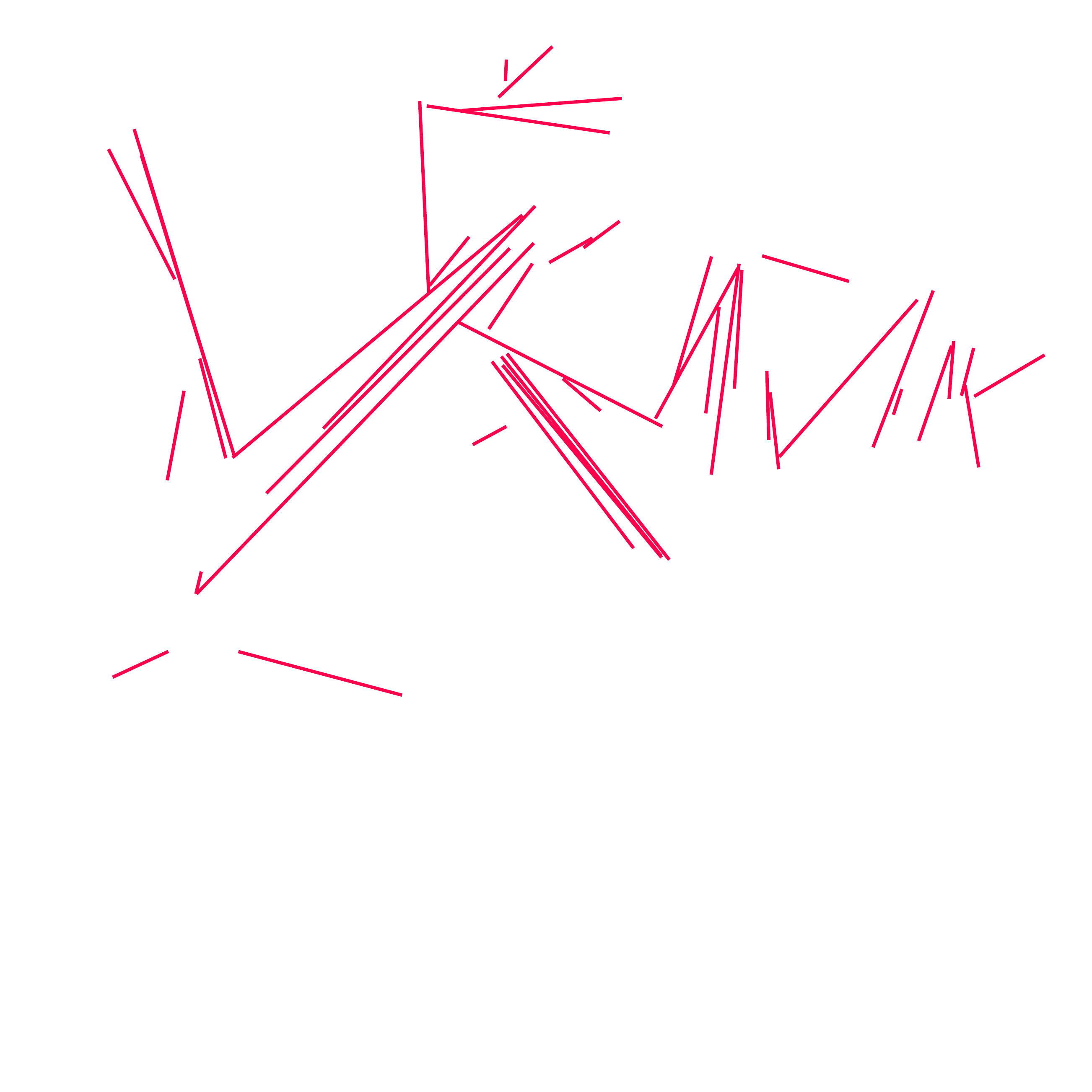}
  \includegraphics[scale=.063]{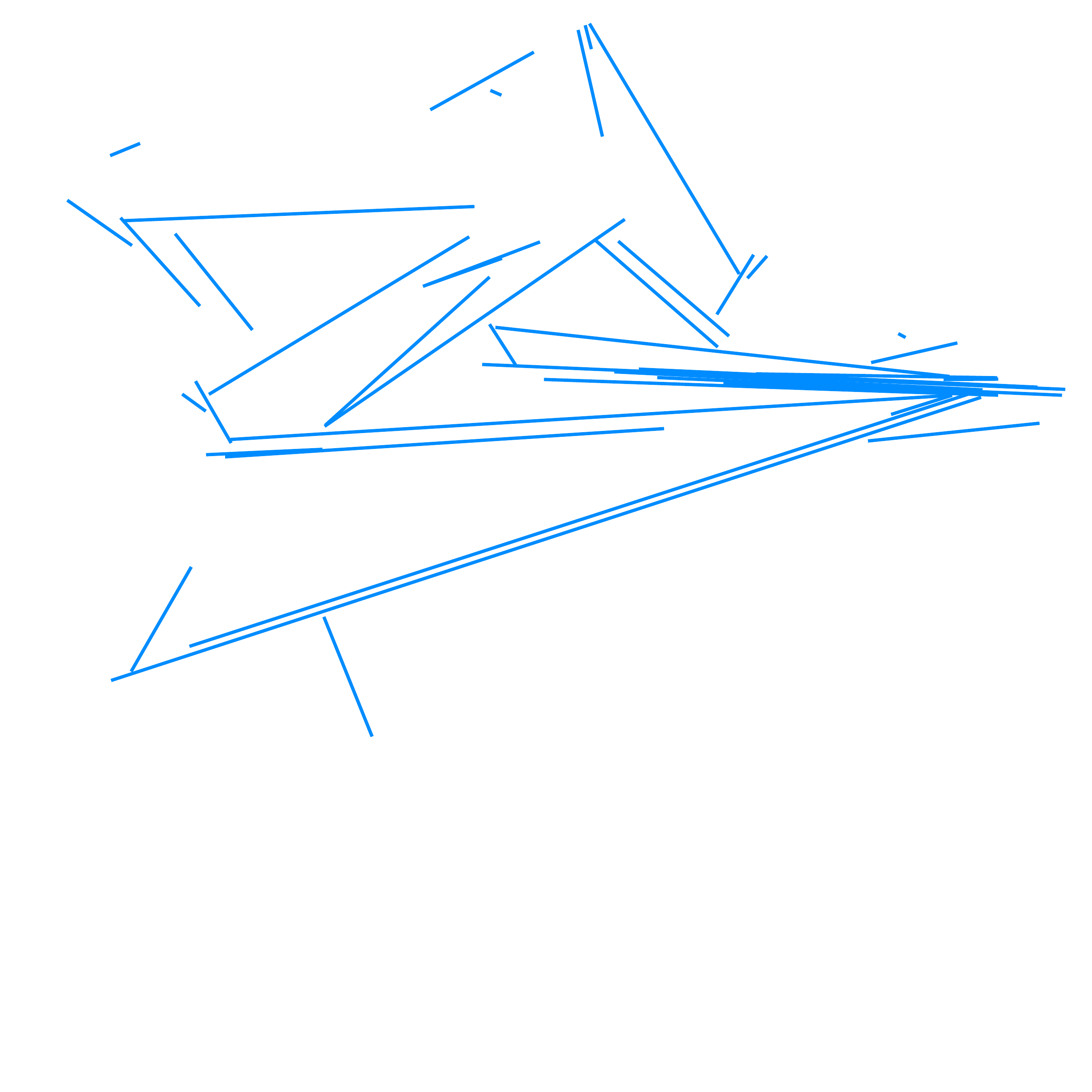}
  \includegraphics[scale=.063]{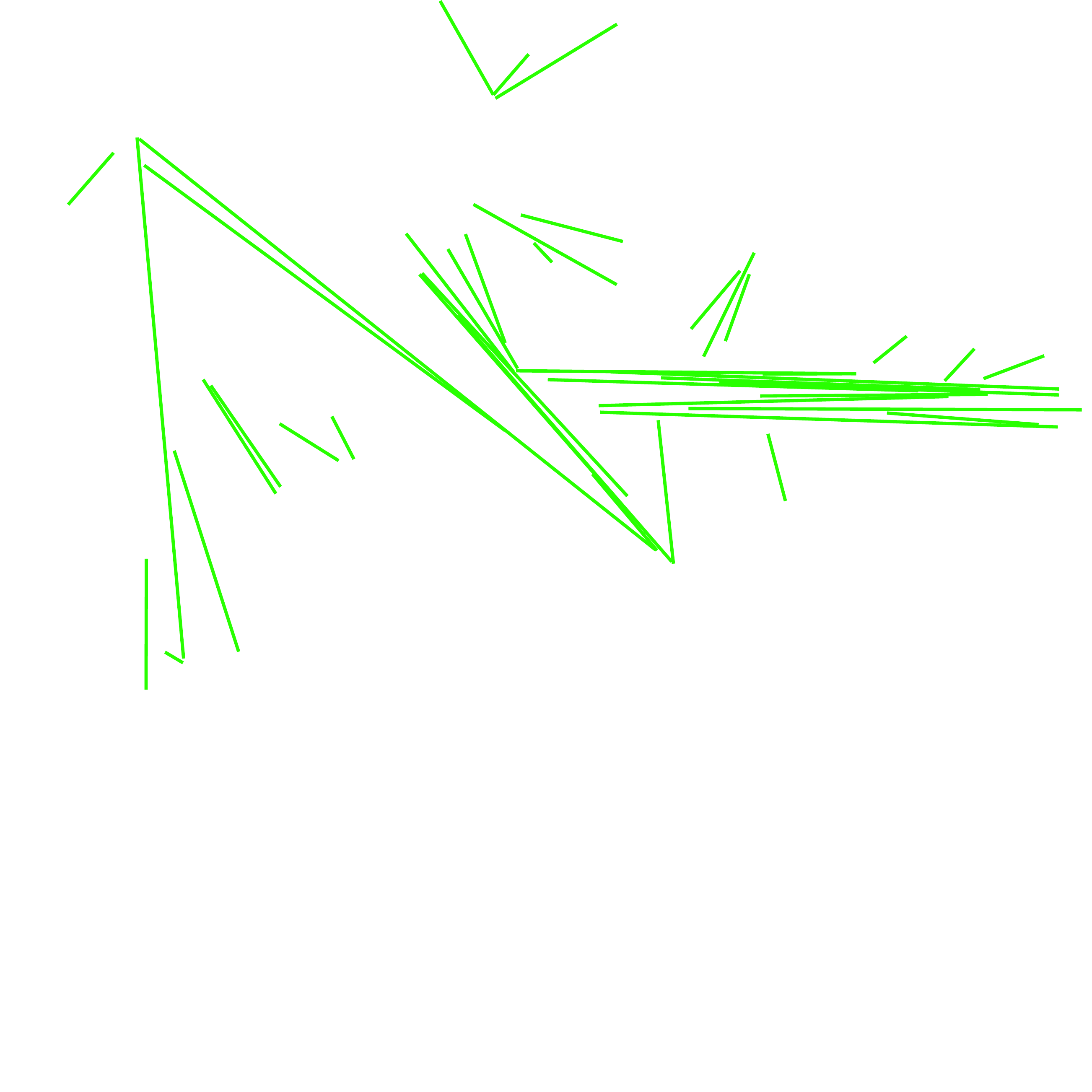}
  \includegraphics[scale=.063]{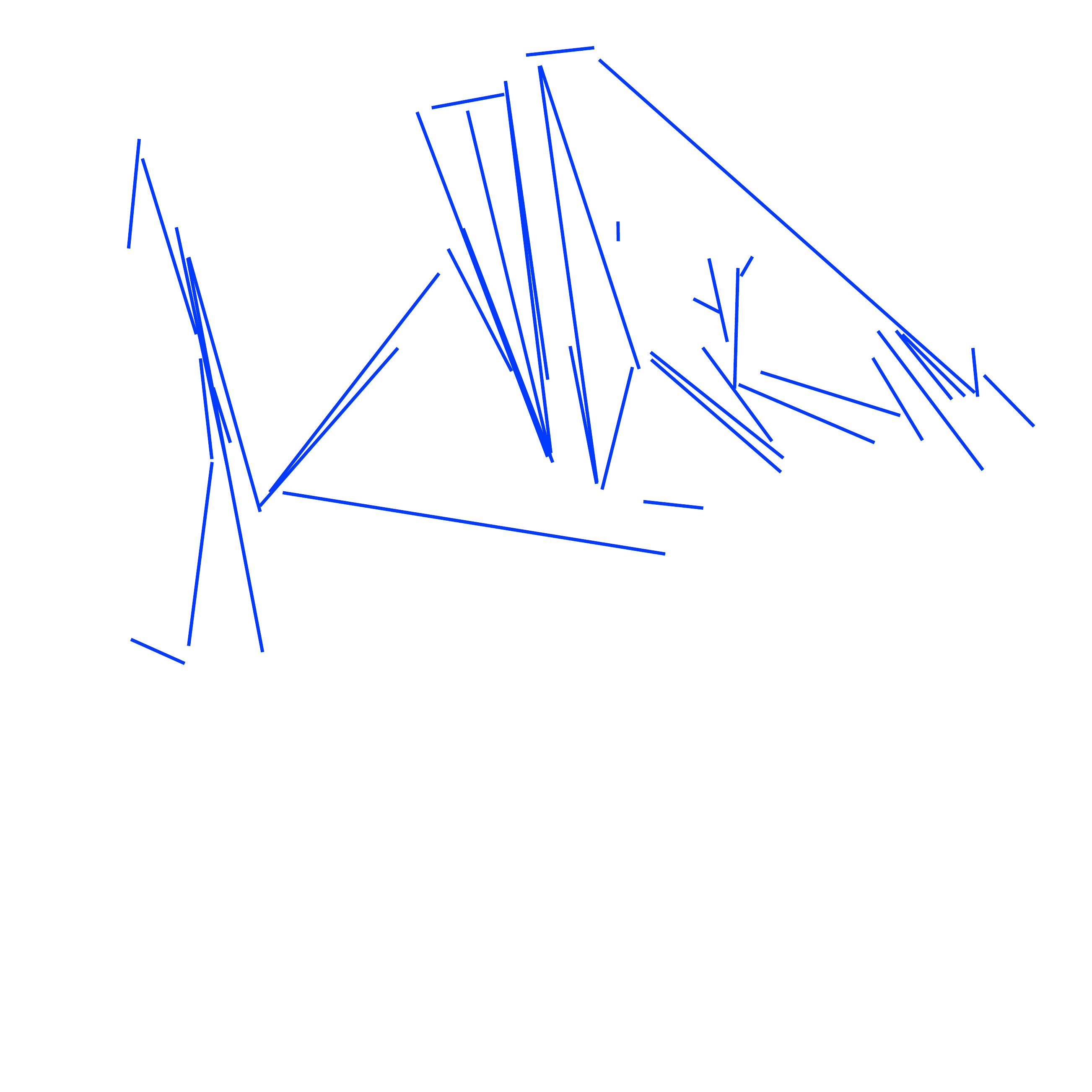}
  \includegraphics[scale=.063]{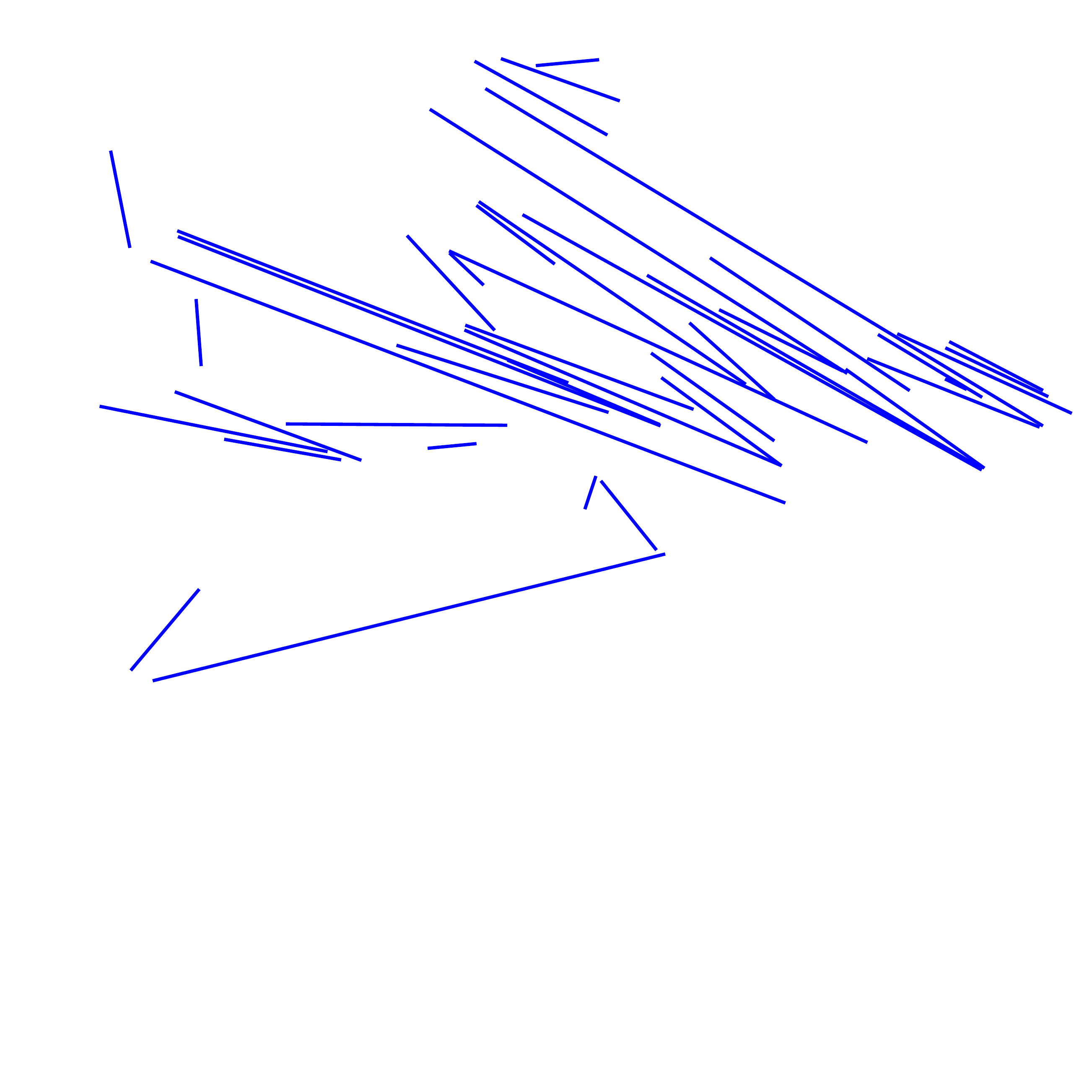}
  \includegraphics[scale=.063]{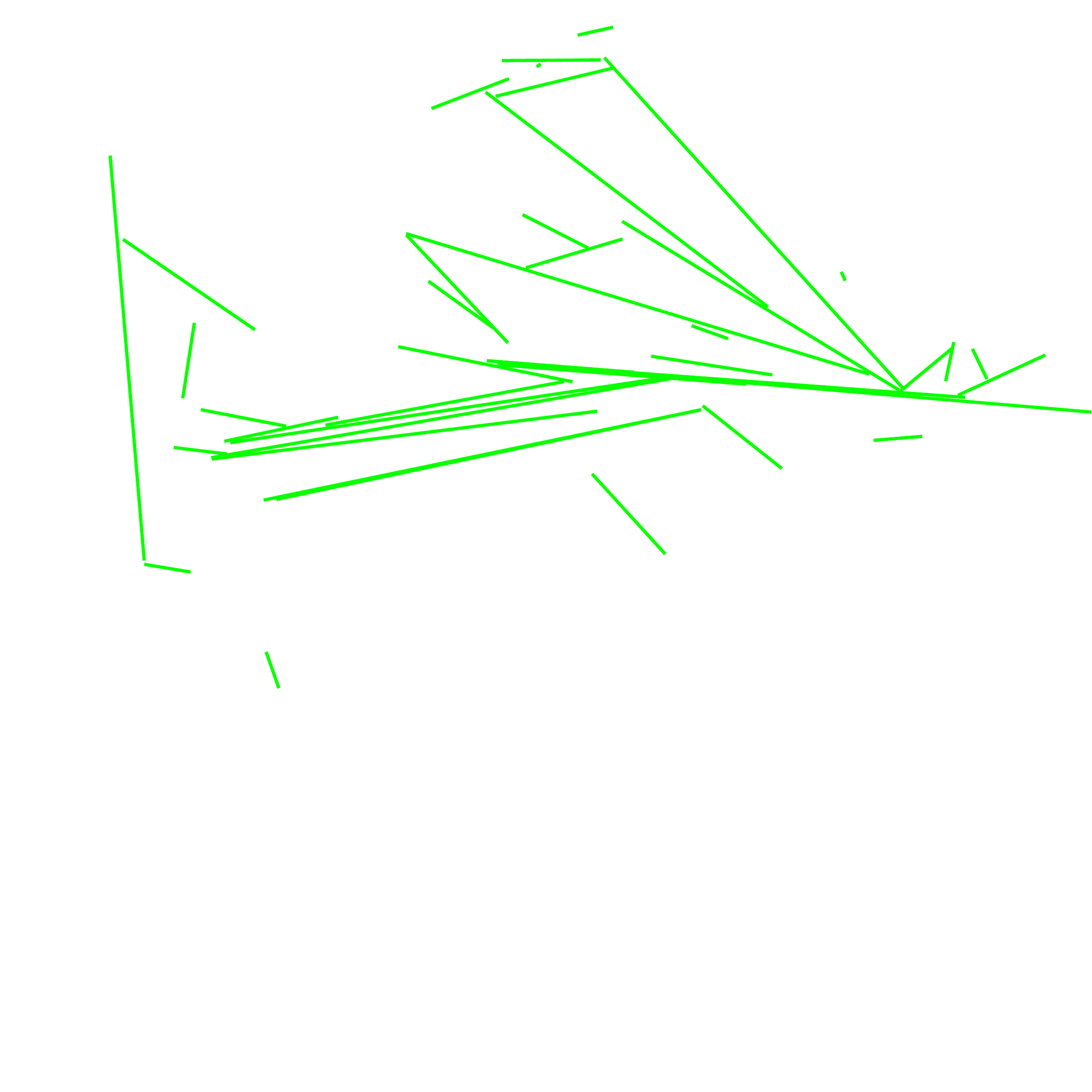}
  \includegraphics[scale=.063]{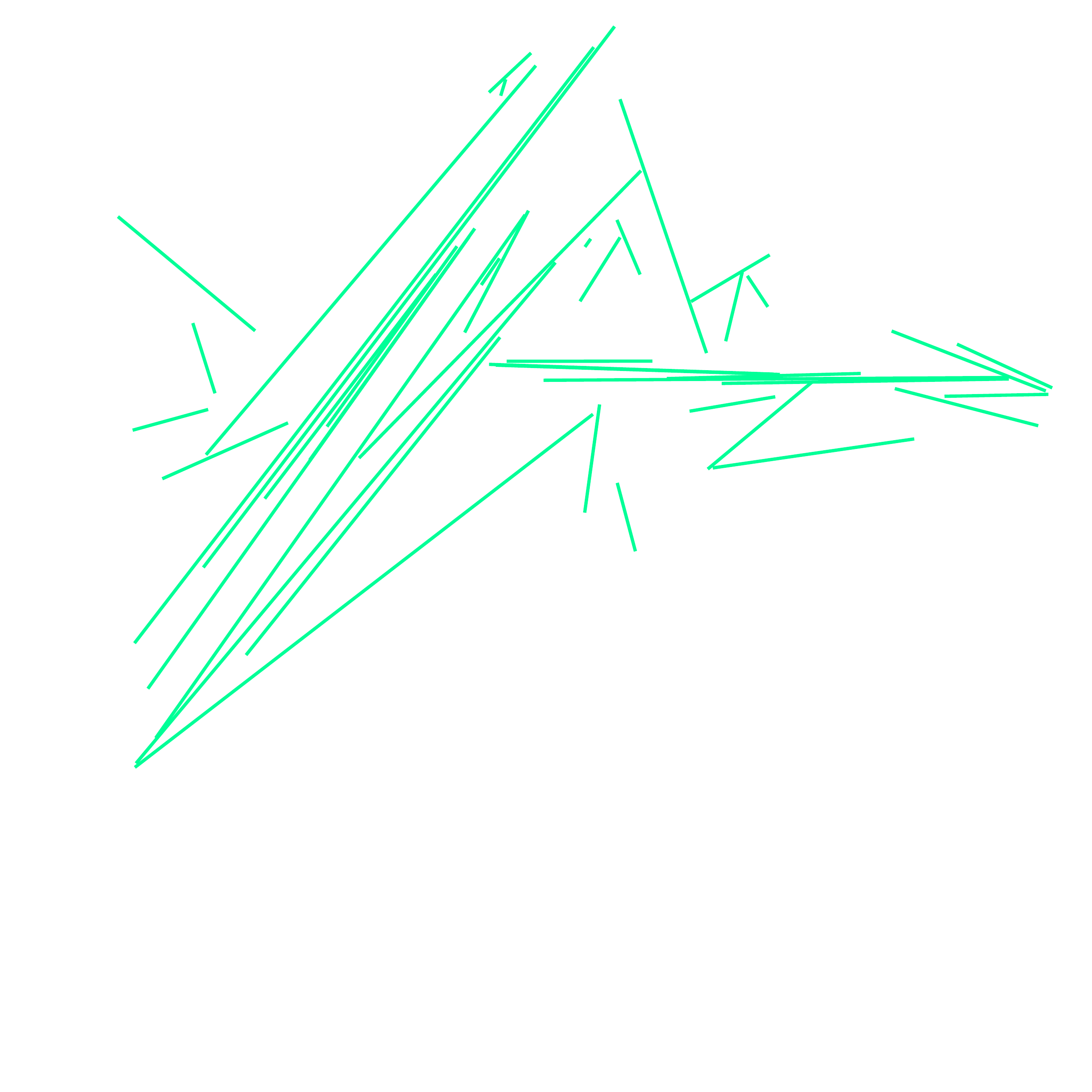}
  \includegraphics[scale=.063]{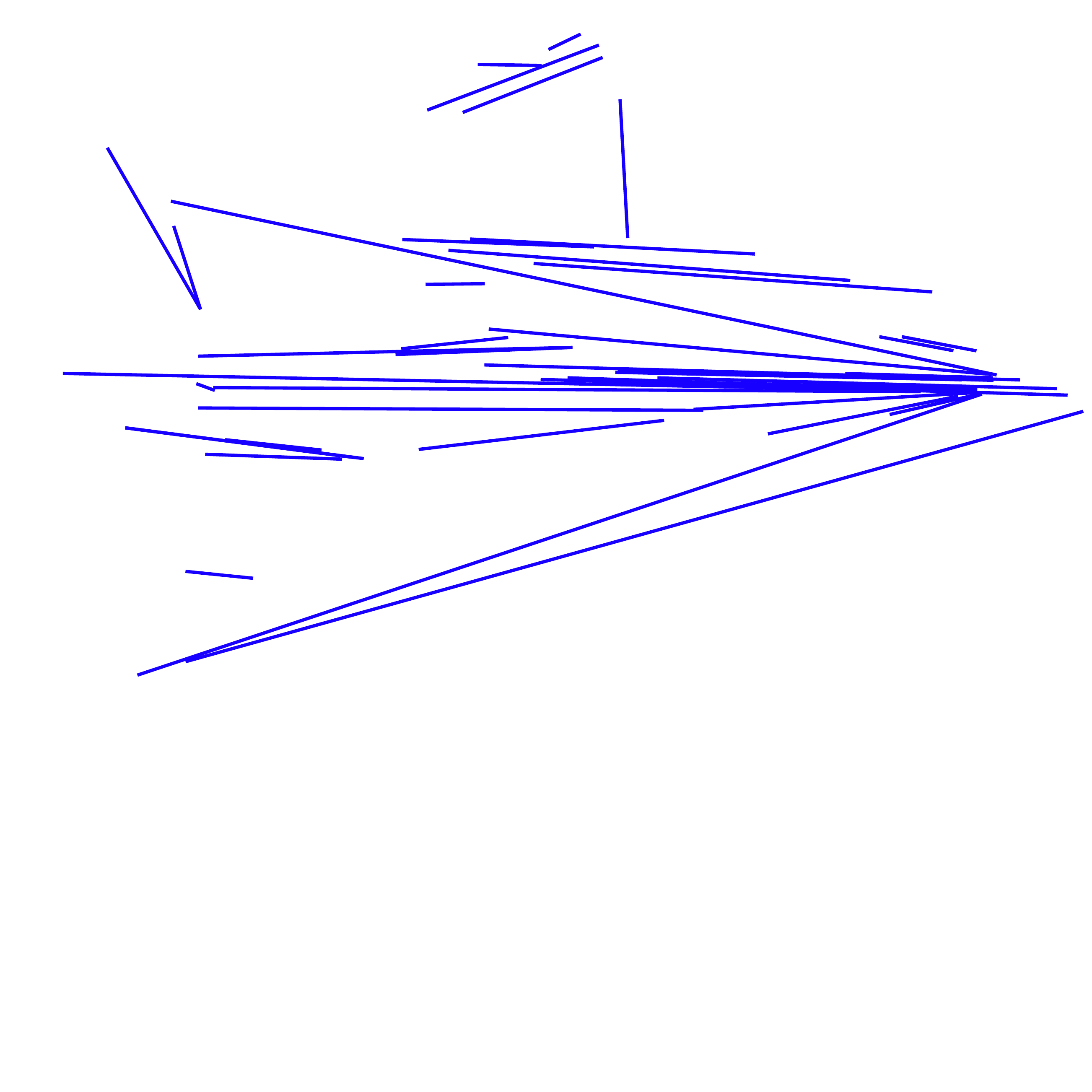}
  \includegraphics[scale=.063]{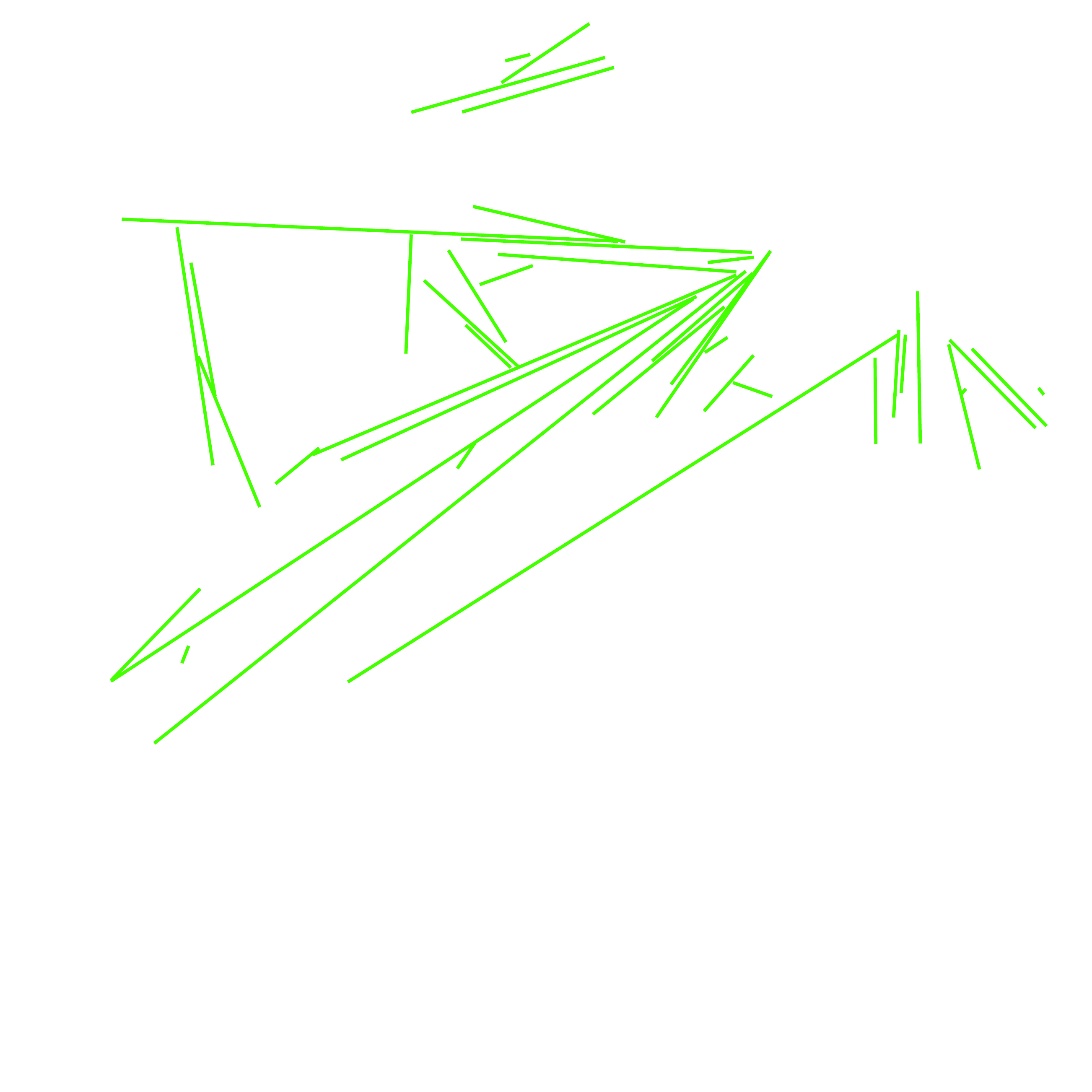}
  \includegraphics[scale=.063]{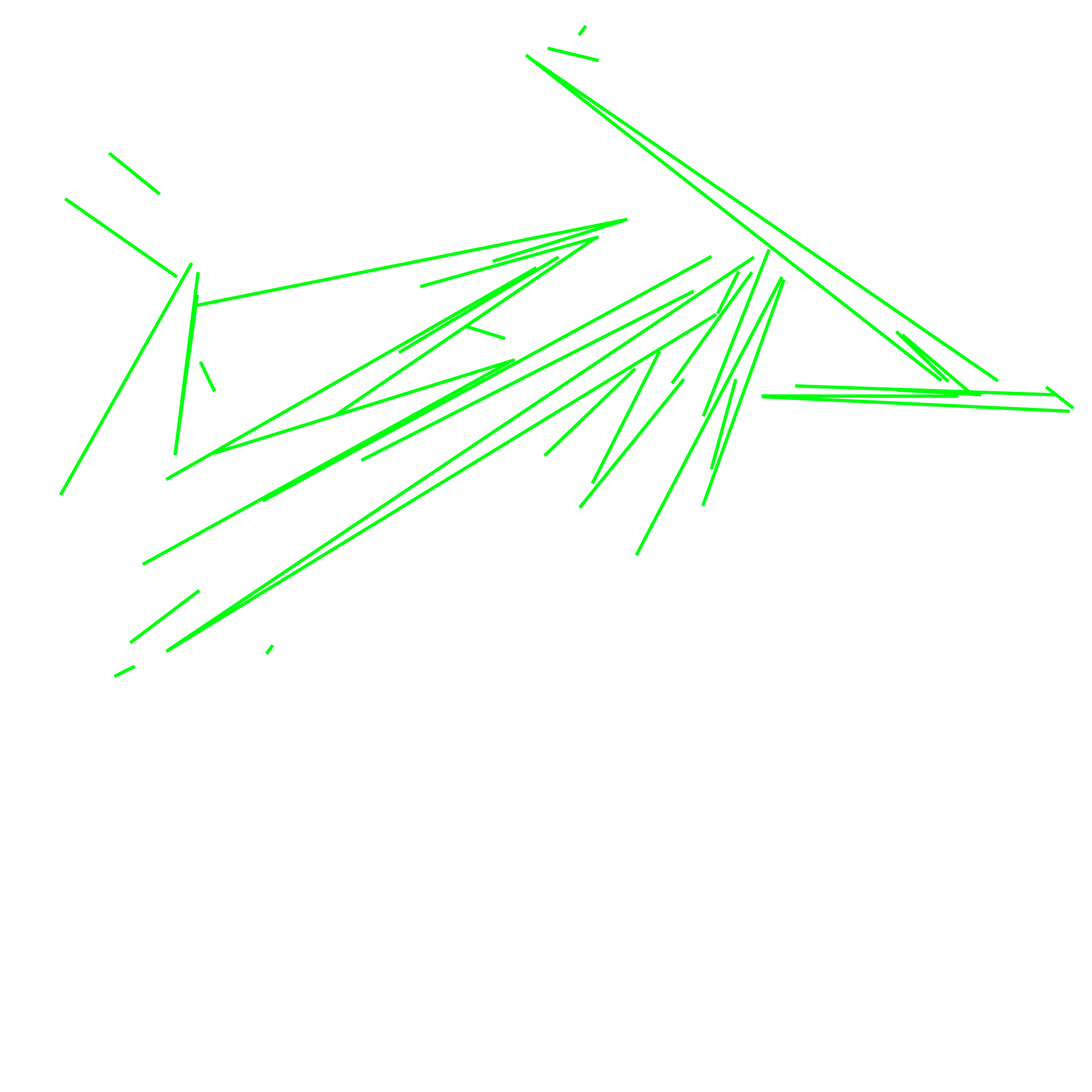}
  \includegraphics[scale=.063]{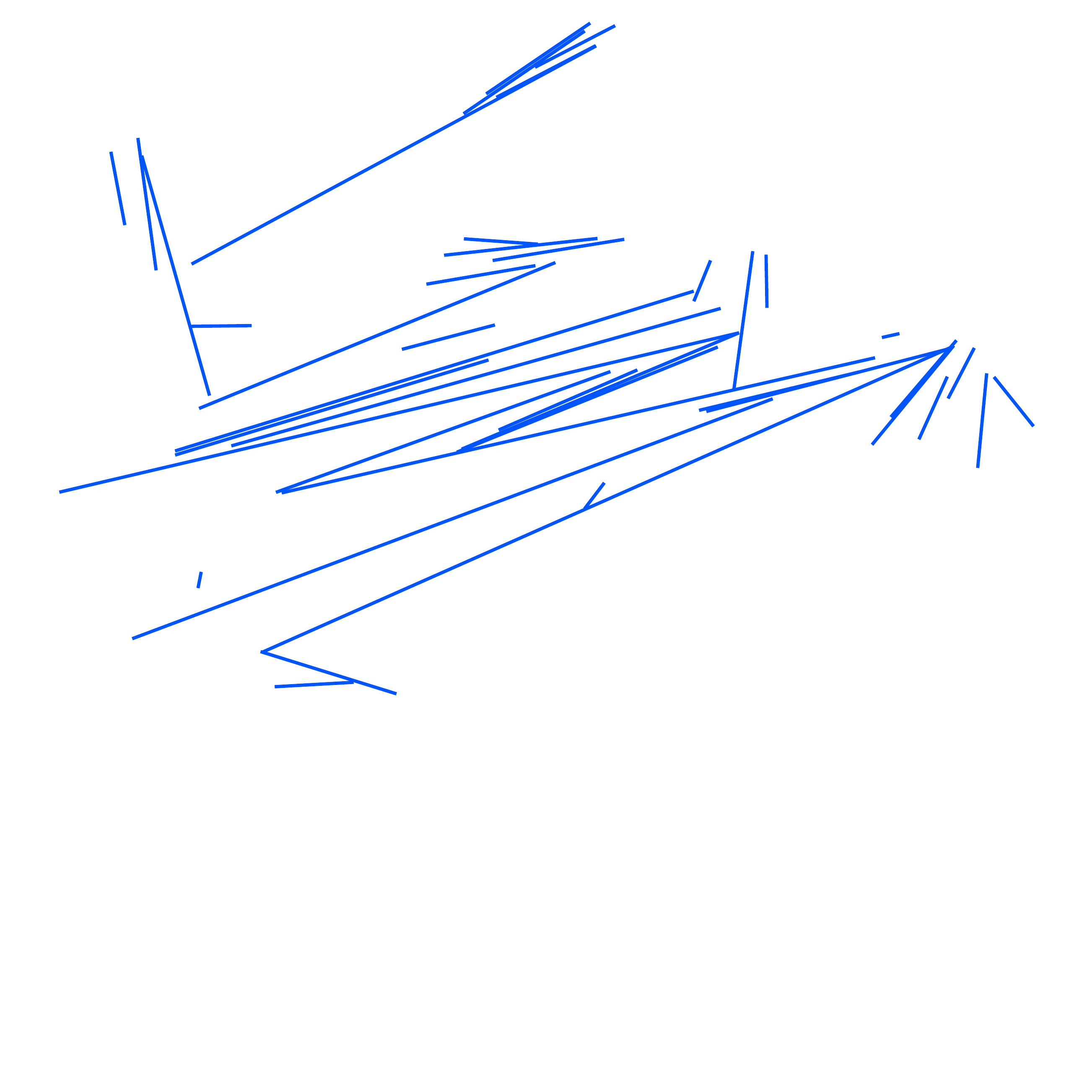}
  \includegraphics[scale=.063]{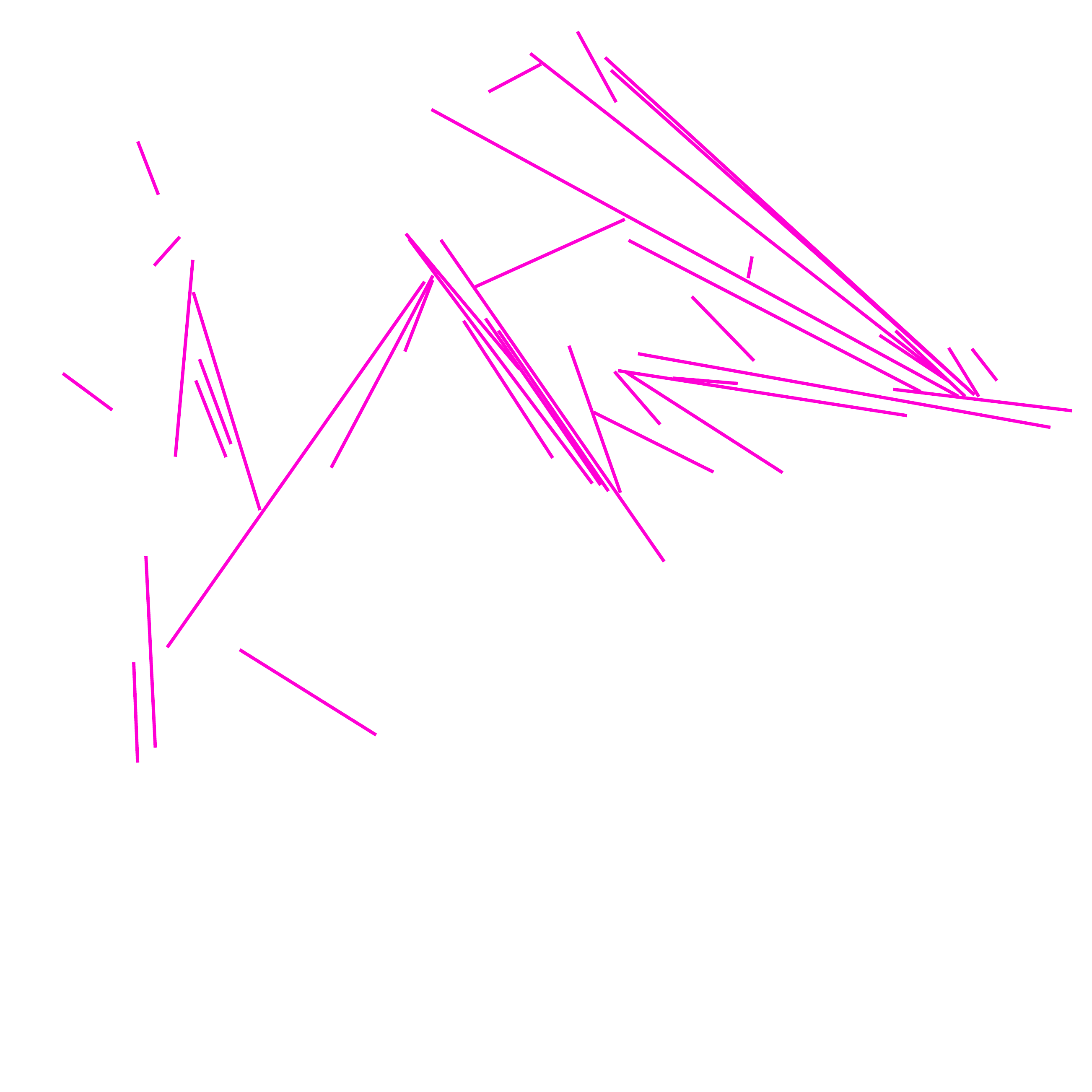}
  \includegraphics[scale=.063]{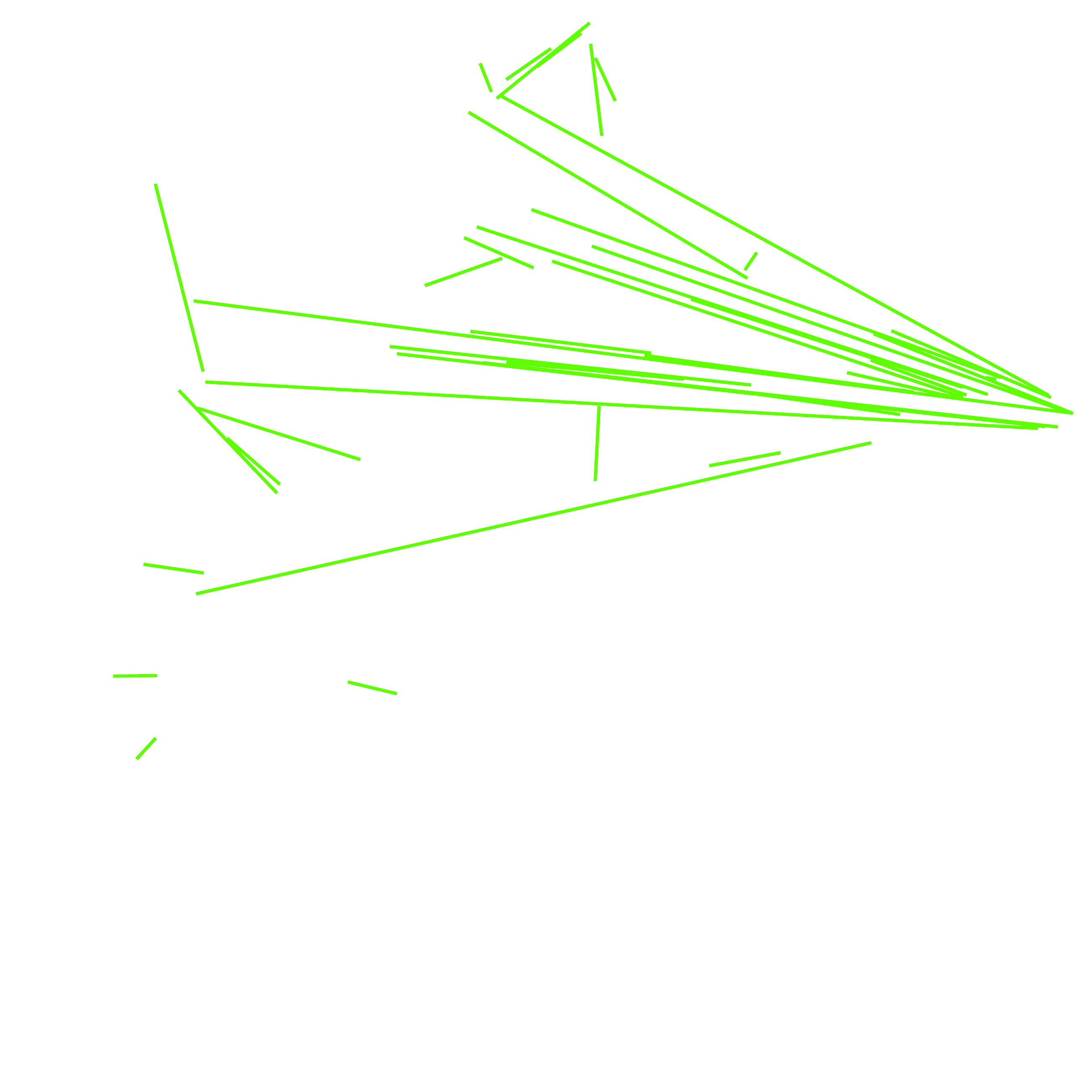}
  \includegraphics[scale=.063]{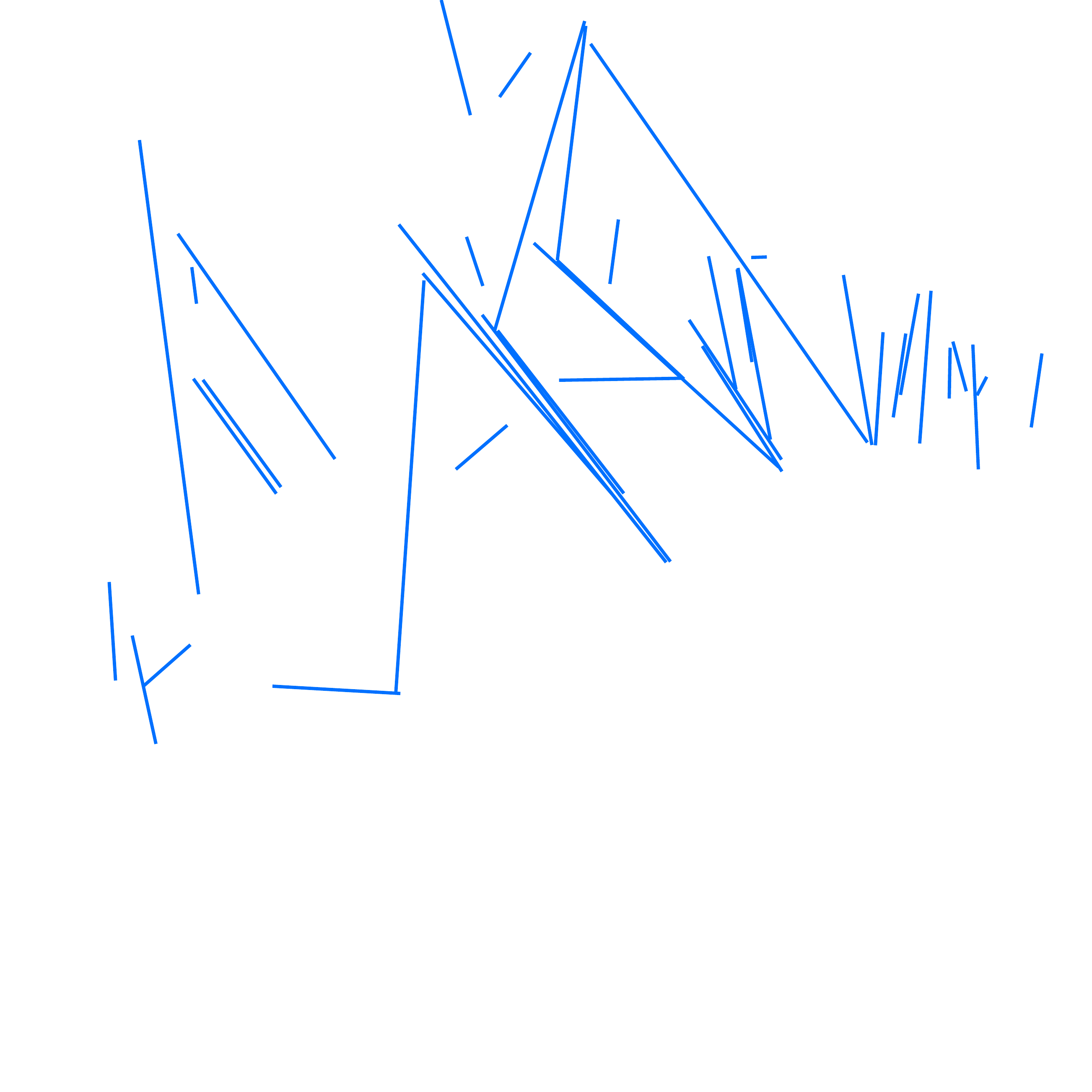}
  \includegraphics[scale=.063]{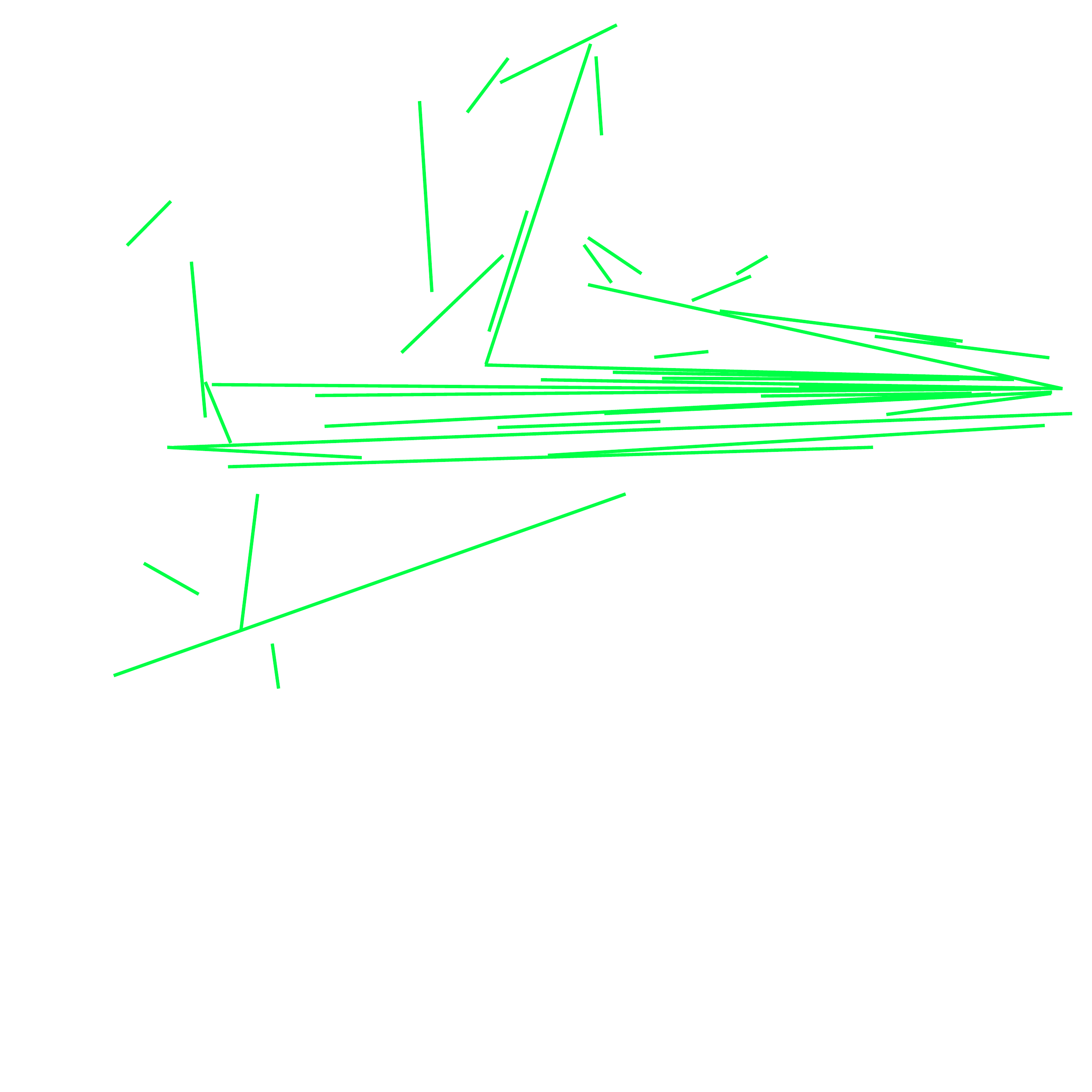}
  \includegraphics[scale=.063]{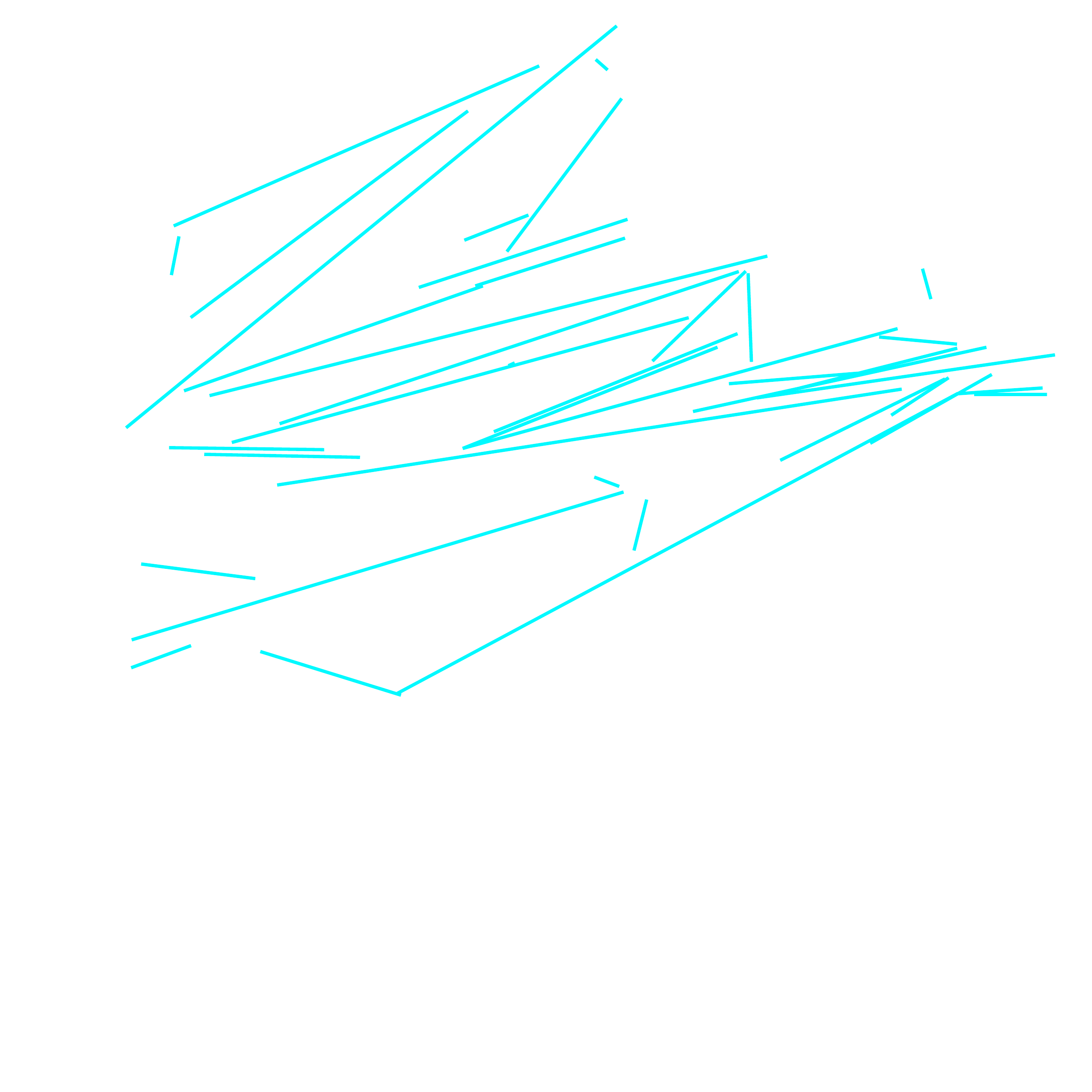}
  \includegraphics[scale=.063]{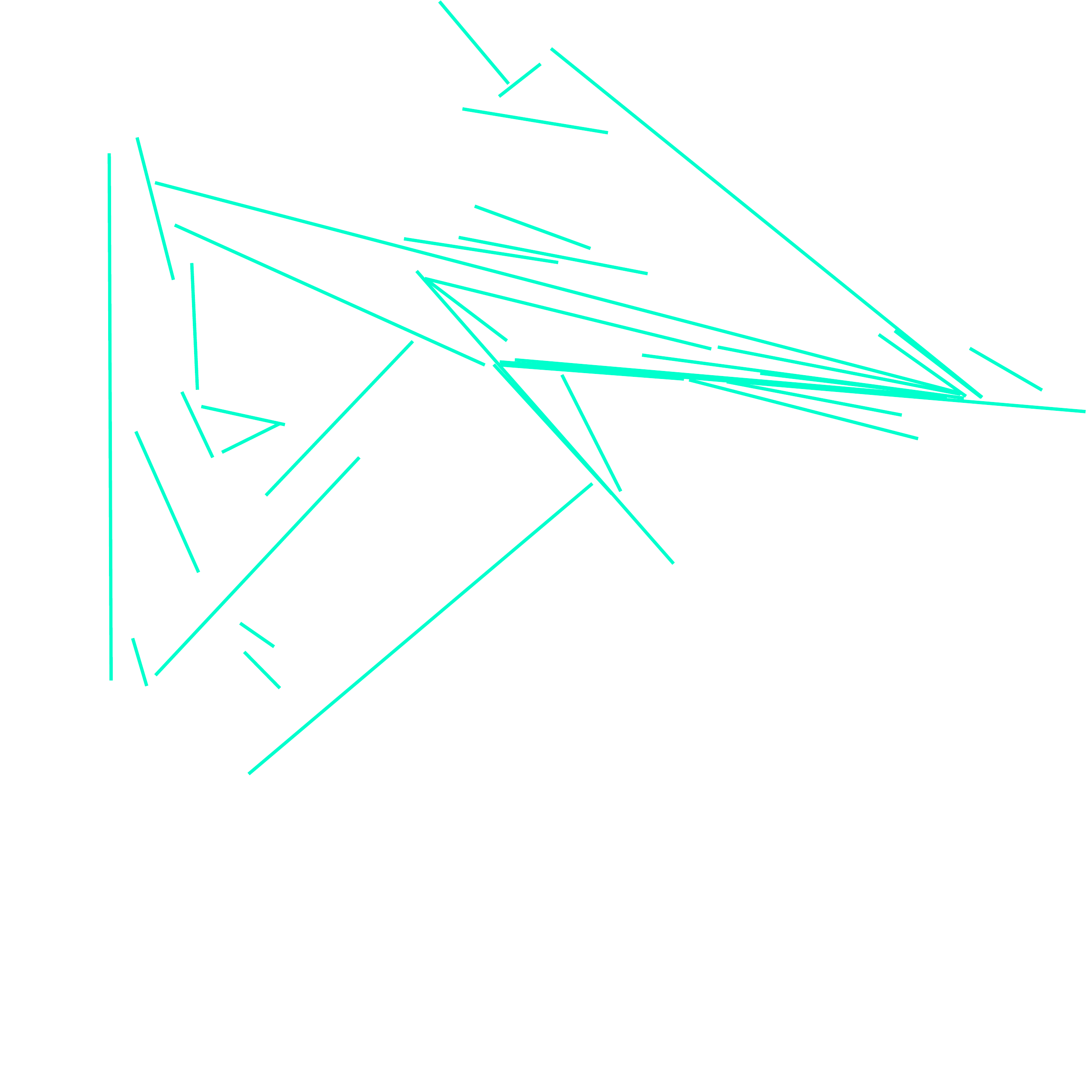}
  \includegraphics[scale=.063]{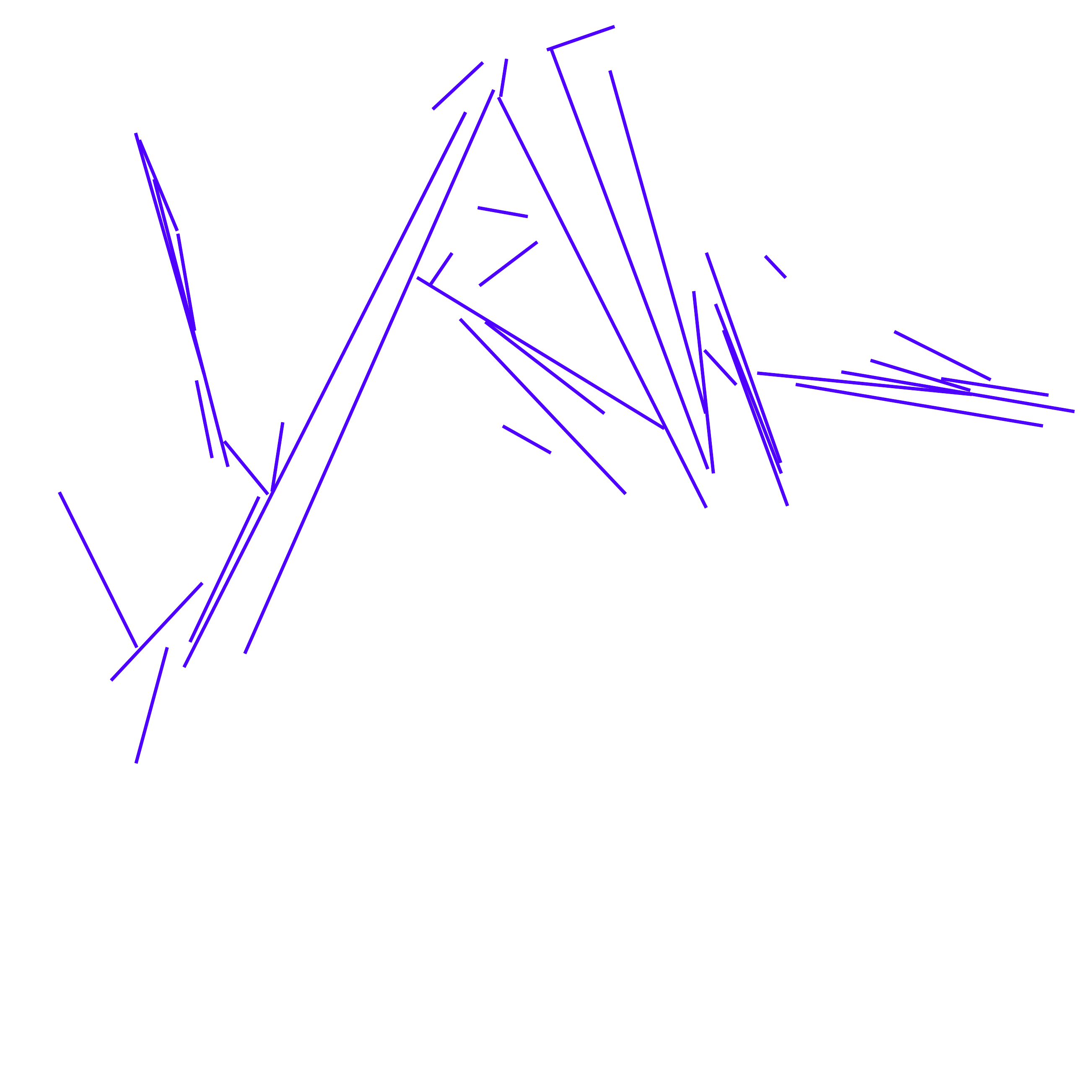}
  \includegraphics[scale=.063]{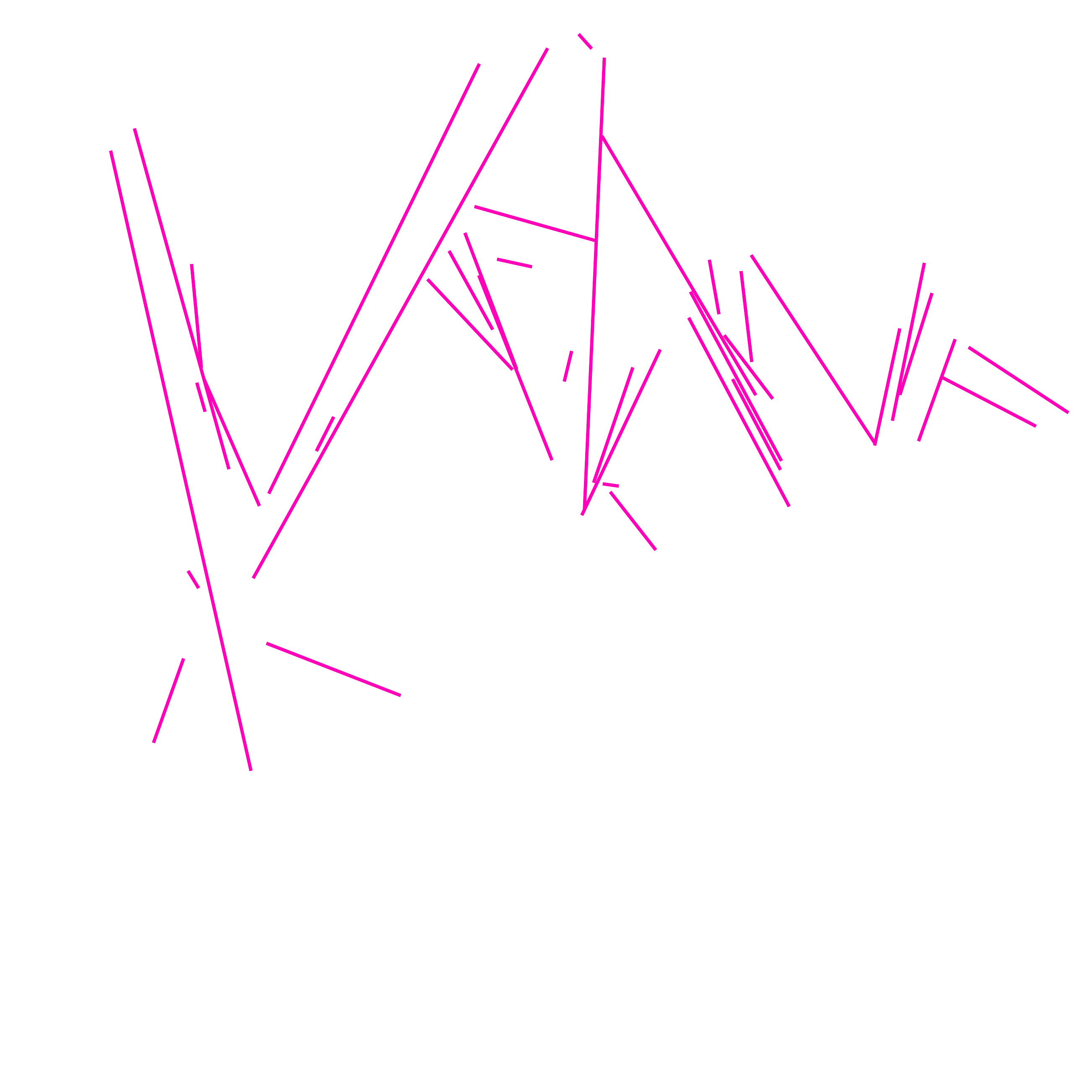}
  \includegraphics[scale=.063]{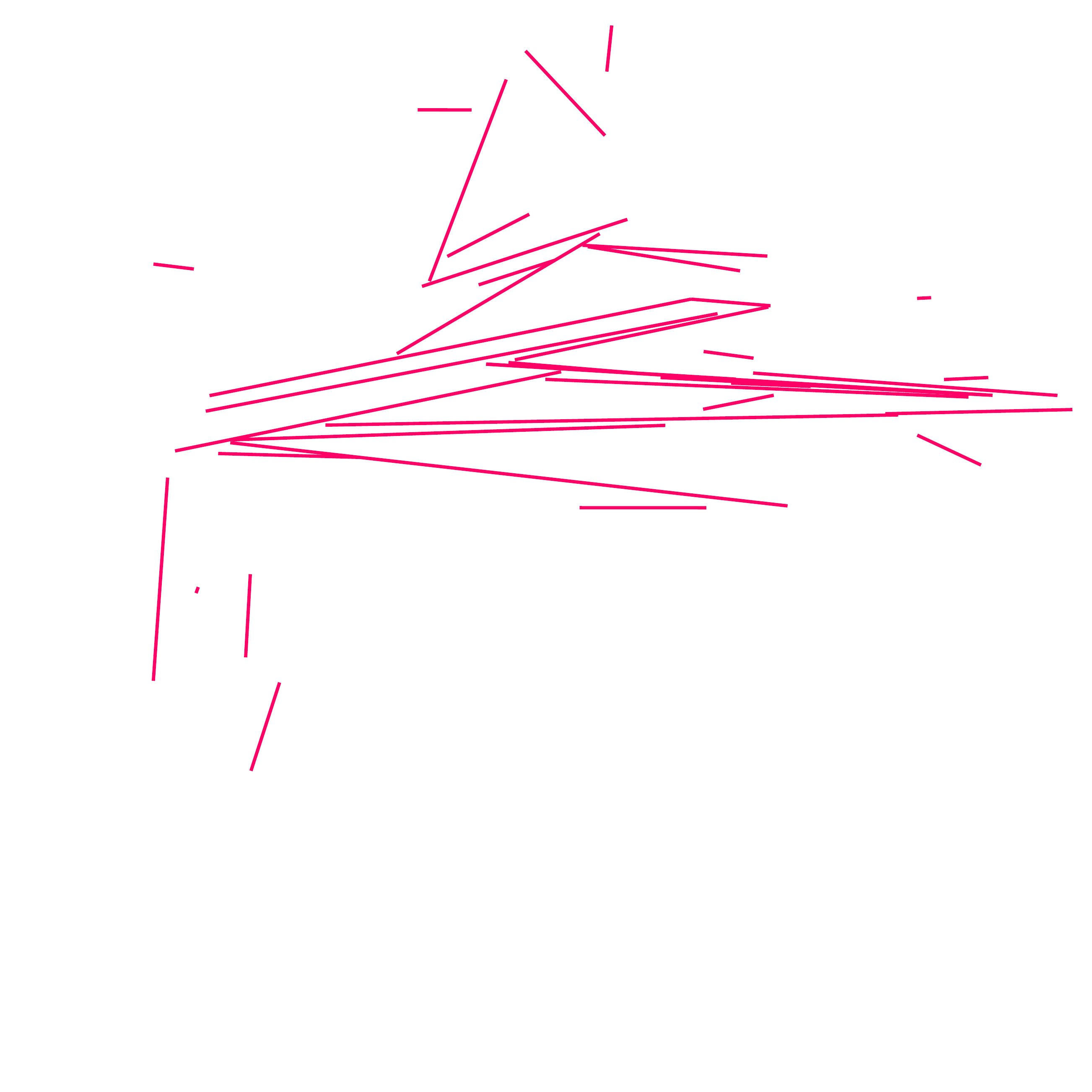}
  \includegraphics[scale=.063]{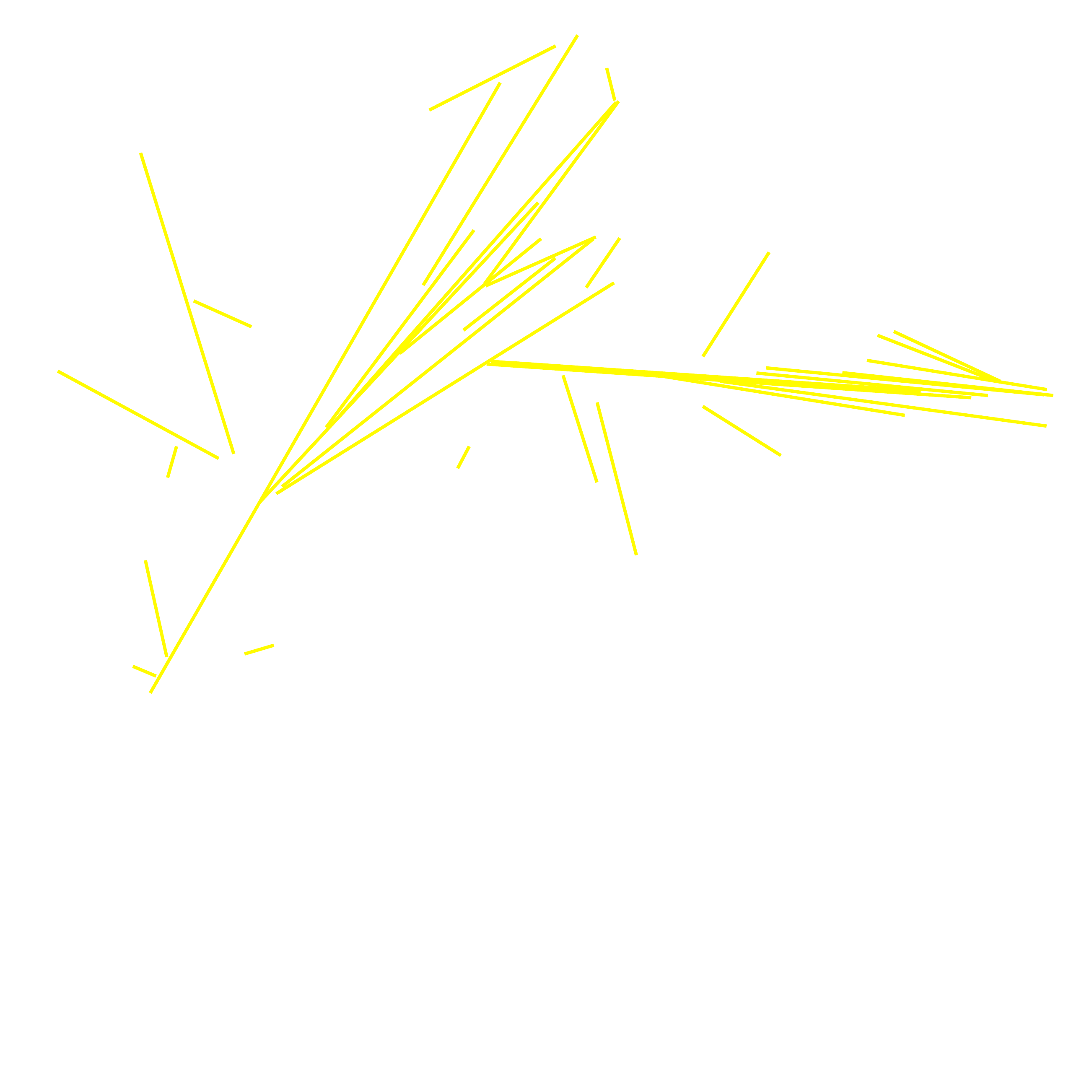}
  \includegraphics[scale=.063]{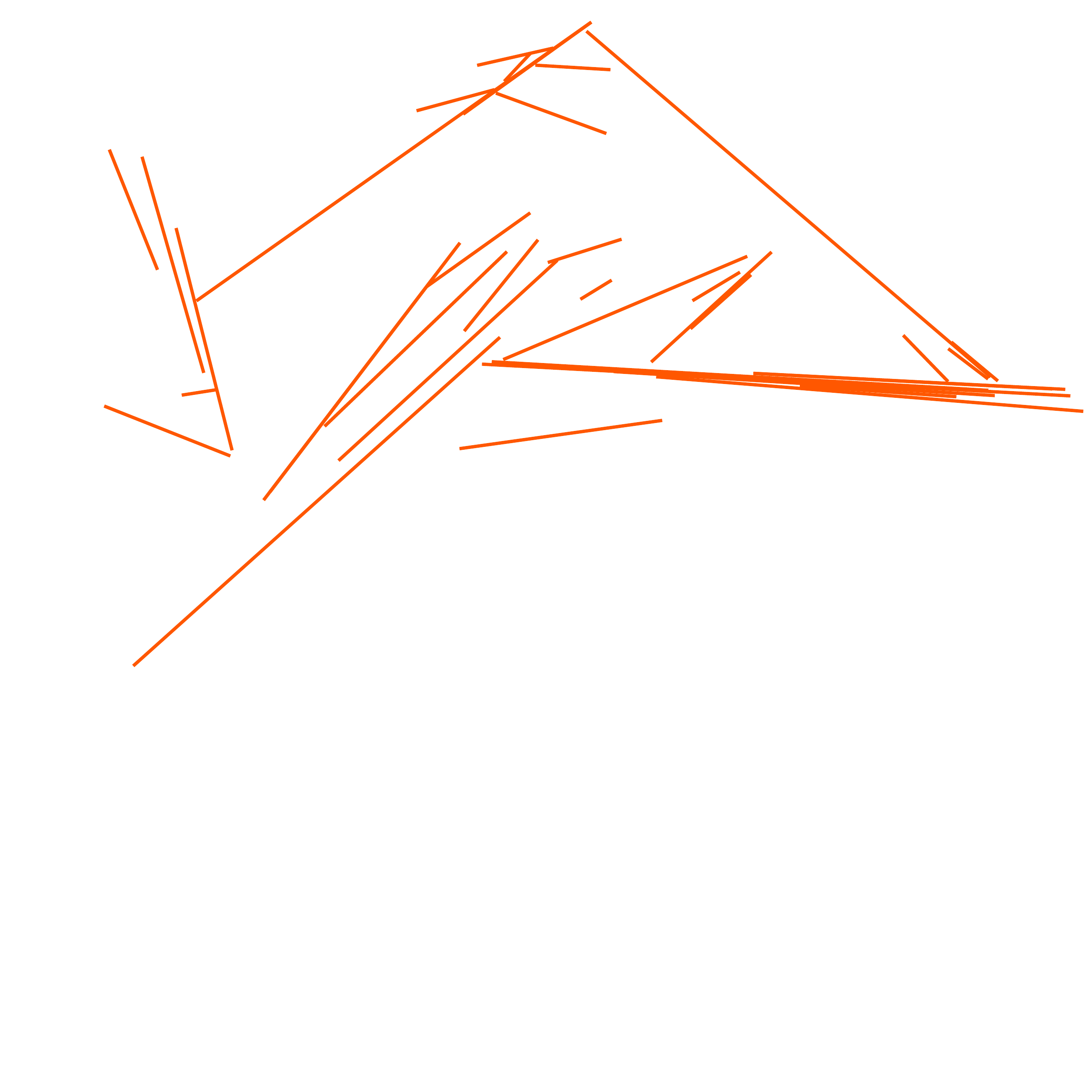}
  \includegraphics[scale=.063]{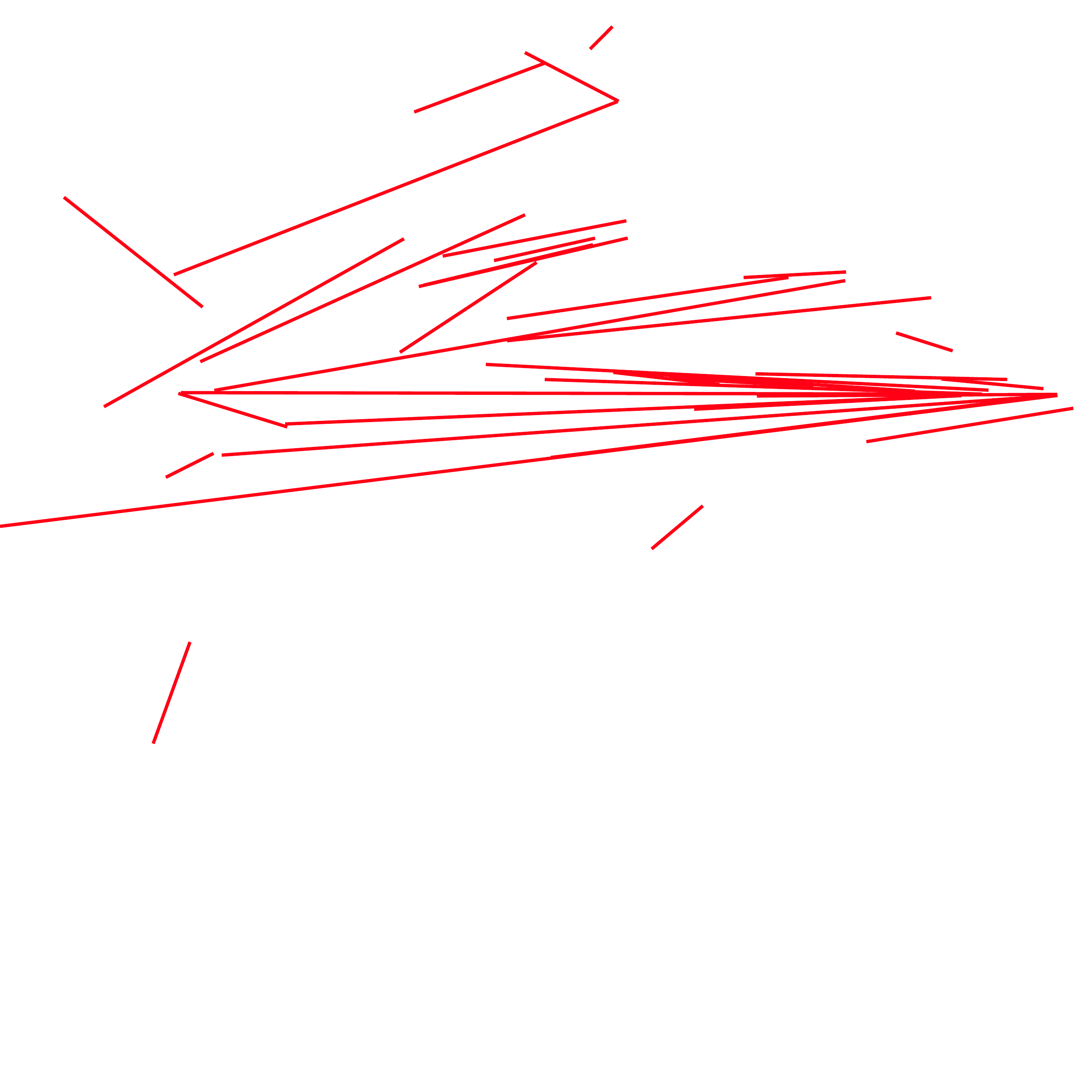}
  \includegraphics[scale=.063]{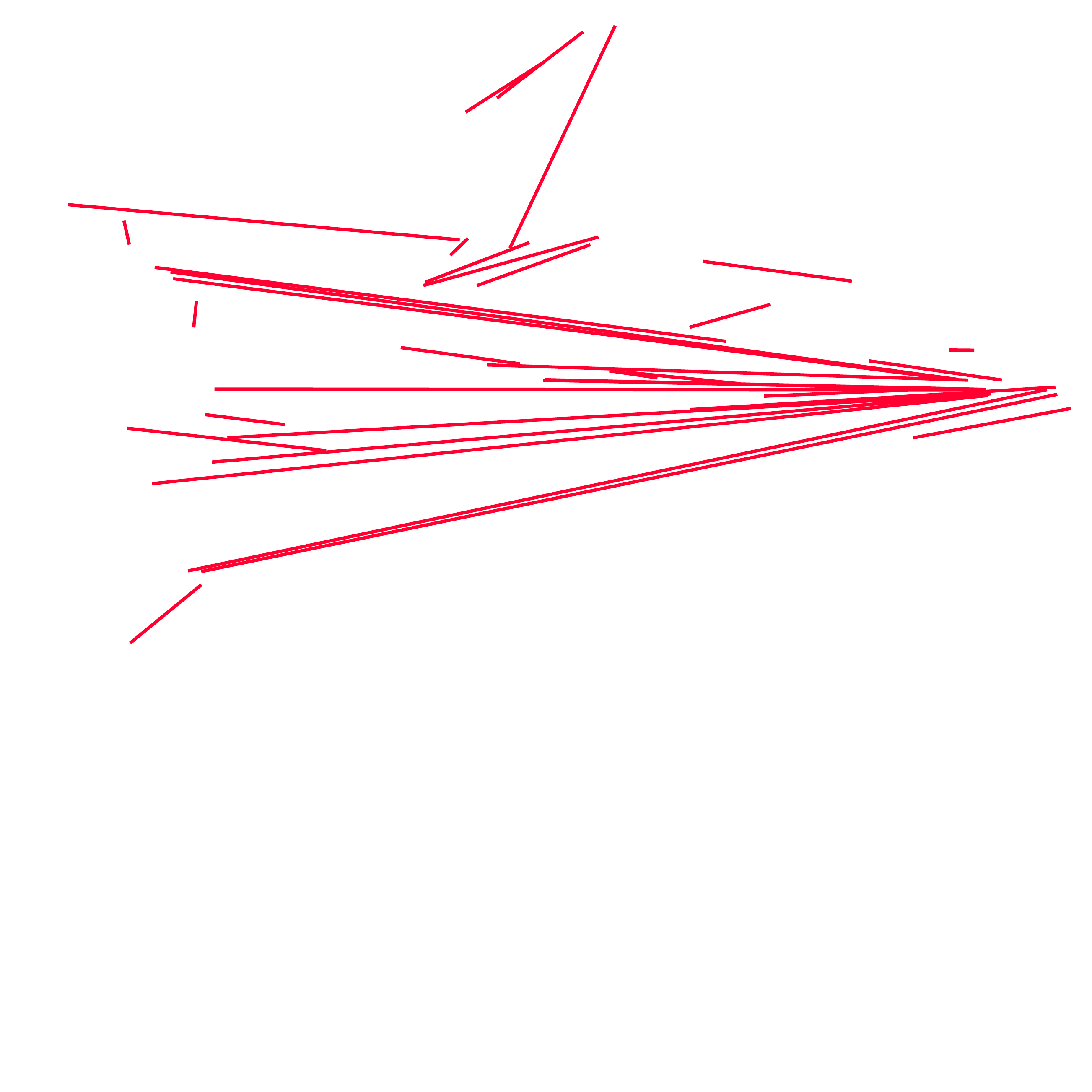}
  \includegraphics[scale=.063]{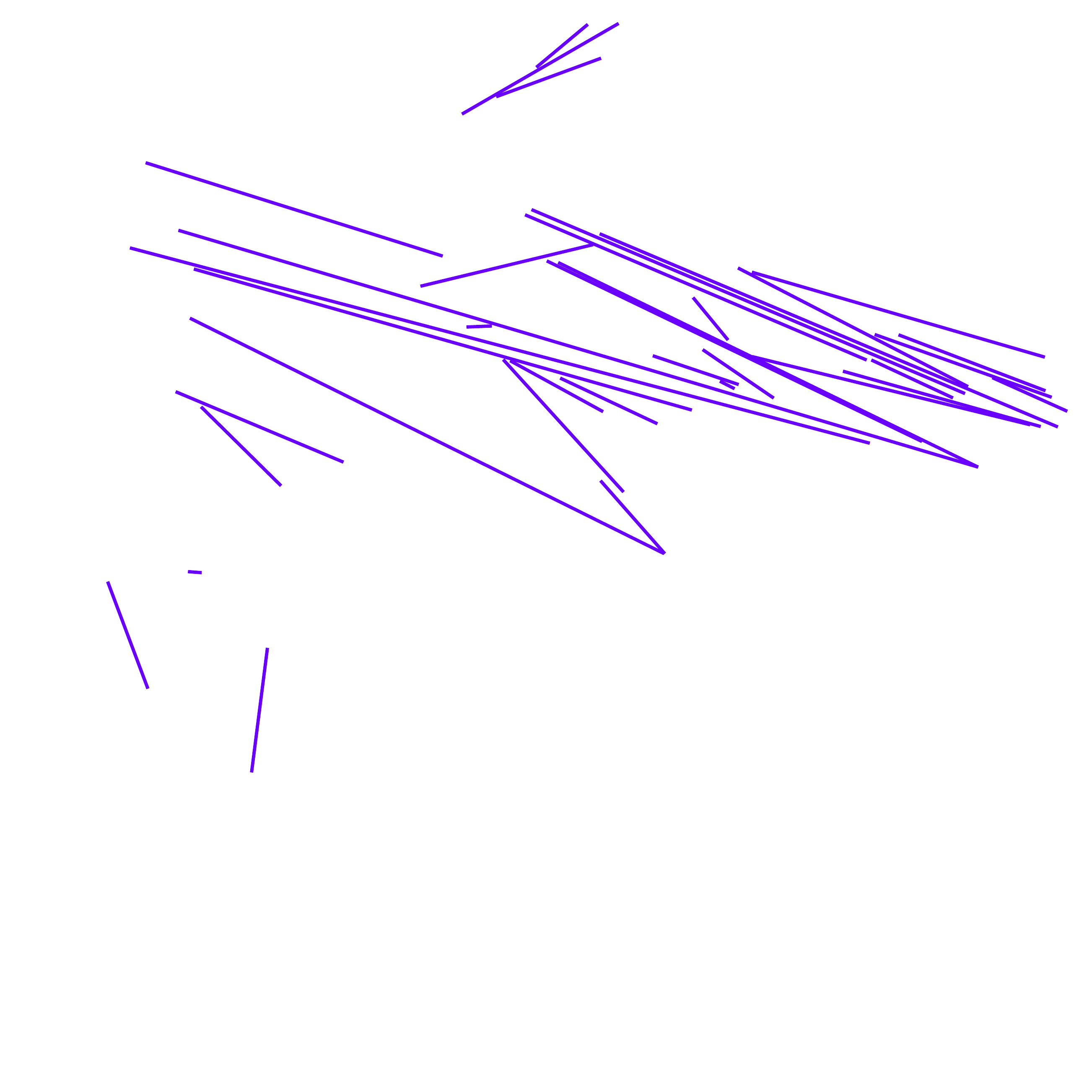}
  \includegraphics[scale=.063]{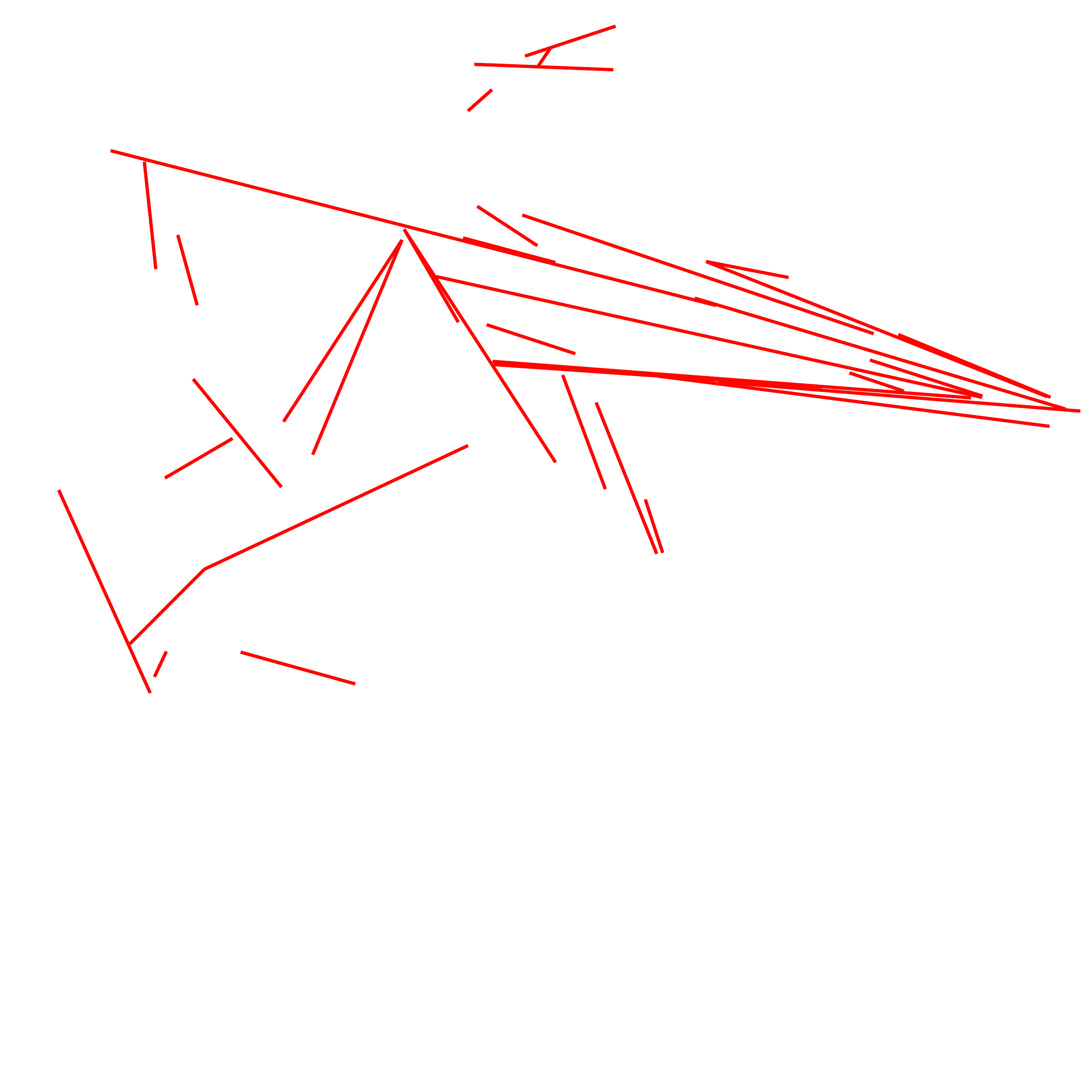}
  \includegraphics[scale=.063]{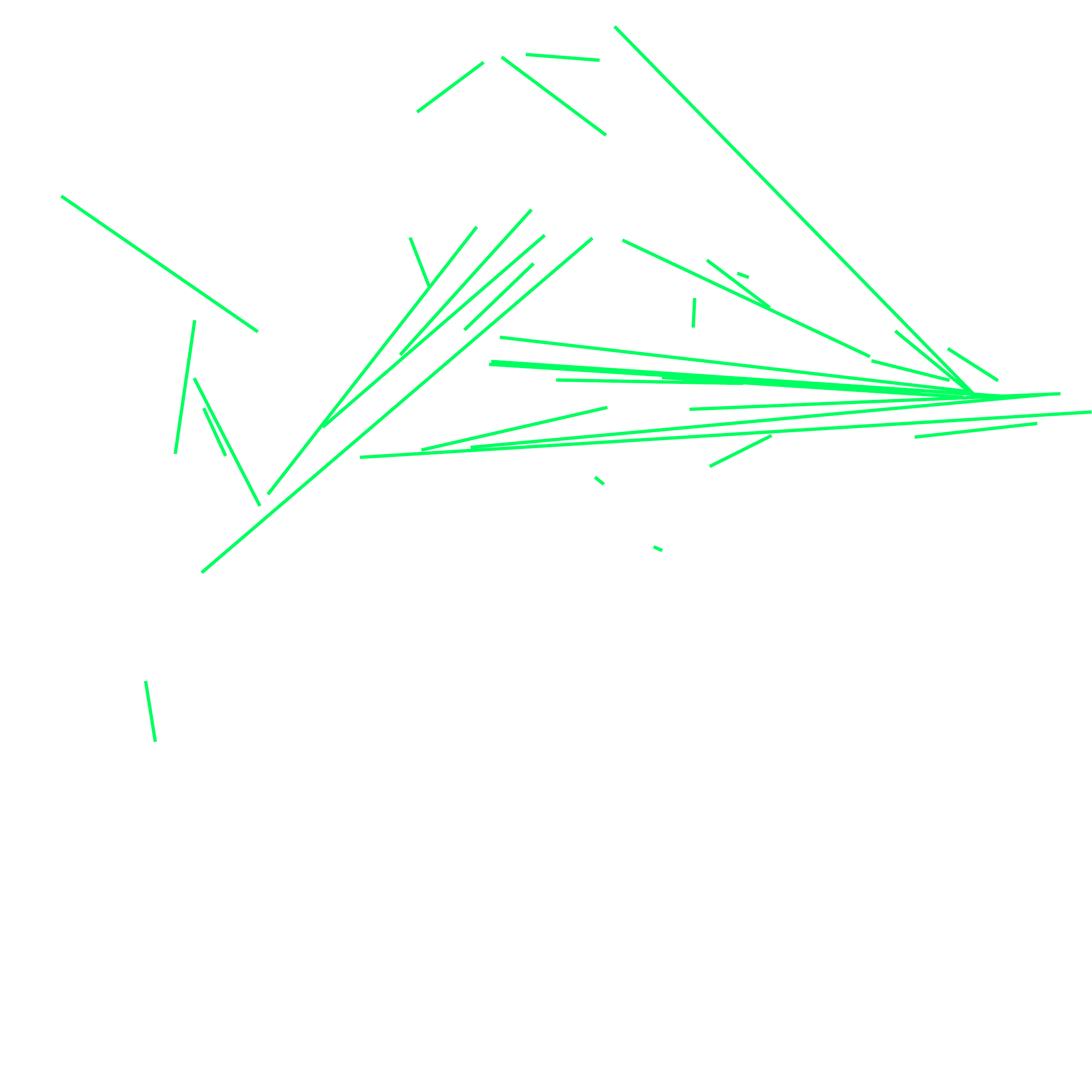}
  \caption{A partition of the input graph of the CG:SHOP2022 instance vispecn2518 into 57 plane graphs.
  It is the smallest instance of the challenge with 2518 segments. On top left, you see all 57 colors together. On top right, you see a clique of size 57, hence the solution is optimal. Each of the 57 colors is then presented in small figures.}
  \label{f:intro}
\end{figure}

The CG:SHOP challenge (Computational Geometry: Solving Hard Optimization Problems) is an annual geometric optimization competition, whose first edition took place in 2019. The 2022 edition proposed a problem called \emph{minimum partition into plane subgraphs}. 
The input is a graph $G$ embedded in the plane with edges drawn as straight line segments, and the goal is to partition the set of edges into a small number of plane graphs (Fig.~\ref{f:intro}) \cite{survey2022fekete}.
This goal can be formulated as a vertex coloring problem on a graph $G'$ defined as follows. The vertices of $G'$ are the segments defining the edges of $G$, and the edges of $G'$ correspond to pairs of \emph{crossing} segments (segments that intersect only at a common endpoint are not considered crossing).

The three top-ranking teams (Lasa, Gitastrophe, and Shadoks) on the CG:SHOP 2022 challenge all used a common approach called \emph{conflict optimization} ~\cite{Lasa,Git,Sha} while the fourth team used a SAT-Boosted Tabu Search~\cite{Sch22}. Conflict optimization is a technique used by Shadoks to obtain the first place in the  CG:SHOP 2021 challenge for low-makespan coordinated motion planning~\cite{CFG22}, and the main ideas of the technique lent themselves well to the 2022 challenge. Next, we describe the conflict optimizer as a metaheuristic to solve constraint satisfaction problems (CSP)~\cite{Tsa93}. We start by describing a CSP.

A CSP is a triple of
\begin{itemize}
    \item \emph{variables} $X = (x_1,\ldots,x_n)$,
    \item \emph{domains} $\mathcal{D} = (D_1,\ldots,D_n)$, and
    \item \emph{constraints} $\mathcal{R}$.
\end{itemize}
Each variable $x_i$ must be assigned a \emph{value} in the corresponding domain $D_i$ such that all constraints are satisfied. In general, the constraints may forbid arbitrary subsets of values. We restrict our attention to a particular type of constraints (\emph{binary CSP}), which only involve pairs of assignments. A \emph{partial evaluation} is an assignment of a subset of the variables, called \emph{evaluated}, with the remaining variables called \emph{non-evaluated}. All constraints involving a non-evaluated variable are satisfied by default. We only consider assignments and partial assignments that satisfy all constraints.

The conflict optimizer iteratively modifies a partial evaluation with the goal of emptying the set $S$ of non-evaluated variables, at which point it stops. At each step, a variable $x_i$ is removed from $S$. If there exists a value $x \in D_i$ that satisfies all constraints, then we assign the value $x$ to the variable $x_i$. Otherwise, we proceed as follows. For each possible value $x \in D_i$, we consider the set $K(i,x)$ of variables (other than $x_i$) that are part of constraints violated by the assignment $x_i = x$. We assign to $x_i$ the value $x$ that minimizes
\[\sum_{x_j \in K(i,x)} w(j),\]
where $w(j)$ is a weight function to be described later. The variables $x_j \in K(i,x)$ become non-evaluated and added to $S$.

The weight function should be such that $w(j)$ increases each time $x_j$ is added to $S$, in order to avoid loops that keep moving the same variables back and forth from $S$. Let $q(j)$ be the number of times $x_j$ became non-evaluated. A possible weight function is $w(j) = q(j)$. More generally, we can have $w(j) = q(j)^p$ for some exponent $p$ (typically between $1$ and $2$).
Of course, several details of the conflict optimizer are left open. For example, which element to choose from $S$, whether some random noise should be added to $w$, and the decision to restart the procedure from scratch after a certain time.

The CSP as is, does not apply to optimization problems. However, we can, impose a maximum value $k$ of the objective function in order to obtain a CSP. 
The conflict optimizer was introduced in a low makespan coordinated motion planning setting. In that setting, the variables are the robots, the domains are their paths (of length at most $k$) and the constraints forbid collisions between two paths.
In the graph coloring setting, the domains are the $k$ colors of the vertices and the constraints forbid adjacent vertices from having the same color.

The conflict optimizer can be adapted to non-binary CSP, but in that case multiple variables may be unassigned for a single violated constraint. The strategy has some resemblance to the similarly named \emph{min-conflicts algorithm}~\cite{MJPL92}, but notable differences are that a partial evaluation is kept instead of an invalid evaluation and the weight function that changes over time.

While the conflict optimization strategy is simple, there are different ways to apply it to the graph coloring problem. The goal of the paper is to present how the top three teams applied it or complemented it with additional strategies. We compare the relative benefits of each variant on the instances given in the CG:SHOP 2022 challenge. We also compare them to baselines on some instances issued from graph coloring benchmarks.

The paper is organized as follows. Section~\ref{s:conflict} presents the details of the conflict optimization strategy applied to graph coloring. In the three sections that follow, the three teams Lasa, Gitastrophe, and Shadoks present the different parameters and modified strategies that they used to make the algorithm more efficient for the CG:SHOP 2022 challenge. The last section is devoted to the experimental results.

\subsection{Literature Review}

The study of graph coloring goes back to the 4-color problem (1852) and it has been intensively studied since the 1970s (see~\cite{JTB11,Lew15} for surveys). Many heuristics have been proposed~\cite{GPJ99,tabucol,MMI72,MIZ06}, as well as exact algorithms~\cite{Epp02,GSF12,LMM06}. We briefly present two classes of algorithms: greedy algorithms and exact algorithms.

\paragraph{Greedy algorithms.} 
These algorithms are used to find good quality initial solutions in a short amount of time.
The classic greedy heuristic considers the vertices in arbitrary order and colors each vertex with the smallest non-conflicting color.
The two most famous modern greedy heuristics are \emph{DSATUR}~\cite{bre79} and \emph{Recursive Largest First} (\emph{RLF})~\cite{leighton1979graph}.
At each step (until all vertices are colored), DSATUR selects the vertex $v$ that has the largest number of different colors in its neighbourhood. Ties are broken by selecting a vertex with maximum degree. The vertex $v$ is colored with the smallest non-conflicting color. 
\emph{RLF} searches for a large independent set $I$, assigns the vertices $I$ the same color, removes $I$ from $G'$, and repeats until all vertices are colored.

\paragraph{Exact algorithms.}
Some exact methods use a branch-and-bound strategy, for example extending the DSATUR heuristic by allowing it to backtrack~\cite{san2012new,furini2017improved}.
Another type of exact method (branch-and-cut-and-price) decomposes the vertex coloring problem into an iterative resolution of two sub-problems~\cite{mehrotra1996column,GSF12,furini2012exact}.
The ``master problem'' maintains a small set of valid colors using a set-covering formulation.
The ``pricing problem'' finds a new valid coloring that is promising by solving a maximum weight independent set problem.
Exact algorithms are usually able to find the optimal coloring for graphs with a few hundred vertices.
However, even the smallest CG:SHOP 2022 competition instances involve at least a few thousands vertices.

\section{Conflict Optimization for Graph Coloring} \label{s:conflict}

Henceforth, we will only refer to the intersection conflict graph $G'$ induced by the instance.
Vertices will refer to the vertices $V(G')$,
and edges will refer to the edges $E(G')$.
Our goal is to partition the vertices using a minimum set of $k$ color classes $\mathcal C = \{C_1,\ldots,C_k\}$,
where no two vertices in the same color class $C_i$ are incident to a common edge.

\subsection{Conflict Optimization}\label{ss:conflict}

We consider the classical problem of coloring the vertices of a graph $G' = (V(G'), E(G'))$.
We assume that an initial solution $\mathcal C = \{C_1,\ldots,C_k\}$ has been previously computed (the choice of the initial solution does not seem to impact the quality of the final solution produced by the conflict optimizer). The goal of the conflict optimizer is to reduce the number of colors of $\mathcal C$ by one. When (and if) the conflict optimizer terminates, it will give such a solution. However, after a certain amount of time or when a certain situation arrives, we may decide to abort the  execution of the conflict optimizer without any solution, and perhaps try again.

Throughout the execution, we maintain a partial coloring, which is a valid coloring for a subset of the vertices. The complementary subset of uncolored vertices is called the \emph{conflict set} and denoted $S$.
The conflict optimizer proceeds as follows:

\begin{enumerate}
    \item Pick a color class $C_i$ to be eliminated. 
        Uncolor all vertices in $C_i$ and make $S \leftarrow C_i$.
        A valid vertex-coloring is maintained for the set $V(G') \setminus S$.
        If $S$ is empty, we have a valid vertex coloring of $G'$ which uses one fewer color.

    \item Pick and remove an element $v$ from $S$.
    For each color class, compute the
    \emph{conflict score} with $v$.
    The conflict score of a color class $C_j$ is
    \begin{equation} \label{eq:score} score(C_j)= f(C_j)  \sum_{\substack{u \in C_j\\(u, v) \in E(G')}} w(u)
    \end{equation} 
    where the weight $w(u)$ is a variable depending on the the number of times
    that $u$ has been removed from the conflict set $S$ 
    in previous iterations, and where $f(C_j)$ is a random variable adding randomness in the process.
    
    \item Pick the color class $C_j$ with the lowest conflict score.
    Uncolor all vertices in $C_j$ which are adjacent to $v$
    and add those vertices to $S$. This step is slightly modified when the BDFS option detailed in the later is activated. In this case, the algorithm does not put in the conflict set $S$ all the vertices in conflict with $S$. Some of them are recolored easily so that they do not enter in the conflict set $S$.
    Insert $v$ into $C_j$.
    \item Repeat steps 2 and 3 until the set $S$ is empty. 
\end{enumerate}

The three teams provided different variants of the algorithm by playing with different options of the optimizer.

\begin{itemize}
    \item[(a)] The first option is the choice of the initial color $C_i$ to be eliminated at the first step of the loop. It is random for Gitastrophe, and the smallest color class for Shadoks and Lasa variants.
    \item[(b)] The second option is the way to choose the element $v$ from $S$ in step 2. Random for Gitastrophe, a fifo queue for Shadoks, and the element that provides the least total conflict score after its removal for Lasa.
    \item[(c)] The third option is the choice of the weight function $w(\cdot)$ defined on the vertices. Different functions can be used, all depending on the parameter $q(u)$ that is defined as the number of times that a vertex $u$ has been removed from $S$.    
    Lasa uses $w(u) = 1+q(u)$.
    Gitastrophe uses $w(u) = 1+q(u)^2$.
    Shadoks uses $w(u) = 1+q(u)^p$ with $p \in [1, 2]$. Shadoks also add a threshold $q_{\max}$ with $w(u) = \infty$ if $q(u) > q_{\max}$. Gitastrophe also has such a threshold, but instead uses it as a heuristic to abort the execution and start again.
    \item[(d)] The fourth option is the choice of $f(C_i)$. Lasa and Gitastrophe simply set $f(C_i) = 1$, while Shadoks use a Gaussian random variable with average $1$ for $f(C_i)$. The right amount of randomness, controlled by the variance $\sigma$, has a significant impact on the search time. 
    \item[(e)] The fifth option is that Shadoks add a Bounded Depth-First Search (BDFS) option which detects vertices that can be recolored easily. These vertices are recolored immediately, instead of entering $S$, and consequently does not suffer an increase in the value of $q(\cdot)$.
\end{itemize}

Some extra options are useful in order to drive the computation.  

\begin{itemize}
    \item Restart: The computation is restarted from step 2 if the size of the conflict set $S$ becomes too large because the coloring of $V(G')\setminus S$ has deteriorated too much to come back to a valid coloring. 
    \item Multistart: Shadoks also use a multistart option to restart from step 1 with a random eliminated color $C_i$ and a color shuffle.
\end{itemize}

The different parameters, options and complementary strategies used by each team are described in the next three sections.

\section{Lasa Team} \label{s:LucPascal}

\subsection{Finding Initial Solutions}

Lasa team used two approaches to find initial solutions:
\begin{enumerate}
    \item {\bf DSATUR} is the classical graph coloring algorithm presented in Section \ref{s:intro}.
    \item {\bf Orientation greedy} is almost the only algorithm where the geometry of the segments is used. 
    If segments are almost parallel, it is likely that they do not
    intersect (thus forming an independent set). This greedy algorithm first sorts the segments by orientation,
    ranging from $-\frac{\pi}{2}$ to $\frac{\pi}{2}$. For each segment
    in this order, the algorithm tries to color it using the first available
    color. If no color has been found, a new color  is created for coloring the considered segment. This algorithm is
    efficient, produces interesting initial solutions and takes into
    account the specificities of the competition.
\end{enumerate}

\subsection{Conflict Optimization}

\paragraph{TABUCOL inspired neighbourhood}
One classical approach for the vertex coloring involves allowing solutions with conflicting vertices (two adjacent vertices with the same color).
It was introduced in 1987~\cite{tabucol} and called TABUCOL.
It starts with an initial solution, removes a color (usually the one with the least number of vertices), and assigns
uncolored vertices with a new color among the remaining ones.
This is likely to lead to some conflicts (\emph{i.e.} two adjacent vertices sharing a same color).
The local search scheme selects a conflicting vertex, and tries to swap its color, choosing the new coloring that minimises the number of conflicts.
If it reaches a state with no conflict, it provides a  solution with one color less than the initial solution.
The process is repeated until the stopping criterion is met.
While the original TABUCOL algorithm includes a ``tabu-list'' mechanism to avoid cycling, it is not always sufficient, and requires some hyper-parameter tuning in order to obtain a good performance on a large variety of instances. To overcome this issue, we use a neighbourhood, but replace the ``tabu-list'' by the conflict optimizer scheme presented above.

\paragraph{PARTIALCOL inspired neighbourhood}
PARTIALCOL another local search algorithm solving the vertex coloring problem was introduced in 2008.
This algorithm proposes a new local search scheme that allows partial coloring (thus allowing uncolored vertices).
The goal is to minimize the number of uncolored vertices.
Similarly to TABUCOL, PARTIALCOL starts with an initial solution, removes one color (unassigning its vertices), and performs local search iterations until no vertex is left uncolored.
When coloring a vertex, the adjacent conflicting vertices are uncolored.
Then, the algorithm repeats the process until all vertices are colored, or the stopping criterion is met.
This neighbourhood was also introduced alongside a tabu-search procedure. The tabu-search scheme is also replaced by a conflict-optimization scheme. Note that this neighbourhood was predominantly used by the other teams.

\section{Gitastrophe} \label{s:Gitastrophe}

\subsection{Solution Initialization}
\label{subsec:GitInit}
The gitastrophe team uses the traditional greedy algorithm of Welsh and Powell \cite{Wel67}
to obtain initial solutions: 
order the vertices in decreasing order of degree,
and assign each vertex the minimum-label color not used by its neighbors. During the challenge
Gitastrophe attempted to use different orderings for the greedy algorithm,
such as sorting by the slope of the line segment associated with each vertex (as the orientation greedy initialization presented in Section \ref{s:LucPascal}),
and also tried numerous other strategies.
Ultimately, after running the solution optimizer for approximately the same amount of time,
all initializations resulted in an equal number of colors.

\subsection{Modifications to the Conflict Optimizer}
Taking inspiration from memetic algorithms,
which alternate between an intensification and a diversification stage,
the algorithm continually switched between a phase using the above conflict score,
and one minimizing only the number of conflicts. Thus during the conflict-minimization phase, the random variables $f(C_j)$ and $w(u)$ are both fixed equal to $1$ leading to a conflict score  \[ score(C_j)=\sum_{u \in C_j, (u, v) \in E(G')} 1.\]
Each phase lasted for $10^5$ iterations.
Adding the conflict-minimization phase gave minor improvements to some of the challenge instances.

\section{Shadoks} \label{s:Shadoks}

In this section, we describe the choices used by the Shadoks team for the options described in Section~\ref{ss:conflict}.

\paragraph{Option (a)} The Shadoks generally chose to eliminate the color with the smallest number of elements. However, if the multistart option is toggled on, then a random color is used each time.

\paragraph{Option (b)} The conflict set $S$ is stored in a queue. The Shadoks tried other strategies, but found that the queue gives the best results.

\begin{figure}[ht] 
  \centering
  \includegraphics[scale=.42]{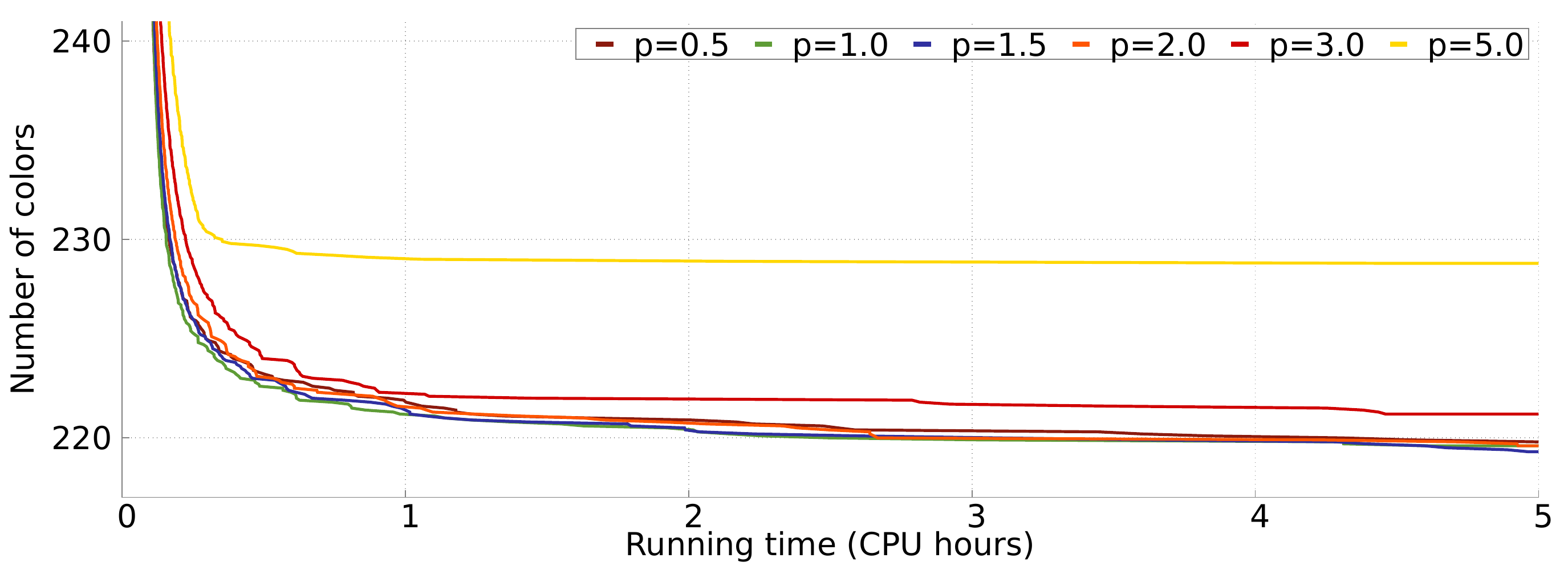}
  \caption{Number of colors over time for the instance \texttt{vispecn13806} using different values  $p$. The algorithm uses $\sigma = 0.15$, easy vertices, $q_{\max} = 59022$, but does not use the BDFS nor any clique.}
  \label{f:p}
\end{figure}

\paragraph{Option (c)} The weight function used is $w(u) = 1 + q(u)^p$, mostly with $p=1.2$. 
The effect of the parameter $p$ is shown in Fig.~\ref{f:p}.  Notice that in all figures, the number of colors shown is the average of ten executions of the code using different random seeds.

If $q(u)$ is larger than a threshold $q_\max$, the Shadoks set $w(u)=\infty$ so that the vertex $u$ never reenters $S$. If at some point an uncolored vertex $v$ is adjacent to some vertex $u$ of infinite weight in every color class, then the conflict optimizer is restarted. When restarting, the initial coloring is shuffled by moving some vertices from their initial color class to a new one. 

\begin{figure}[ht] 
  \centering
    \includegraphics[scale=.42]{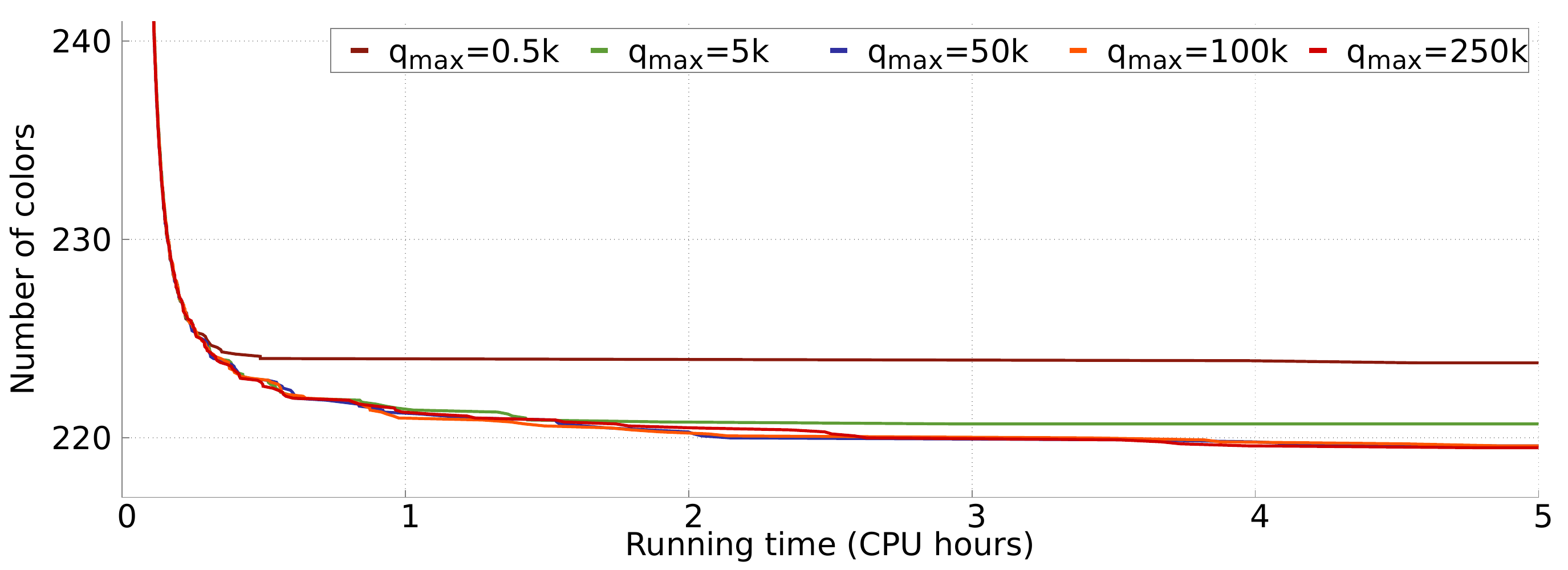}
  \caption{Number of colors over time with different values of $q_{\max}$ obtained on the instance \texttt{vispecn13806}. Parameters are $\sigma = 0.15$, $p=1.2$, no clique knowledge, and no BDFS.}
  \label{f:maxqueue}
\end{figure}

Looking at Fig.~\ref{f:maxqueue}, the value of $q_\max$ does not seem to have much influence as long as it is not too small. Throughout the challenge the Shadoks almost exclusively used
$q_{\max} = 2000 \cdot (75000/\edges)^2$, where $\edges$ is the number of vertices. This value roughly ensures a restart every few hours.

\begin{figure}[ht] 
  \centering
  \includegraphics[scale=.42]{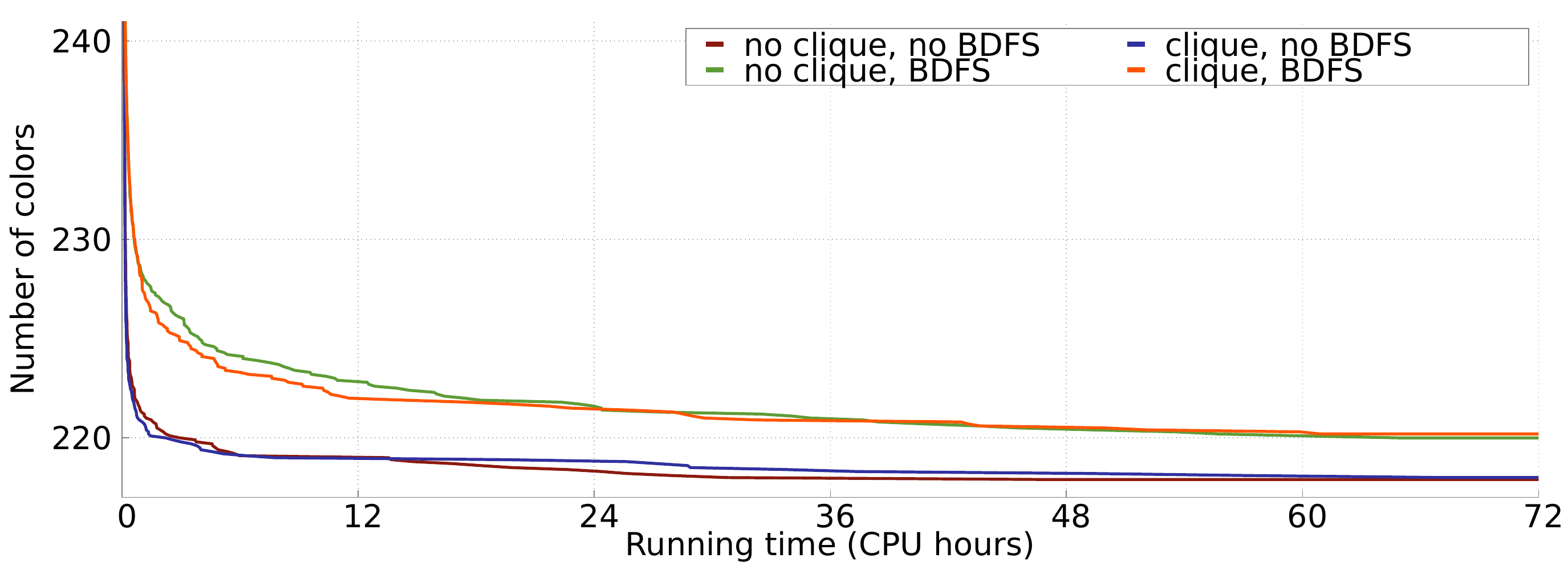}
  \caption{Number of colors over time with and without clique knowledge and BDFS obtained on the instance \texttt{vispecn13806}. Parameters are $\sigma = 0.15$, $p=1.2$, and $q_{\max} = 1 500 000$.}
  \label{f:dfs-info}
\end{figure}

If the clique option is toggled on, each vertex $u$ in the largest known clique has $w(u) = \infty$. The impact of the clique option on the computation is shown in Fig.~\ref{f:dfs-info}. The idea is that since each vertex of the clique must have a different color, it is useless to change their color. The algorithm works by recoloring the other vertices. During the challenge, the Shadoks used several methods to produce large cliques, including simulated annealing and mixed integer programming.

\begin{figure}[ht] 
  \centering
   \includegraphics[scale=.42]{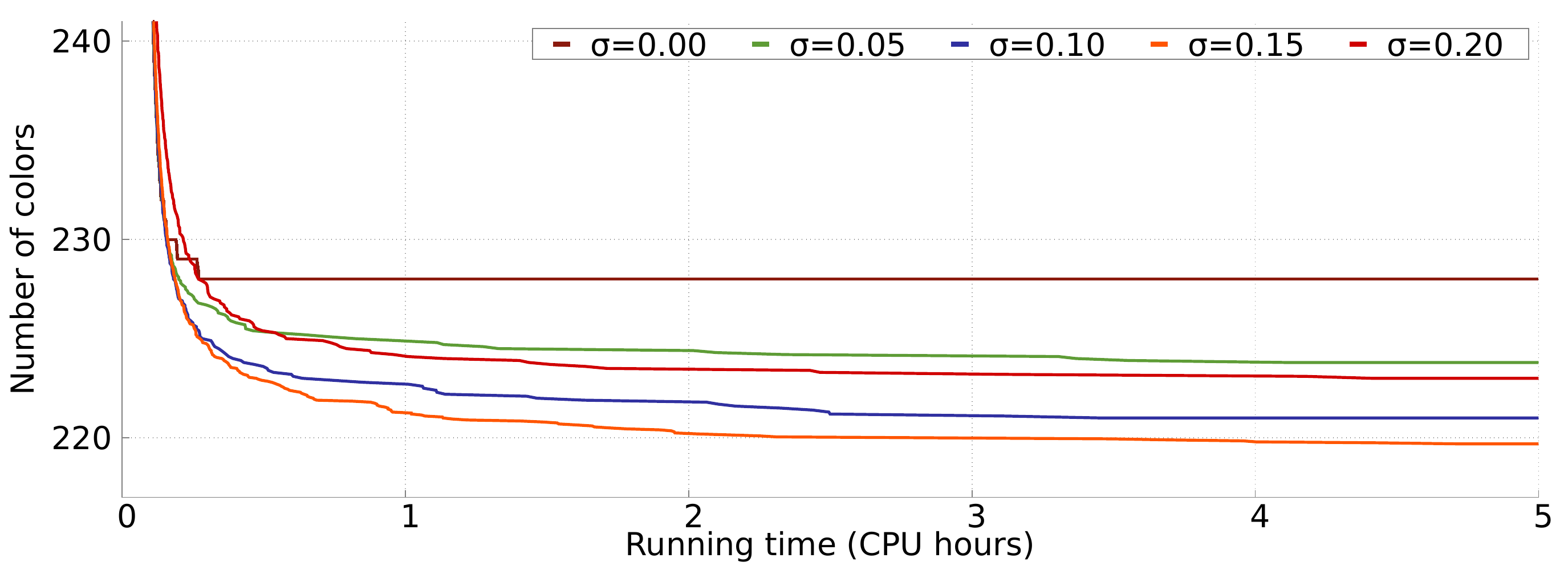}
  \caption{Number of colors over time for the instance \texttt{vispecn13806} for different values of $\sigma$. In both figures the algorithm uses $p=1.2$, easy vertices, $q_{\max} = 59022$, but does not use the BDFS nor any clique. For $\sigma \geq 0.25$, no solution better than 248 colors is found.
}
  \label{f:psigma}
\end{figure}

\paragraph{Option (d)} The Shadoks use the function $f$ as a Gaussian random variable of mean $1$ and variance $\sigma$. A good default value is $\sigma = 0.15$. The effect of the variance is shown in Fig.~\ref{f:psigma}. Notice that setting $\sigma = 0$ gives much worse results.

\paragraph{Option (e)} The goal of BDFS is to further optimize very good solutions that the conflict optimizer is not able to improve otherwise. Fig.~\ref{f:dfs-info} shows the influence of  BDFS. While on this figure, the advantages of BDFS cannot be noticed, its use near the end of the challenge improved about 30 solutions.

The \emph{bounded depth-first search} (BDFS) algorithm tries to improve the dequeuing process. The goal is to prevent a vertex in conflict with some adjacent colored vertices from entering in the conflict set. At the first level, the algorithm searches for a recoloring of some adjacent vertices which allows us to directly recolor the conflict vertex. If no solution is found, the algorithm could recolor some vertices at larger distances from the conflict vertex.
To do so, a local search  is performed by trying to recolor vertices at a bounded distance from the conflict vertex in the current partial solution. 

The BDFS algorithm has two parameters: \emph{adjacency bound} $a_{\max}$ and \emph{depth} $d$. In order to recolor a vertex $v$, BDFS gets the set $\mathcal{C}$ of color classes with at most $a_{\max}$ neighbors of $v$. If a class in $\mathcal{C}$ has no neighbor of $v$, $v$ is assigned to $C$. Otherwise, for each class $C \in \mathcal{C}$, BDFS tries to recolor the vertices in $C$ which are adjacent to $v$ by recursively calling itself with depth $d-1$. At depth $d=0$ the algorithm stops trying to color the vertices.

During the challenge the Shadoks used BDFS with parameters $a_{\max} = 3$ and $d = 3$. The depth was increased to $5$ (resp. $7$) when the number of vertices in the queue was $2$ (resp. $1$).

\paragraph{Degeneracy order} \label{ss:easy}
Given a target number of colors $\colors$, we call \emph{easy vertices} a set of vertices $Y$ such that, if the remainder of the vertices of $G'$ are colored using $\colors$ colors, then we are guaranteed to be able to color all vertices of $G'$ with $\colors$ colors. This is obtained using the degeneracy order $Y$.
To obtain $Y$ we iteratively remove from the graph a vertex $v$ that has at most $\colors - 1$ neighbors, appending $v$ to the end of $Y$. We repeat until no other vertex can be added to $Y$. Notice that, once we color the remainder of the graph with at least $\colors$ colors, we can use a greedy coloring for $Y$ in order from last to first without increasing the number of colors used. Removing the easy vertices reduces the total number of vertices, making the conflict optimizer more effective. The Shadoks always toggle this option on (the challenge instances contain from $0$ to $23\%$ easy vertices).

\section{Results} \label{s:results}

We provide the results of the experiments performed with the code from the three teams on two classes of instances. First, we present the results on some selected CG:SHOP 2022 instances. These instances are intersection graphs of line segments. Second, we execute the code on graphs that are not intersection graphs, namely the classic DIMACS graphs~\cite{dimacs}, comparing the results of our conflict optimizer implementations to previous solutions. The source code for the three teams is available at:

\begin{itemize}
    \item Lasa: \url{https://github.com/librallu/dogs-color}
    \item Gitastrophe: \url{ https://github.com/jacketsj/cgshop2022-gitastrophe}
    \item Shadoks: \url{https://github.com/gfonsecabr/shadoks-CGSHOP2022}
\end{itemize}

\subsection{CG:SHOP 2022 Instances}

We selected 14 instances (out of 225) covering the different types of instances given in the CG:SHOP 2022 challenge. The results are presented in Table~\ref{tab:results}. For comparison, we executed the HEAD~\cite{head} code on some instances using the default parameters. The table shows the smallest number of colors for which HEAD found a solution. We ran HEAD for 1 hour of repetitions for each target number of colors on a single CPU core (the HEAD solver takes the target number of colors as a parameter and we increased this parameter one by one).
At the end of the challenge, 8 colorings computed by Lasa, 11 colorings computed by Gitastrophe, and 23 colorings computed by Shadoks over 225 instances have been proved optimal (their number of colors is equal to the size of a clique).

\begin{table}[ht]
\caption{Several CG:SHOP 2022 results. We compare the size of the largest known clique to the smallest coloring found by each team on a selection of 14 CG:SHOP 2022 instances.}
\label{tab:results}
\centering
\begin{tabular}{|l | c | c | c | c | c | c |}
\hline
Instance & Clique & Best & HEAD~\cite{head} & \hspace{0.48cm}Lasa\hspace{0.5cm} & Gitastrophe & \hspace{0.23cm}Shadoks\hspace{0.25cm} \\
\hline
\texttt{rvisp5013}       &  46 &  \textbf{49} & 59 & 50  & \textbf{49}  & \textbf{49}  \\
\texttt{rsqrpecn8051}    & 173 & \textbf{175} & 207 & 177 & 176 & \textbf{175} \\
\texttt{vispecn13806}    & 77  & \textbf{218} & 283 & 224 & 221 & \textbf{218} \\
\texttt{rsqrp14364}      & 134 & \textbf{136} & 174 & 137 & 137 & \textbf{136} \\
\texttt{vispecn19370}    & 169 & \textbf{192} & 266 & 197 & 194 & \textbf{192} \\
\texttt{rvisp24116}      &  97 & \textbf{104} & 166 & 110 & 105 & \textbf{104} \\
\texttt{visp26405}       &  78 & \textbf{81} & 112 & 83  & \textbf{81}  & \textbf{81}  \\
\texttt{sqrp28863}       & \textbf{190} & \textbf{190} & 297 & 191 & 191 & \textbf{190} \\
\texttt{visp38574}       & 118 & \textbf{133} & 199 & 138 & 134 & \textbf{133} \\ 
\texttt{sqrpecn45700}    & 460 & \textbf{462} & & 465 & 465 & \textbf{462} \\
\texttt{reecn51526}      & 308 & \textbf{310} & & 315 & 312 & \textbf{310} \\
\texttt{vispecn58391}    & 305 & \textbf{367} & & 380 & 369 & \textbf{367} \\
\texttt{vispecn65831}    & 357 & \textbf{439} & & 453 & 440 & \textbf{439} \\
\texttt{sqrp72075}       & 264 & \textbf{269} & & 272 & 271 & \textbf{269}\\
\hline
\end{tabular}
\end{table}

\begin{figure}[ht] 
  \centering
   \includegraphics[width=0.8\textwidth]{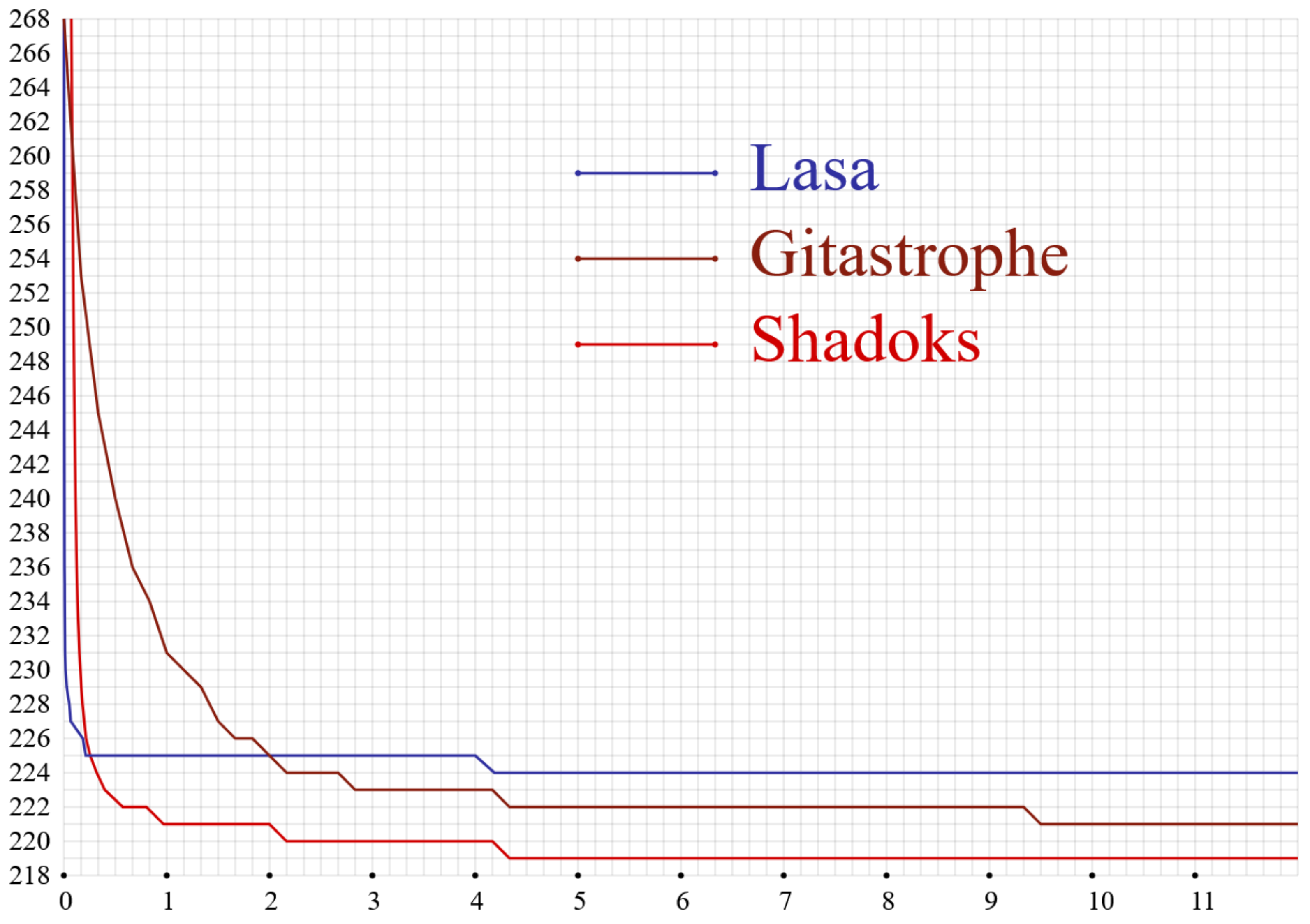}
  \caption{Number of colors over time (in hours) for the instance \texttt{vispecn13806}.}
  \label{f:comparison}
\end{figure}

In order to compare the efficiency of the algorithms, we executed the different implementations on the CG:SHOP instance \texttt{vispecn13806}. The edge density of this graph is 19\%, the largest clique that we found has 177 vertices and the best coloring found during the challenge uses 218 colors. Notice that \texttt{vispecn13806} is the same instance used in other Shadoks experiments in Section~\ref{s:Shadoks}. Notice also that HEAD algorithm provides 283 colors after one hour compared to less than 240 colors for the conflict optimizers.
We ran the three implementations on three different servers and compared the results shown in Figure~\ref{f:comparison}. For each implementation, the $x$ coordinate is the running time in hours, while the $y$ coordinate is the smallest number of colors found at that time.

\subsection{Results on DIMACS Graphs}

We tested the implementation of each team on the DIMACS instances~\cite{dimacs} to gauge the performance of the conflict optimizer on other classes of graphs. We compared our results to the best known bounds and to the state of the art coloring algorithms HEAD~\cite{head} and QACOL~\cite{qacol,qacol2}.

The time limit for Lasa's algorithms is 1 hour. CWLS is Lasa's conflict optimizer with the neighbourhood presented in TABUCOL~\cite{tabucol}, while PWLS is the optimizer with the neighbourhood presented in PARTIALCOL~\cite{blochliger2008graph}.
Gitastrophe  algorithm ran 10 minutes after which the number of colors no longer decreases. Shadoks algorithm ran for 1 hour without the BDFS option (results with BDFS are worse).

Results are presented in Table~\ref{tab:dimacs}. We only kept the difficult DIMACS instances. For the other instances, all the results match the best known bounds.
The DIMACS instances had comparatively few edges (on the order of thousands or millions);
the largest intersection graphs considered in the CG:SHOP challenge had over 1.5 billion edges.

\begin{table}[ht]
\caption{Comparison of our method with state-of-the-art graph coloring algorithms. The conflict optimizer underperforms except on the geometric graphs \texttt{r*} and \texttt{dsjr*}.}
\label{tab:dimacs}
\centering
\begin{tabular}{|l|c|c|c|c|c|c|c|c|}
\hline
Instance & Best & HEAD & QACOL & Lasa & Lasa & Gitastrophe & Shadoks  \\
Instance & & \cite{head} & \cite{qacol,qacol2} & CWLS & PWLS & &   \\
\hline
\texttt{dsjc250.5  }  &\textbf{28}    & \textbf{28} & \textbf{28}   & \textbf{28}  & 29  & 29  & \textbf{28} \\
\texttt{dsjc500.1  }  &\textbf{12}    & \textbf{12} & \textbf{12}   & 13  & 13  & 13  & 13 \\
\texttt{dsjc500.5  }  &\textbf{47}    & \textbf{47} & 48   & 49  & 51  & 52  & 50 \\
\texttt{dsjc500.9  }  &\textbf{126}   & \textbf{126} & \textbf{126} & \textbf{126} & 130 & 130 & 128 \\
\texttt{dsjc1000.1 }  &\textbf{20}    & \textbf{20} & \textbf{20}   & 21  &  22 & 21  & 21 \\
\texttt{dsjc1000.5 }  &\textbf{82}    & \textbf{82} & \textbf{82}   & 89   &  94  & 93  & 91 \\
\texttt{dsjc1000.9 }  &\textbf{222}   & \textbf{222} & \textbf{222} & 223 & 240 & 235 & 231 \\
\texttt{r250.5     }  &\textbf{65}    & \textbf{65} & \textbf{65}   & \textbf{65}  & \textbf{65}  & \textbf{65}  & \textbf{65} \\
\texttt{r1000.1c   }  &\textbf{98}    & \textbf{98} & \textbf{98}   & \textbf{98}  & \textbf{98}  & \textbf{98}  & \textbf{98} \\
\texttt{r1000.5    }  &\textbf{234}   & 245 & 238 & \textbf{234} & \textbf{234} & \textbf{234} & 237 \\
\texttt{dsjr500.1c }  &\textbf{84}    & 85 & 85   & 85 & 85 & 85  & 85 \\
\texttt{dsjr500.5  }  &\textbf{122}   & - & \textbf{122}   & \textbf{122}& \textbf{122}& \textbf{122} & \textbf{122} \\
\texttt{le450\_25c }  &\textbf{25}    & \textbf{25} & \textbf{25}   & 26 & 26 & 26  & 26 \\
\texttt{le450\_25d }  &\textbf{25}    & \textbf{25} & \textbf{25}  & 26 & 26 & 26  & 26 \\
\texttt{flat300\_28\_0} &28  & 31 & 31   & 31 & 32 & 33  & 32 \\
\texttt{flat1000\_50\_0} &\textbf{50} & \textbf{50} & -    & \textbf{50} & \textbf{50} & 91  & 54 \\
\texttt{flat1000\_60\_0} &\textbf{60} & \textbf{60} & -    & \textbf{60} & 92 & 93  & 90 \\
\texttt{flat1000\_76\_0} &\textbf{81} & \textbf{81} & \textbf{81}   & 88 & 93 & 92  & 90 \\
\texttt{C2000.5}        &\textbf{145}  & 146 & \textbf{145} & 165 & 173 & 173 & 168 \\
\texttt{C4000.5}        &\textbf{260} & 266 & 259 & 311 & 320 & 317 & 312 \\
\hline
\end{tabular}
\end{table}

We notice that the conflict optimizer works extremely poorly on random graphs, but it is fast and appears to perform well on geometric graphs (r250.5, r1000.1c, r1000.5, dsjr500.1c and dsjr500.5), matching the best-known results~\cite{dlmcol}.
Interestingly, these geometric graphs are not intersection graphs as in the CG:SHOP challenge, but are generated based on a distance threshold.
On the DIMACS graphs, Lasa implementation shows better performance than the other implementations.

\section{Acknowledgments}

We would like to thank the challenge organizers and other competitors for their time, feedback, and making this whole event possible.

The Shadoks would like to thank Hélène Toussaint, Raphaël Amato, Boris Lonjon, and William Guyot-Lénat from LIMOS, as well as the Qarma and TALEP teams and Manuel Bertrand from LIS, who continue to make the computational resources of the LIMOS and LIS clusters available to our research.

The work of Loïc Crombez has been sponsored by the French government research program ``Investissements d'Avenir'' through the IDEX-ISITE initiative 16-IDEX-0001 (CAP 20-25).
The work of Guilherme D. da Fonseca is supported by the French ANR PRC grant ADDS (ANR-19-CE48-0005).
The work of Yan Gerard is supported by the French ANR PRC grants ADDS (ANR-19-CE48-0005), ACTIVmap (ANR-19-CE19-0005) and by the French government IDEX-ISITE initiative 16-IDEX-0001 (CAP 20-25).
The work of Aldo Gonzalez-Lorenzo is supported by the French ANR PRC grant COHERENCE4D (ANR-20-CE10-0002).
The work of Pascal Lafourcade is supported by the French ANR PRC grant MobiS5 (ANR-18-CE39-0019), DECRYPT (ANR-18-CE39-0007), SEVERITAS (ANR-20-CE39-0005) and by the French government IDEX-ISITE initiative 16-IDEX-0001 (CAP 20-25).
The work of Luc Libralesso is supported by the French ANR PRC grant DECRYPT (ANR-18-CE39-0007).


\bibliographystyle{plainurl}
\bibliography{references}

\end{document}